\documentclass[fleqn,usenatbib]{mnras}

\usepackage[margin=5pt]{subfig}
\usepackage{comment}
\usepackage{longtable}
\usepackage{caption}
\usepackage{threeparttable}
\usepackage{booktabs}
\usepackage[figuresright]{rotating}
\usepackage{textcase}
\newcommand{\be}{\begin{equation}}
\newcommand{\ee}{\end{equation}}
\newcommand{\bea}{\begin{eqnarray}}
\newcommand{\eea}{\end{eqnarray}}
\newcommand{\hunit}{$\rm{km \ s^{-1} \ Mpc^{-1}}$}
\newcommand{\lcdm}{$\Lambda$CDM}
\newcommand{\pcdm}{$\phi$CDM}
\usepackage{array}
\makeatletter
\newcommand{\thickhline}{%
    \noalign {\ifnum 0=`}\fi \hrule height 1pt
    \futurelet \reserved@a \@xhline
}
\newcolumntype{"}{@{\hskip\tabcolsep\vrule width 1pt\hskip\tabcolsep}}
\makeatother

\newcommand{\hii}{H\,\textsc{ii}}
\newcommand{\Om}{\Omega_{\rm m0}}
\newcommand{\Ok}{\Omega_{\rm k0}}
\newcommand{\wX}{w_{\rm X}}
\newcommand{\om}{$\Omega_{\rm m0}$}
\newcommand{\ok}{$\Omega_{\rm k0}$}
\newcommand{\wx}{$w_{\rm X}$}
\newcommand{\obh}{\Omega_{b}h^2}
\newcommand{\och}{\Omega_{c}h^2}
\newcommand{\onh}{\Omega_{\nu}h^2}
\newcommand{\obhs}{$\Omega_{b}h^2$}
\newcommand{\ochs}{$\Omega_{c}h^2$}
\newcommand{\onhs}{$\Omega_{\nu}h^2$}

\usepackage[T1]{fontenc}
\usepackage{scalerel}
\usepackage{tikz}
\usetikzlibrary{svg.path}

\definecolor{orcidlogocol}{HTML}{A6CE39}
\tikzset{
  orcidlogo/.pic={
    \fill[orcidlogocol] svg{M256,128c0,70.7-57.3,128-128,128C57.3,256,0,198.7,0,128C0,57.3,57.3,0,128,0C198.7,0,256,57.3,256,128z};
    \fill[white] svg{M86.3,186.2H70.9V79.1h15.4v48.4V186.2z}
                 svg{M108.9,79.1h41.6c39.6,0,57,28.3,57,53.6c0,27.5-21.5,53.6-56.8,53.6h-41.8V79.1z M124.3,172.4h24.5c34.9,0,42.9-26.5,42.9-39.7c0-21.5-13.7-39.7-43.7-39.7h-23.7V172.4z}
                 svg{M88.7,56.8c0,5.5-4.5,10.1-10.1,10.1c-5.6,0-10.1-4.6-10.1-10.1c0-5.6,4.5-10.1,10.1-10.1C84.2,46.7,88.7,51.3,88.7,56.8z};
  }
}

\newcommand\orcidicon[1]{\href{https://orcid.org/#1}{\mbox{\scalerel*{
\begin{tikzpicture}[yscale=-1,transform shape]
\pic{orcidlogo};
\end{tikzpicture}
}{|}}}}

\usepackage{hyperref} 

\DeclareRobustCommand{\VAN}[3]{#2}
\let\VANthebibliography\thebibliography
\def\thebibliography{\DeclareRobustCommand{\VAN}[3]{##3}\VANthebibliography}

\usepackage{graphicx}	
\usepackage{amsmath}	
\usepackage{amssymb}	
\usepackage{fnpct}

\title[GRB cosmological parameter constraints]{Standardizing Platinum Dainotti-correlated gamma-ray bursts, and using them with standardized Amati-correlated gamma-ray bursts to constrain cosmological model parameters}

 \author[Cao, Dainotti, \& Ratra]{
 Shulei Cao$^{\orcidicon{0000-0003-2421-7071}}$,$^{1}$\thanks{E-mail: shulei@phys.ksu.edu}
 Maria Dainotti$^{\orcidicon{0000-0003-4442-8546}}$,$^{2,3}$\thanks{E-mail: maria.dainotti@nao.ac.jp}
 Bharat Ratra$^{\orcidicon{0000-0002-7307-0726}1}$\thanks{E-mail: ratra@phys.ksu.edu}
 \\
 $^{1}$Department of Physics, Kansas State University, 116 Cardwell Hall, Manhattan, KS 66506, USA\\
 $^{2}$National Astronomical Observatory of Japan, 2-21-1 Osawa, Tokyo 181-8588, Japan\\
 $^{3}$Space Science Institute, Boulder, CO 80301, USA\\
 }

\date{Accepted XXX. Received YYY; in original form ZZZ}

\pubyear{2021}

\begin{document}
\label{firstpage}
\pagerange{\pageref{firstpage}--\pageref{lastpage}}
\maketitle

\begin{abstract}
We show that the Platinum gamma-ray burst (GRB) data compilation, probing the redshift range $0.553 \leq z \leq 5.0$, obeys a cosmological-model-independent three-parameter fundamental plane (Dainotti) correlation and so is standardizable. While they probe the largely unexplored $z \sim 2.3-5$ part of cosmological redshift space, the GRB cosmological parameter constraints are consistent with, but less precise than, those from a combination of baryon acoustic oscillation (BAO) and Hubble parameter [$H(z)$] data. In order to increase the precision of GRB-only cosmological constraints, we exclude common GRBs from the larger Amati-correlated A118 data set composed of 118 GRBs and jointly analyze the remaining 101 Amati-correlated GRBs with the 50 Platinum GRBs. This joint 151 GRB data set probes the largely unexplored $z \sim 2.3-8.2$ region; the resulting GRB-only cosmological constraints are more restrictive, and consistent with, but less precise than, those from $H(z)$ + BAO data.
\end{abstract}


\begin{keywords}
cosmological parameters -- dark energy -- cosmology: observations -- gamma-ray bursts
\end{keywords}


\section{Introduction} \label{sec:intro}

Currently accelerated cosmological expansion and other cosmological observations are reasonably well accommodated in the spatially-flat \lcdm\ model \citep{peeb84} with $\sim 70\%$ of the current cosmological energy budget being a time-independent cosmological constant ($\Lambda$), $\sim 25\%$ being non-relativistic cold dark matter (CDM), and most of the remaining $\sim 5\%$ being non-relativistic baryonic matter \citep[see, e.g.][]{Farooq_Ranjeet_Crandall_Ratra_2017, scolnic_et_al_2018, planck2018b, eBOSS_2020}. In this paper we also study cosmological models with a little spatial curvature or dynamical dark energy, since the observations do not rule out such models, and since some sets of measurements seem mutually incompatible when analyzed in the spatially-flat \lcdm\ model \citep[see, e.g.][]{DiValentinoetal2021a, PerivolaropoulosSkara2021}. 

It is still unclear whether this incompatibility is evidence against the spatially-flat \lcdm\ model or is caused by unidentified systematic errors in one of the established cosmological probes or by evolution of the parameters themselves with the redshift \citep{Dainottietal2021a, Dainottietal2022}. Newer, alternate cosmological probes could help alleviate this issue. Recent examples of such probes include reverberation-mapped quasar (QSO) measurements that reach to redshift $z \sim 1.9$ \citep{Czernyetal2021, Zajaceketal2021, Yuetal2021, Khadkaetal_2021a, Khadkaetal2021c}, \hii\ starburst galaxy measurements that reach to $z \sim 2.4$ \citep{Mania_2012, Chavez_2014, G-M_2019, GM2021, Caoetal_2020, Caoetal_2021c, Caoetal_2021a, Johnsonetal2021, Mehrabietal2022}, QSO angular size measurements that reach to $z \sim 2.7$ \citep{Cao_et_al2017a, Ryanetal2019, Caoetal_2020, Caoetal_2021a, Zhengetal2021, Lian_etal_2021}, QSO flux measurements that reach to $z \sim 7.5$ \citep{RisalitiLusso2015, RisalitiLusso2019, KhadkaRatra2020a, KhadkaRatra2020b, KhadkaRatra2021, KhadkaRatra2022, Lussoetal2020, Yangetal2020, ZhaoXia2021, Lietal2021, Lian_etal_2021, Rezaeietal2021, Luongoetal2021},\footnote{The most recent \cite{Lussoetal2020} QSO flux compilation assumes a UV--X-ray correlation model that is invalid above a significantly lower redshift, $z \sim 1.5-1.7$, so these QSOs can only be used to derive lower-$z$ cosmological constraints \citep{KhadkaRatra2021, KhadkaRatra2022}.} and the main subject of this paper, gamma-ray burst (GRB) measurements that reach to $z \sim 8.2$ \citep{Amati2008, Amati2019, CardoneCapozzielloDainotti2009, Cardoneetal2010, samushia_ratra_2010, Dainottietal2013a, Dainottietal2013b, Postnikovetal2014, Wangetal2015, Wang_2016, Wangetal_2021, Dirirsa2019, KhadkaRatra2020c, Huetal_2021, Daietal_2021, Demianskietal_2021, Khadkaetal_2021b, Luongoetal2021, galaxies9040077, Caoetal_2021a}[Dainotti et al. 2011a,b, Apj, Mnras]. Some of these probes might eventually allow for a reliable extension of the Hubble diagram to $z \sim 3-4$, well beyond the reach of Type Ia supernovae. GRBs have been detected to $z \sim 9.4$ \citep{Cucchiaraetal2011}, and might be detectable to $z = 20$ \citep{LambReichart2000}, so in principle GRBs could act as a cosmological probe to higher redshifts than 8.2.

\citet{CaoKhadkaRatra2021} recently used the A118 two-parameter Amati-correlated GRB data set \citep{Khadkaetal_2021b} and the MD-LGRB, GW-LGRB, and MD-SGRB two-parameter Dainotti-correlated GRB data sets \citep{Wangetal_2021, Huetal_2021} to constrain cosmological parameters.\footnote{The Amati correlation relates the rest-frame peak photon energy and the rest-frame isotropic radiated energy \citep{Amati2008} and the two-dimensional Dainotti correlation relates the luminosity at the end of the plateau phase and the rest-frame end-time of plateau emission \citep{Dainottietal2008, Dainottietal2010, Dainottietal2011, Dainottietal2013a, Dainottietal2015, Dainottietal2017}. As discussed below, in this paper we use three-dimensional Dainotti and two-dimensional Amati correlation GRBs.} The circularity problem was circumvented by simultaneously constraining cosmological model parameters and GRB correlation parameters (see \citealp{Dainottietal2013b} for a more extended discussion) and the cosmological-model-independence of the GRB correlation parameters shows that these GRBs are standardizable \citep{KhadkaRatra2020c, CaoKhadkaRatra2021}. Not only does the simultaneous fitting method circumvent the circularity problem, it also allows for the derivation of unbiased GRB-only constraints (unlike the constraints derived from GRBs that have been calibrated by using other data, that are correlated with the calibrating data), that can be straightforwardly used to compare with other constraints derived from other data, such as $H(z)$ + BAO data as we have done here. The A118, MD-LGRB, GW-LGRB, and MD-SGRB GRBs provide cosmological constraints that are mostly compatible with those determined using better-established cosmological probes \citep{CaoKhadkaRatra2021}.

Here, we use the new Platinum compilation of 50 long GRBs, spanning $0.553 \leq z \leq 5.0$, that obey the three-parameter fundamental plane (Dainotti) correlation between the peak prompt luminosity, the luminosity at the end of the plateau emission and its rest frame duration \citep{Dainottietal2016,Dainottietal2017,Dainottietal2020} to constrain cosmological model parameters and GRB correlation parameters. The platinum sample is listed in Table \ref{tab:P50} of Appendix \ref{sec:appendix}. For this data set, measured quantities for a GRB are redshift $z$, characteristic time scale $T^{*}_{X}$, which marks the end of the plateau emission, the measured $\gamma$-ray energy flux $F_{X}$ at $T^{*}_{X}$ and the prompt peak flux $F_{\rm peak}$ over a 1 s interval, and X-ray spectral index of the plateau phase $\beta^{\prime}$. This sample spans the redshift range $0.553 \leq z \leq 5.0$. We find that the Platinum GRBs are standardizable through the Dainotti correlation and they also provides cosmological parameter constraints compatible with those from better-established cosmological probes, as well as with those derived from A118 GRB data. We also combine the Platinum data set with the 101 non-overlapping Amati-correlated GRBs (A101) from the A118 data set to perform a joint (Platinum + A101) analysis. We find that this joint GRB data set provides slightly more restrictive cosmological constraints (in which the twice-as-large A101 data set dominates the statistics and so plays a more dominant role) that are consistent with those from a combined analysis of baryon acoustic oscillation (BAO) and Hubble parameter [$H(z)$] data. However, the cosmological constraints from Platinum, A118, A101, and Platinum + A101 data are less restrictive (precise) than those from $H(z)$ + BAO data because there are more parameters to be constrained in the GRB cases but not yet enough precise-enough data points to determine more precise GRB constraints. A joint analysis of the $H(z)$ + BAO + Platinum + A101 data results in slightly more restrictive cosmological constraints relative to those from just $H(z)$ + BAO data.

Our paper is organized as follows. We summarize the cosmological models we use in Sec.\ \ref{sec:model} and outline the data sets adopted in Sec.\ \ref{sec:data}. We then describe our analyses methods in Sec.\ \ref{sec:analysis} and discuss results in Sec.\ \ref{sec:results}. We summarize our conclusions in Sec.\ \ref{sec:conclusion}.

\section{Cosmological models}
\label{sec:model}

We constrain cosmological model parameters and GRB correlation parameters in six spatially-flat and non-flat dark energy cosmological models.\footnote{For recent constraints on spatial curvature see \citet{73}, \citet{Chenetal2016}, \citet{Ranaetal2017}, \citet{Oobaetal2018a, Oobaetal2018b}, \citet{Yuetal2018}, \citet{ParkRatra2019a, ParkRatra2019b}, \citet{Wei2018}, \citet{DESCollaboration2019}, \citet{Lietal2020}, \citet{Handley2019}, \citet{EfstathiouGratton2020}, \citet{DiValentinoetal2021b}, \citet{Vagnozzietal2020, Vagnozzietal2021}, \citet{KiDSCollaboration2021}, \citet{ArjonaNesseris2021}, \citet{Dhawanetal2021}, \citet{Renzietal2021}, \citet{Gengetal2021}, and references therein.} The essential cosmological quantity for constraining purposes, for data we use, is the Hubble parameter, $H(z, \textbf{\emph{p}})\equiv H_0E(z, \textbf{\emph{p}})$, with $H_0$ being the Hubble constant and $E(z, \textbf{\emph{p}})$ the expansion rate as a function of redshift $z$ and the cosmological parameters $\textbf{\emph{p}}$. Here we consider one massive and two massless neutrino species, with the effective number of relativistic neutrino species $N_{\rm eff} = 3.046$ and the total neutrino mass $\sum m_{\nu}=0.06$ eV. The non-relativistic neutrino physical energy density parameter is $\onh=\sum m_{\nu}/(93.14\ \rm eV)$, where $h$ is the reduced Hubble constant in units of 100 \hunit. With the baryonic (\obhs) and cold dark matter (\ochs) physical energy density parameters as free cosmological parameters to be constrained, the derived non-relativistic matter density parameter is therefore $\Om = (\onh + \obh + \och)/{h^2}$.

In the flat and non-flat \lcdm\ models, the expansion rate function
\be
\label{eq:EzL}
    E(z) = \sqrt{\Om\left(1 + z\right)^3 + \Ok\left(1 + z\right)^2 + \Omega_{\Lambda}},
\ee
where $\Omega_{\Lambda} = 1 - \Om - \Ok$ is the cosmological constant dark energy density parameter and $\Ok$ is the curvature energy density parameter. In the non-flat \lcdm\ model the free parameters being constrained are $H_0$, \obhs\!, \ochs\!, and \ok, whereas in the flat \lcdm\ model $\Ok = 0$ is implied.

In the flat and non-flat XCDM parametrizations, 
\be
\label{eq:EzX}
    E(z) = \sqrt{\Om\left(1 + z\right)^3 + \Ok\left(1 + z\right)^2 + \Omega_{\rm X}\left(1 + z\right)^{3\left(1 + \wX\right)}},
\ee
where \wx\ and $\Omega_{\rm X} = 1 - \Om - \Ok$ are the equation of state parameter and the dynamical dark energy density parameter of the X-fluid, respectively. In the non-flat XCDM parameterization the free parameters being constrained are $H_0$, \obhs\!, \ochs\!, \ok, and \wx\, whereas in the flat XCDM parametrization $\Ok = 0$ is implied.

In the flat and non-flat \pcdm\ models \citep{peebrat88,ratpeeb88,pavlov13}\footnote{For recent constraints on \pcdm\ see \cite{chen_etal_2017}, \cite{Zhaietal2017}, \cite{ooba_etal_2018b, ooba_etal_2019}, \cite{park_ratra_2018, park_ratra_2019b, park_ratra_2020}, \cite{Sangwanetal2018}, \cite{SolaPercaulaetal2019}, \citet{Singhetal2019}, \cite{UrenaLopezRoy2020}, \cite{SinhaBanerjee2021}, \cite{Xuetal2021}, \cite{deCruzetal2021}, \cite{Jesusetal2021}, and references therein.},
\be
\label{eq:Ezp}
    E(z) = \sqrt{\Om\left(1 + z\right)^3 + \Ok\left(1 + z\right)^2 + \Omega_{\phi}(z,\alpha)},
\ee
where the scalar field, $\phi$, dynamical dark energy density parameter
\be
\label{Op}
\Omega_{\phi}(z,\alpha)=\frac{1}{6H_0^2}\bigg[\frac{1}{2}\dot{\phi}^2+V(\phi)\bigg],
\ee
and can be numerically computed by solving the Friedmann equation \eqref{eq:Ezp} and the equation of motion of the scalar field
\be
\label{em}
\ddot{\phi}+3H\dot{\phi}+V'(\phi)=0.
\ee 
We assume an inverse power-law scalar field potential energy density 
\be
\label{PE}
V(\phi)=\frac{1}{2}\kappa m_p^2\phi^{-\alpha}.
\ee
In these equations an overdot and a prime denote a derivative with respect to time and $\phi$, respectively, $m_p$ is the Planck mass, $\alpha$ is a positive constant, and the constant $\kappa$ is determined by the shooting method in the Cosmic Linear Anisotropy Solving System (\textsc{class}) code \citep{class}. In the non-flat \pcdm\ model the free parameters being constrained are $H_0$, \obhs\!, \ochs\!, \ok, and $\alpha$, whereas in the flat \pcdm\ model $\Ok = 0$ is implied.

\section{Data}
\label{sec:data}

In this paper we analyze two different GRB data sets as well as combinations of them and the joint $H(z)$ + BAO data set. These data sets are summarized in Table \ref{tab:data} and described below.\footnote{In this table and elsewhere, for compactness, we sometimes use Plat. as an abbreviation for the Platinum data set.}

\begin{itemize}

\item[]{\bf Platinum sample}. This includes 50 long GRBs which exhibit a plateau phase with an angle $< 41^\circ$, that do not have a flare during the plateau, and have a plateau with duration longer than 500 s. The first criterion is based on evidence that the plateau angles are Gaussianly distributed and those with angle $> 41^\circ$ are outliers; the second criterion allows one to eliminate cases contaminated by the presence of flaring activity; and, the third criterion allows one to eliminate cases where prompt emission may mask the plateau to the point that the definition of the plateau is uncertain \citep{Willingaleetal2007, Willingaleetal2010}. As discussed below, the Platinum GRBs obey the three-dimensional Dainotti relation. The platinum sample is listed in Table \ref{tab:P50} of Appendix \ref{sec:appendix}. For this data set, measured quantities for a GRB are redshift $z$, characteristic time scale $T^{*}_{X}$, which marks the end of the plateau emission, the measured $\gamma$-ray energy flux $F_{X}$ at $T^{*}_{X}$ and the prompt peak flux $F_{\rm peak}$ over a 1 s interval, and X-ray spectral index of the plateau phase $\beta^{\prime}$. This sample spans the redshift range $0.553 \leq z \leq 5.0$.

\item[]{\bf A118 sample}. This sample include 118 long GRBs, listed in Table 7 of \cite{Khadkaetal_2021b}, that obey the two-dimensional Amati relation. For this data set, measured quantities for a GRB are $z$, rest-frame spectral peak energy $E_{\rm p}$, and measured bolometric fluence $S_{\rm bolo}$, computed in the standard rest-frame energy band $1-10^4$ keV. This sample spans the redshift range $0.3399 \leq z \leq 8.2$. 

\item[]{\bf A101 sample}. The A118 data and the Platinum data have 17 common GRBs that are listed in the footnote of Table \ref{tab:data}. We exclude these common GRBs from the A118 data set to form the A101 data set for joint analyses with the Platinum data set. This sample spans the redshift range $0.3399 \leq z \leq 8.2$.

\item[]{\bf Platinum + A101 sample}. This combination GRB sample includes 151 GRBs. This sample spans the redshift range $0.3399 \leq z \leq 8.2$.

\item[]{$\textbf{ \emph{H(z)}}$ \bf and BAO data}. There are 31 $H(z)$ and 11 BAO measurements that have a redshift range $0.07 \leq z \leq 1.965$ and $0.0106 \leq z \leq 2.33$, respectively. The $H(z)$ data are in Table 2 of \cite{Ryan_1} and the BAO data are in Table 1 of \cite{Caoetal_2021b}. We compare cosmological constraints from $H(z)$ + BAO data with those obtained from the GRB data sets, and also jointly analyze GRB and $H(z)$ + BAO data.

\end{itemize}

\begin{table}
\centering
\resizebox{\columnwidth}{!}{%
\begin{threeparttable}
\caption{Summary of data sets used.}
\label{tab:data}
\begin{tabular}{lcc}
\toprule
Data set & $N$ (Number of points) & Redshift range\\
\midrule
Platinum & 50 & $0.553 \leq z \leq 5.0$ \\
A118 & 118 & $0.3399 \leq z \leq 8.2$ \\
A101\tnote{a} & 101 & $0.3399 \leq z \leq 8.2$ \\
Plat. + A101 & 151 & $0.3399 \leq z \leq 8.2$ \\
\midrule
$H(z)$ & 31 & $0.070 \leq z \leq 1.965$ \\
BAO & 11 & $0.38 \leq z \leq 2.334$ \\
\bottomrule
\end{tabular}
\begin{tablenotes}[flushleft]
\item [a] Excluding from A118 those GRBs in common with Platinum [060418, 080721, 081008, 090418(A), 091020, 091029, 110213(A), 110818(A), 111008(A), 120811C, 120922(A), 121128(A), 131030A, 131105A, 140206A, 150314A, and 150403A].
\end{tablenotes}
\end{threeparttable}%
}
\end{table}

\section{Data Analysis Methodology}
\label{sec:analysis}

\begin{table}
\centering
\resizebox{\columnwidth}{!}{%
\begin{threeparttable}
\caption{Flat priors of the constrained parameters.}
\label{tab:priors}
\begin{tabular}{lcc}
\toprule
Parameter & & Prior\\
\midrule
 & Cosmological Parameters & \\
\midrule
$H_0$\tnote{a} &  & [None, None]\\
\obhs\,\tnote{b} &  & [0, 1]\\
\ochs\,\tnote{c} &  & [0, 1]\\
\ok &  & [-2, 2]\\
$\alpha$ &  & [0, 10]\\
\wx &  & [-5, 0.33]\\
\midrule
 & GRB Correlation Parameters & \\
\midrule
$a$ &  & [-5, 5]\\
$b$ &  & [-5, 5]\\
$C_{o}$ &  & [-50, 50]\\
$\sigma_{\rm int}$ &  & [0, 5]\\
$\beta$ &  & [0, 5]\\
$\gamma$ &  & [0, 300]\\
\bottomrule
\end{tabular}
\begin{tablenotes}[flushleft]
\item [a] \hunit. In all four GRB-only analyses, $H_0$ is set to be 70 \hunit, while in other cases the prior range is irrelevant (unbounded).
\item [b] In all four GRB-only analyses, \obhs\ is set to be 0.0245, i.e. $\Omega_{b}=0.05$.
\item [c] In all four GRB-only analyses, the $\Omega_{c}$ range is adjusted to ensure $\Om\in[0,1]$.
\end{tablenotes}
\end{threeparttable}%
}
\end{table}

Luminosity distance, $D_L$, as a function of $z$ and cosmological parameters $\textbf{\emph{p}}$, is given by
\begin{equation}
  \label{eq:DL}
\resizebox{0.475\textwidth}{!}{%
    $D_L(z, \textbf{\emph{p}}) = 
    \begin{cases}
    \frac{c(1+z)}{H_0\sqrt{\Omega_{\rm k0}}}\sinh\left[\frac{\sqrt{\Omega_{\rm k0}}H_0}{c}D_C(z, \textbf{\emph{p}})\right] & \text{if}\ \Omega_{\rm k0} > 0, \\
    \vspace{1mm}
    (1+z)D_C(z, \textbf{\emph{p}}) & \text{if}\ \Omega_{\rm k0} = 0,\\
    \vspace{1mm}
    \frac{c(1+z)}{H_0\sqrt{|\Omega_{\rm k0}|}}\sin\left[\frac{H_0\sqrt{|\Omega_{\rm k0}|}}{c}D_C(z, \textbf{\emph{p}})\right] & \text{if}\ \Omega_{\rm k0} < 0,
    \end{cases}$%
    }
\end{equation}
where the comoving distance is
\begin{equation}
\label{eq:gz}
   D_C(z, \textbf{\emph{p}}) = c\int^z_0 \frac{dz'}{H(z', \textbf{\emph{p}})},
\end{equation}
$c$ is the speed of light, and $H(z, \textbf{\emph{p}})$ is the Hubble parameter that is described in Sec. \ref{sec:model} for each cosmological model.

For Platinum GRBs the X-ray source rest-frame luminosity $L_{X}$, time $T^{*}_{X}$ at the end of the plateau emission, and the peak prompt luminosity $L_{\rm peak}$ are correlated through the three-parameter fundamental plane relation \citep{Dainottietal2016, Dainottietal2017, Dainottietal2020, Dainottietal2021} 
\begin{equation}
    \label{eq:3D}
    \log L_{X} = C_{o}  + a\log T^{*}_{X} + b\log L_{\rm peak},
\end{equation}
where 
\be
\label{eq:Lx}
    L_{X}=\frac{4\pi D_L^2}{(1+z)^{1-\beta^{\prime}}}F_{X},
\ee
\be
\label{eq:Lpeak}
    L_{\rm peak}=\frac{4\pi D_L^2}{(1+z)^{1-\beta^{\prime}}}F_{\rm peak},
\ee
$C_{o}$ is the intercept parameter, and $a$ and $b$ are the slope parameters, with all three to be determined from data. $F_{X}$ and $F_{\rm peak}$ are the measured $\gamma$-ray energy flux (erg cm$^{-2}$ s$^{-1}$) at $T^{*}_{X}$ and in the peak of the prompt emission over a 1 s interval, respectively, and $\beta^{\prime}$ is the X-ray spectral index of the plateau phase.

We compute $L_{X}$ and $L_{\rm peak}$ as functions of cosmological parameters $\textbf{\emph{p}}$ at the redshift of each GRB by using eqs.\ \eqref{eq:DL}, \eqref{eq:Lx}, and \eqref{eq:Lpeak}. We then compute the natural log of the likelihood function \citep{D'Agostini_2005}
\be
\label{eq:LH_GRB}
    \ln\mathcal{L}_{\rm GRB}= -\frac{1}{2}\Bigg[\chi^2_{\rm GRB}+\sum^{N}_{i=1}\ln\left(2\pi\sigma^2_{\mathrm{tot},i}\right)\Bigg],
\ee
where
\be
\label{eq:chi2_GRB}
    \chi^2_{\rm GRB} = \sum^{N}_{i=1}\bigg[\frac{(\log L_{X,i} - C_{o}  - a\log T^{*}_{X,i} - b\log L_{\mathrm{peak},i})^2}{\sigma^2_{\mathrm{tot},i}}\bigg],
\ee
with
\be
\sigma^2_{\mathrm{tot},i}=\sigma_{\rm int,\,\textsc{p}}^2+\sigma_{{\log L_{X,i}}}^2+a^2\sigma_{{\log T^{*}_{X,i}}}^2+b^2\sigma_{{\log L_{\mathrm{peak},i}}}^2.
\ee
where $\sigma_{\rm int,\,\textsc{p}}$ is the Platinum GRB data intrinsic scatter parameter, that also contains the unknown systematic uncertainty, and $N$ is the number of data points.

For GRBs which obey the Amati correlation, a detailed description of the procedure can be found in Sec.\ 4 of \cite{CaoKhadkaRatra2021}. Here we denote its intrinsic scatter parameter as $\sigma_{\rm int,\,\textsc{a}}$ as opposed to the Platinum one, $\sigma_{\rm int,\,\textsc{p}}$.

Detailed descriptions of the $H(z)$ + BAO data analysis procedure can be found in Sec.\ 4 of \cite{Caoetal_2020} and \cite{Caoetal_2021b}.

The flat priors of the free parameters are listed in Table \ref{tab:priors}. The best-fitting values and posterior distributions of all free parameters are obtained through maximizing the likelihood functions using the Markov chain Monte Carlo (MCMC) code \textsc{MontePython} \citep{Brinckmann2019}, with the physics coded in \textsc{class}. The convergence of the MCMC chains for each free parameter is guaranteed by the Gelman-Rubin criterion ($R-1 < 0.05$). The \textsc{python} package \textsc{getdist} \citep{Lewis_2019} is used to compute the posterior means and uncertainties and plot the marginalized likelihood distributions and contours.

We use the Akaike Information Criterion ($AIC$) and the Bayesian Information Criterion ($BIC$) to compare the goodness of fit of models with different numbers of parameters. Their definitions can be found in \cite{CaoKhadkaRatra2021}. We also compare the goodness of fit using the DIC criterion \citep{KunzTrottaParkinson2006,Amati2019} defined as
\be
\label{eq:DIC}
DIC=-2\ln \mathcal{L}_{\rm max} + 2n_{\rm eff},
\ee
where $n_{\rm eff}=\langle-2\ln \mathcal{L}\rangle+2\ln \mathcal{L}_{\rm max}$ is the number of effectively constrained parameters with brackets representing the average over the posterior distribution. We compute $\Delta AIC$, $\Delta BIC$, and $\Delta DIC$ differences of the other five cosmological models with respect to the flat \lcdm\ reference model. Positive (negative) values of $\Delta AIC$, $\Delta BIC$, or $\Delta DIC$ indicate that the model under study fits the data worse (better) than does the reference model. Relative to the model with minimum $AIC(BIC/DIC)$, $\Delta AIC(BIC/DIC) \in (0, 2]$ is said to be weak evidence against the candidate model, $\Delta AIC(BIC/DIC) \in (2, 6]$ is positive evidence against the candidate model, while $\Delta AIC(BIC/DIC) \in (6, 10] $ is strong evidence against the candidate model, with $\Delta AIC(BIC/DIC)>10$ being very strong evidence against the candidate model.

\section{Results}
\label{sec:results}

\begin{figure*}
\centering
 \subfloat[]{%
    \includegraphics[width=0.5\textwidth,height=0.5\textwidth]{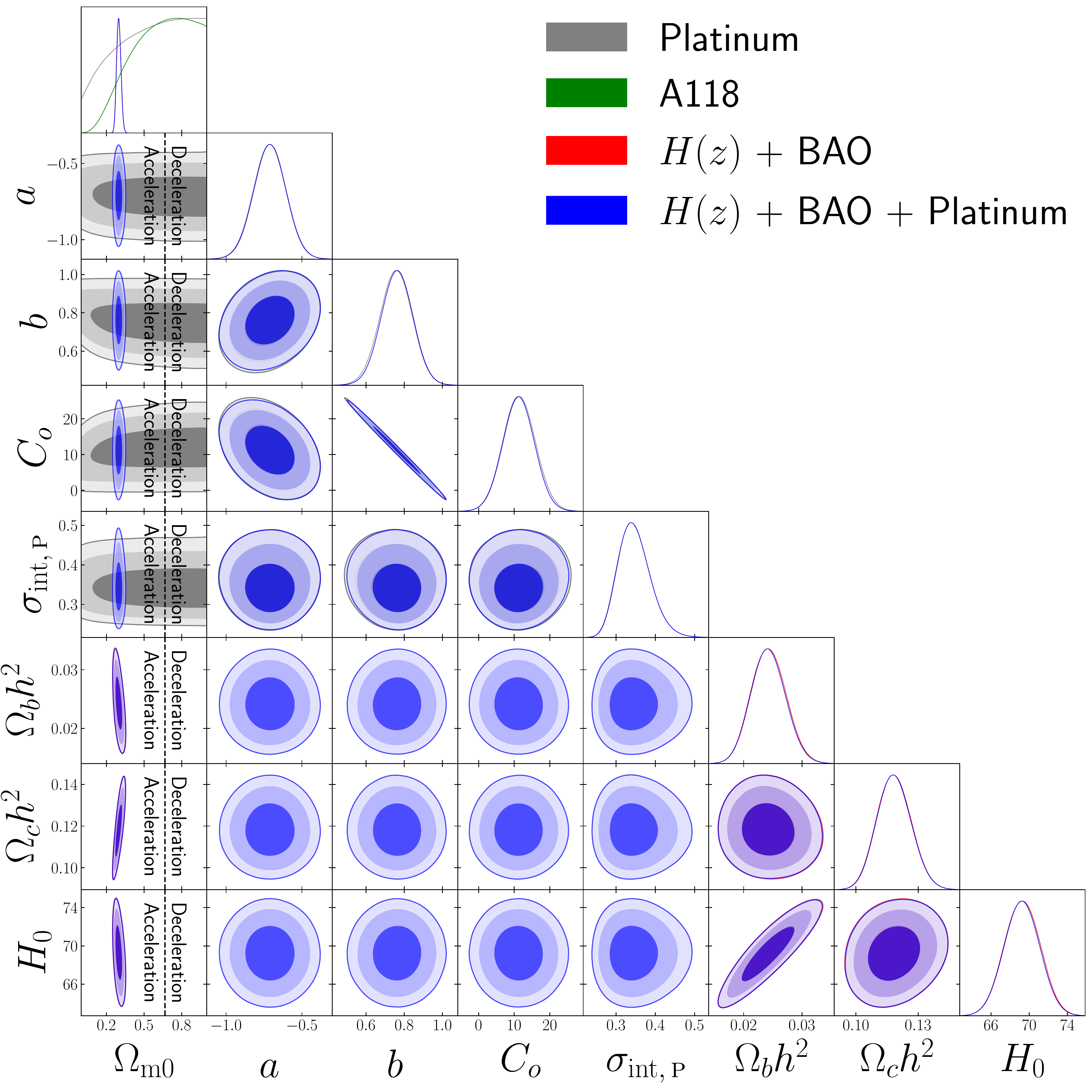}}
 \subfloat[]{%
    \includegraphics[width=0.5\textwidth,height=0.5\textwidth]{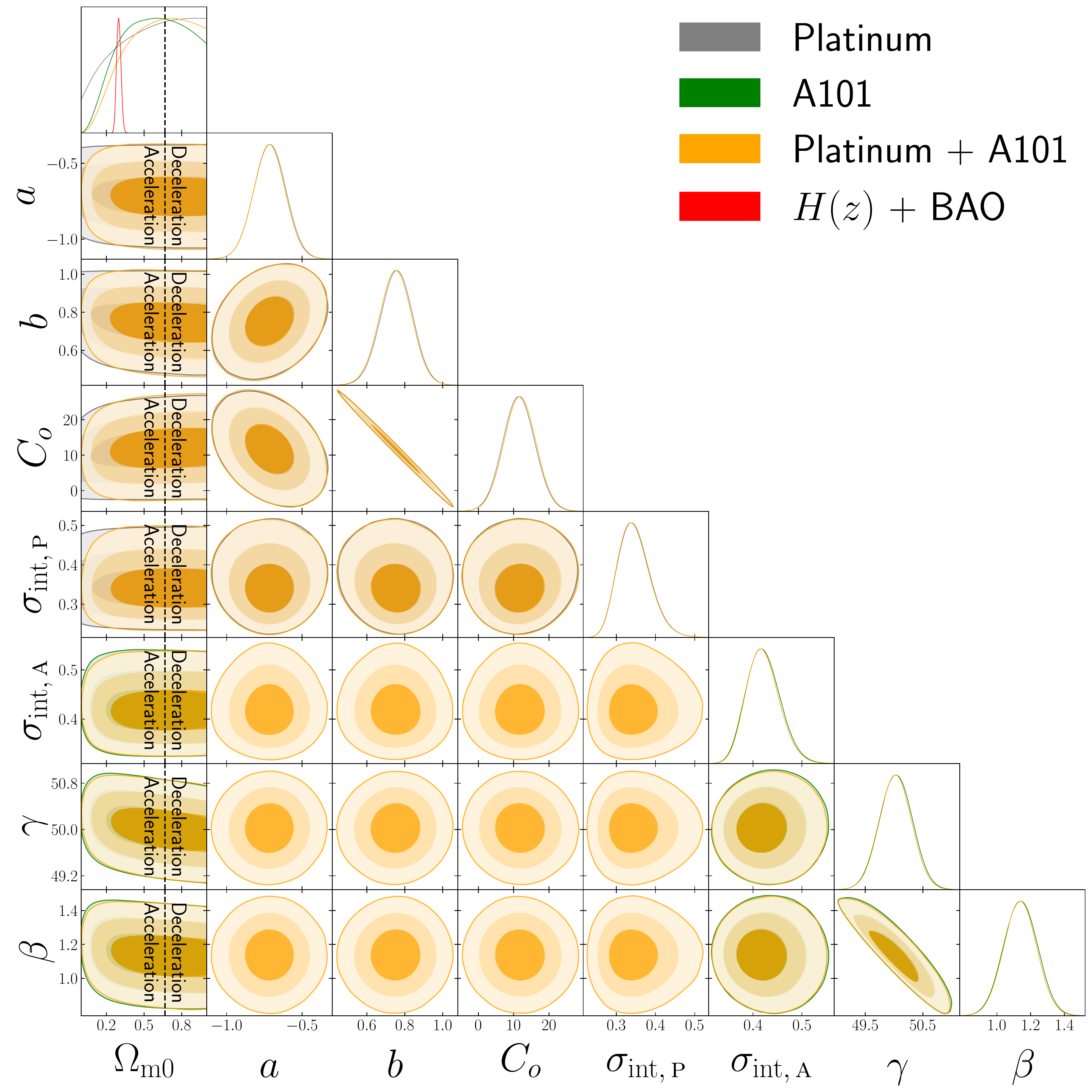}}\\
 \subfloat[]{%
    \includegraphics[width=0.5\textwidth,height=0.5\textwidth]{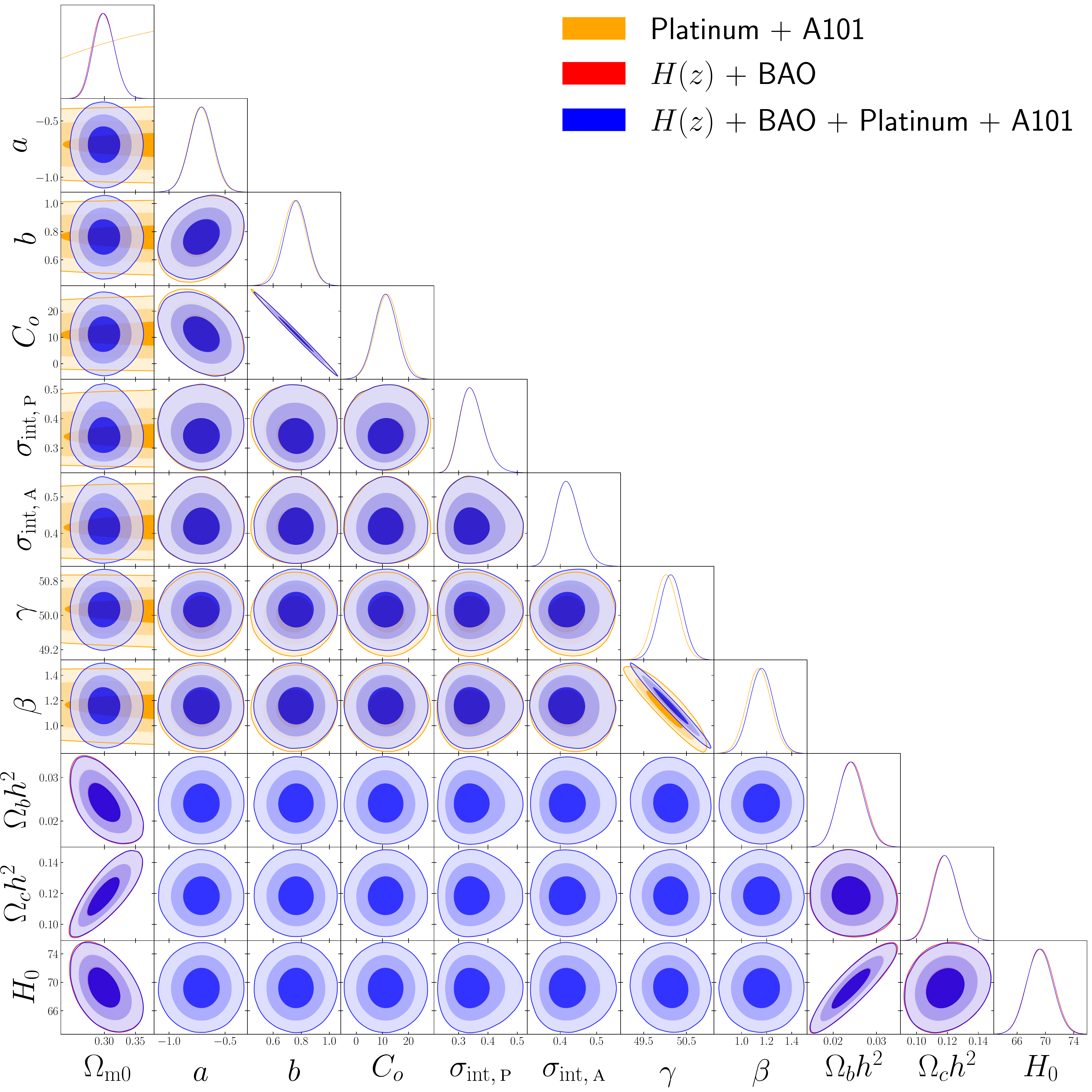}}
 \subfloat[Cosmological parameters zoom in]{%
    \includegraphics[width=0.5\textwidth,height=0.5\textwidth]{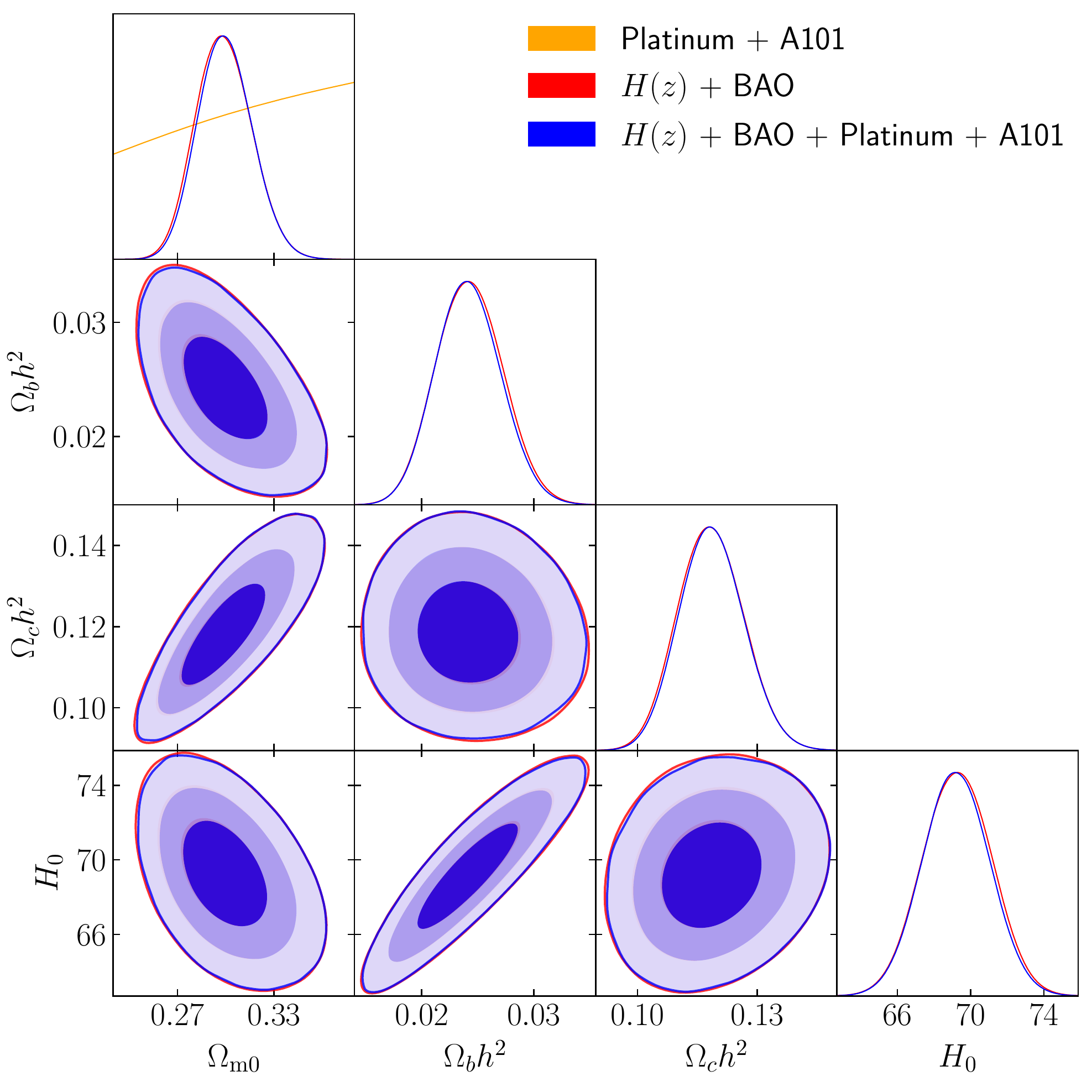}}\\
\caption{One-dimensional likelihood distributions and 1$\sigma$, 2$\sigma$, and 3$\sigma$ two-dimensional likelihood confidence contours for flat \lcdm\ from various combinations of data. The zero-acceleration black dashed lines in panels (a) and (b) divide the parameter space into regions associated with currently-accelerating (left) and currently-decelerating (right) cosmological expansion.}
\label{fig1}
\end{figure*}

\begin{figure*}
\centering
 \subfloat[]{%
    \includegraphics[width=0.5\textwidth,height=0.5\textwidth]{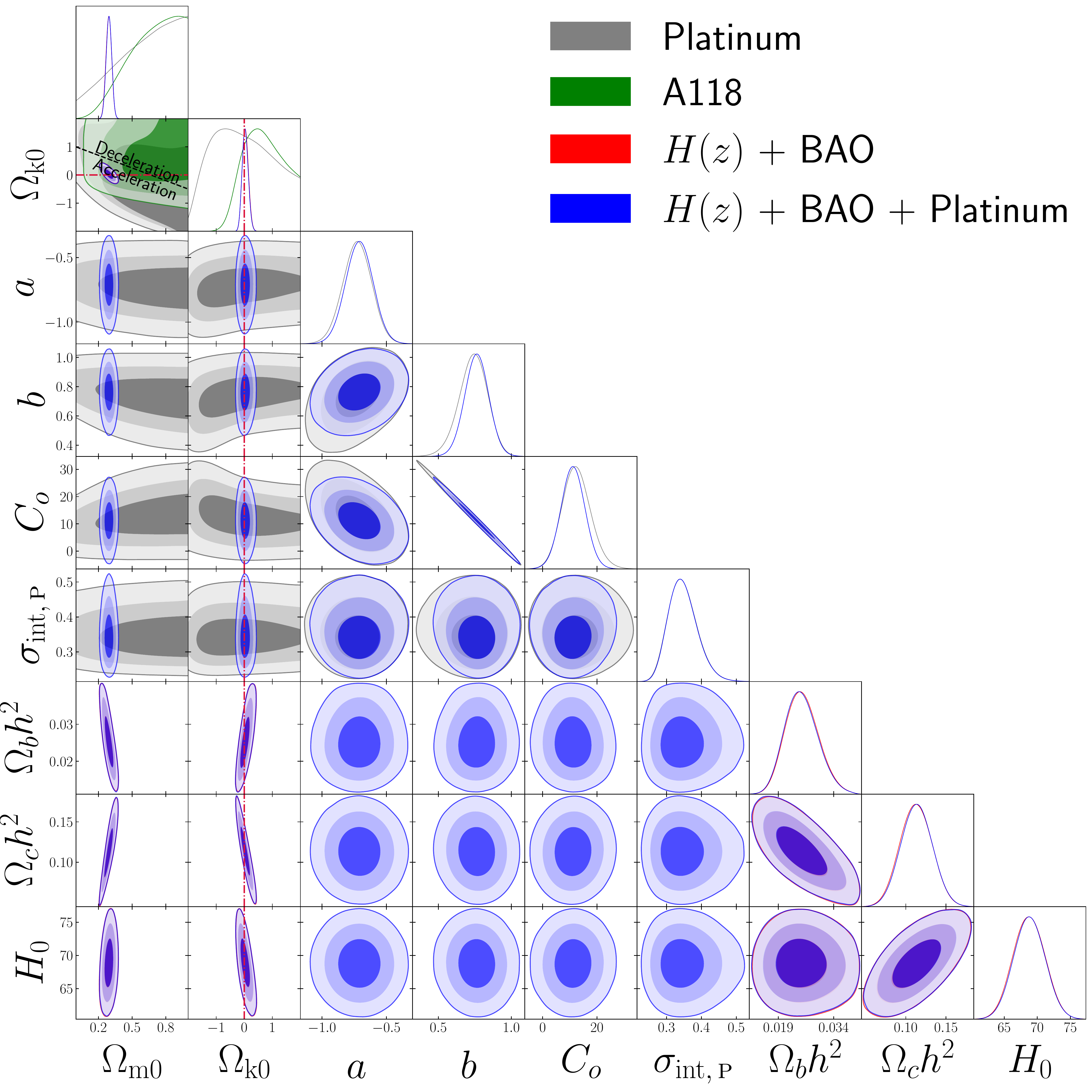}}
 \subfloat[]{%
    \includegraphics[width=0.5\textwidth,height=0.5\textwidth]{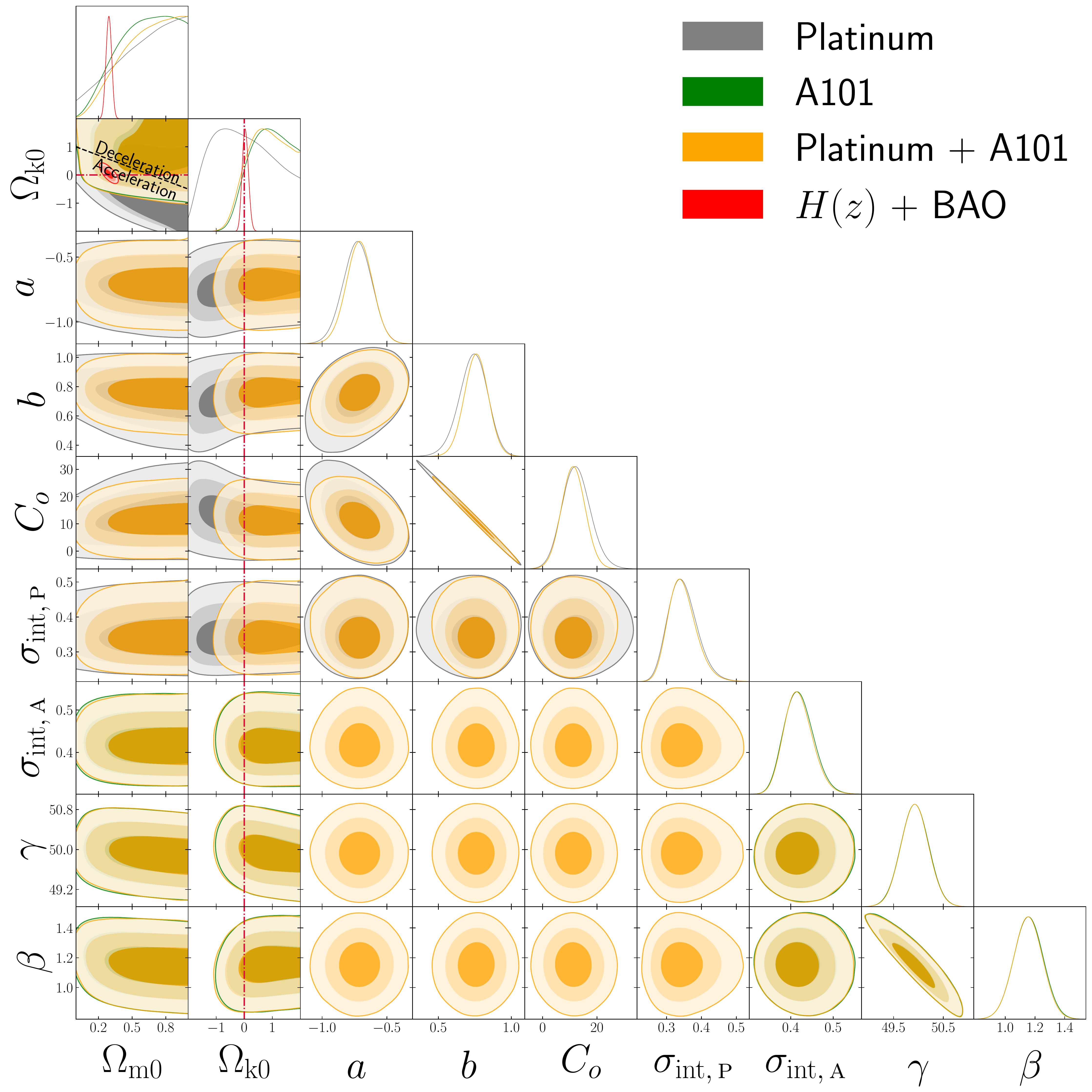}}\\
 \subfloat[]{%
    \includegraphics[width=0.5\textwidth,height=0.5\textwidth]{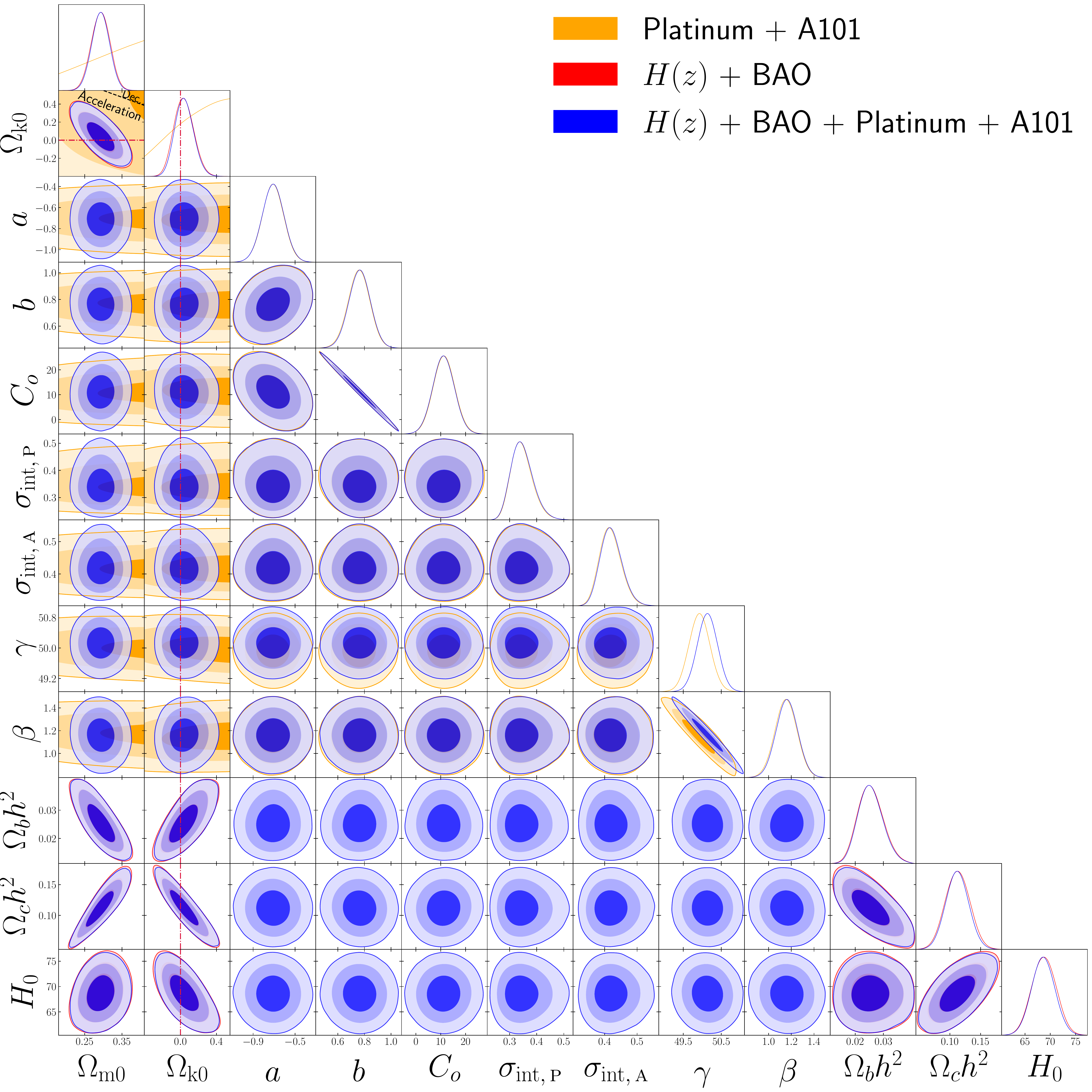}}
 \subfloat[Cosmological parameters zoom in]{%
    \includegraphics[width=0.5\textwidth,height=0.5\textwidth]{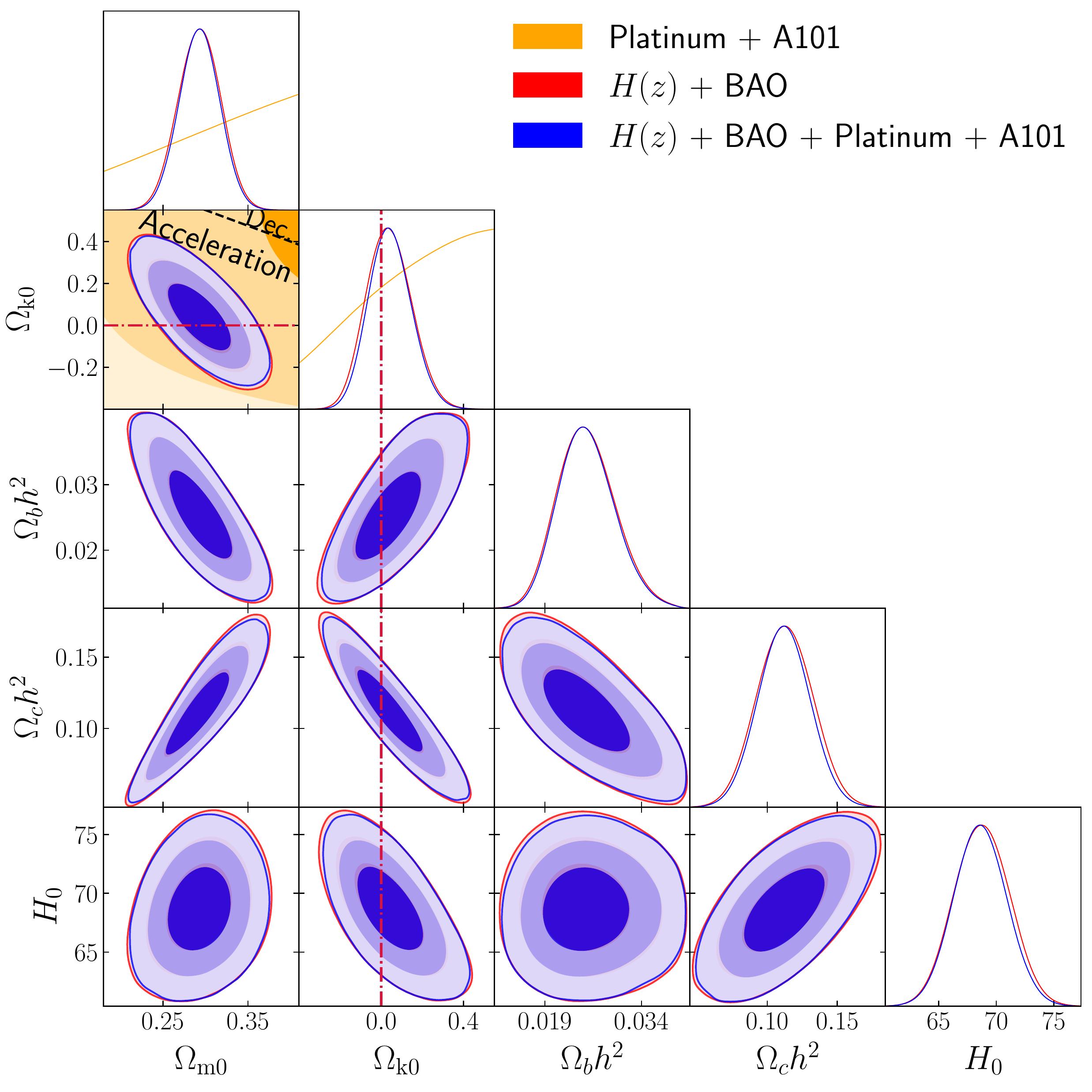}}\\
\caption{Same as Fig.\ \ref{fig1} but for non-flat \lcdm. The zero-acceleration black dashed lines divide the parameter space into regions associated with currently-accelerating (below left) and currently-decelerating (above right) cosmological expansion.}
\label{fig2}
\end{figure*}

\begin{figure*}
\centering
 \subfloat[]{%
    \includegraphics[width=0.5\textwidth,height=0.5\textwidth]{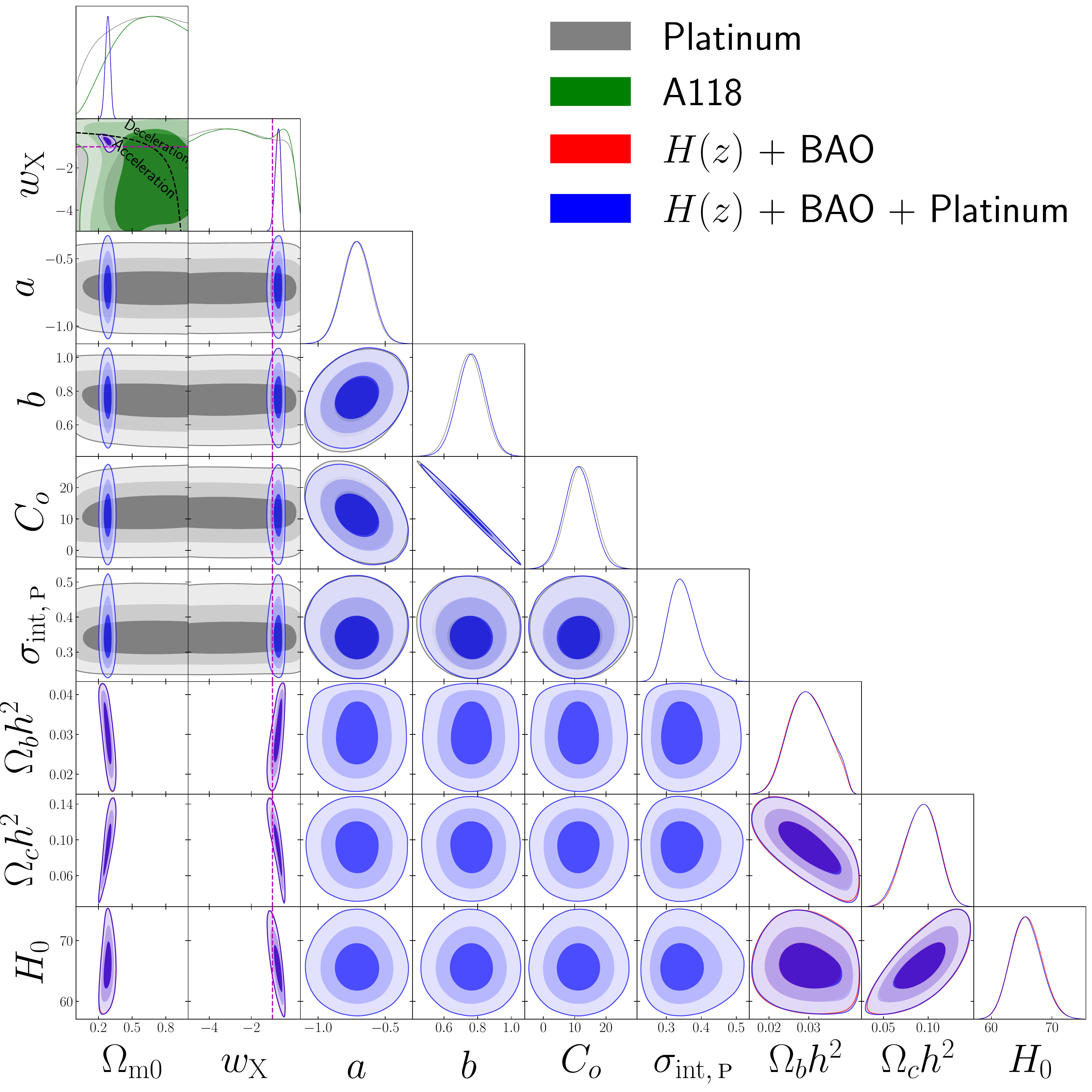}}
 \subfloat[]{%
    \includegraphics[width=0.5\textwidth,height=0.5\textwidth]{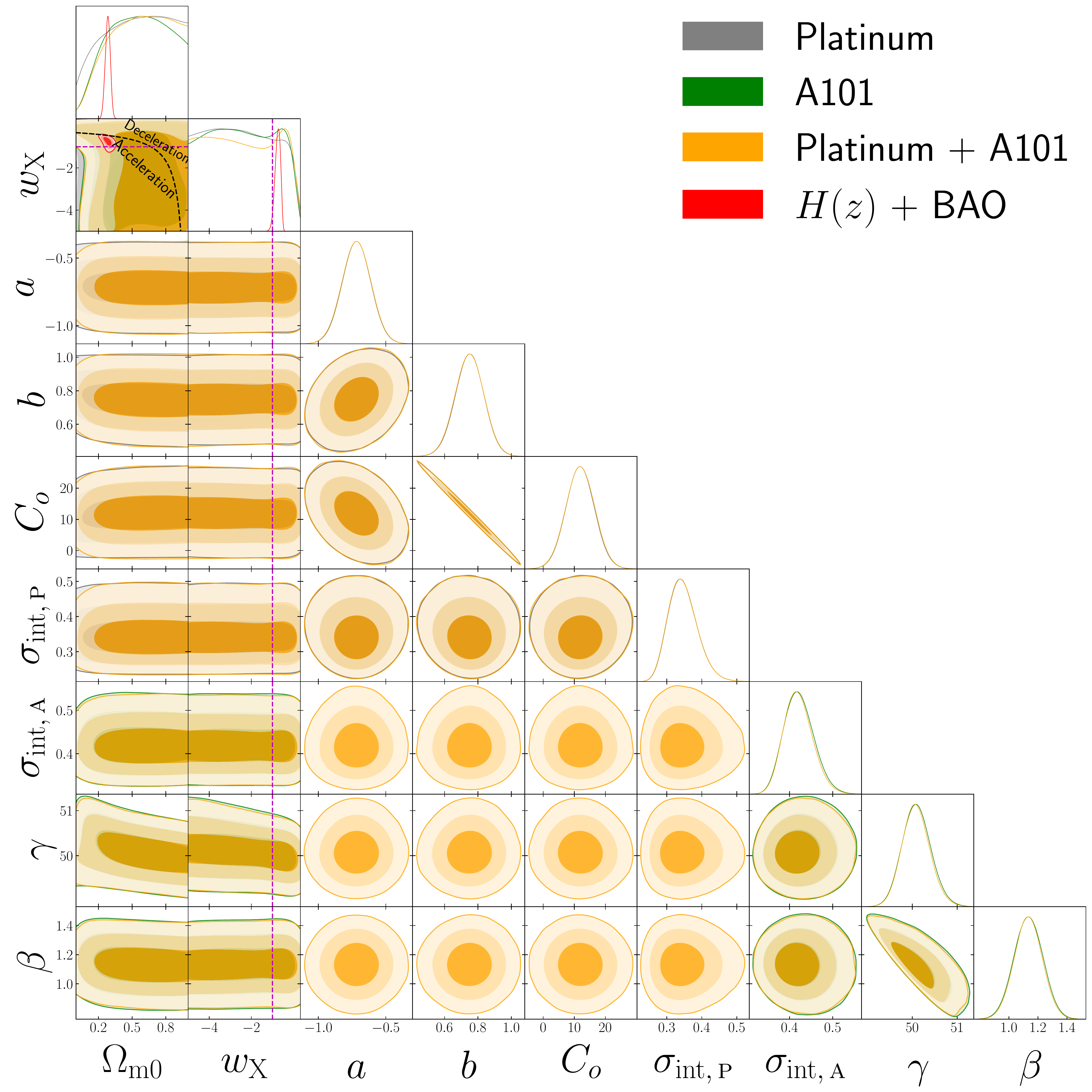}}\\
 \subfloat[]{%
    \includegraphics[width=0.5\textwidth,height=0.5\textwidth]{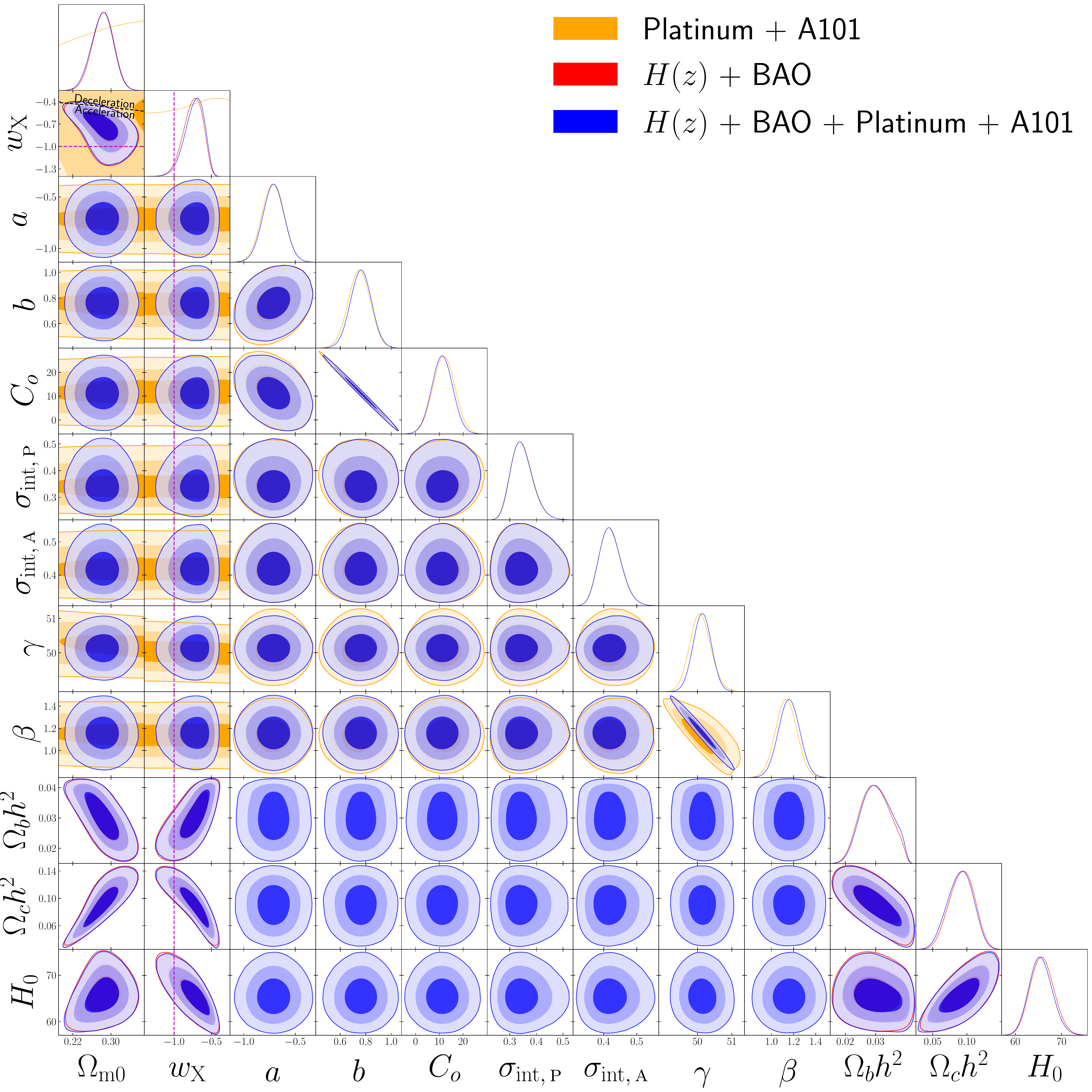}}
 \subfloat[Cosmological parameters zoom in]{%
    \includegraphics[width=0.5\textwidth,height=0.5\textwidth]{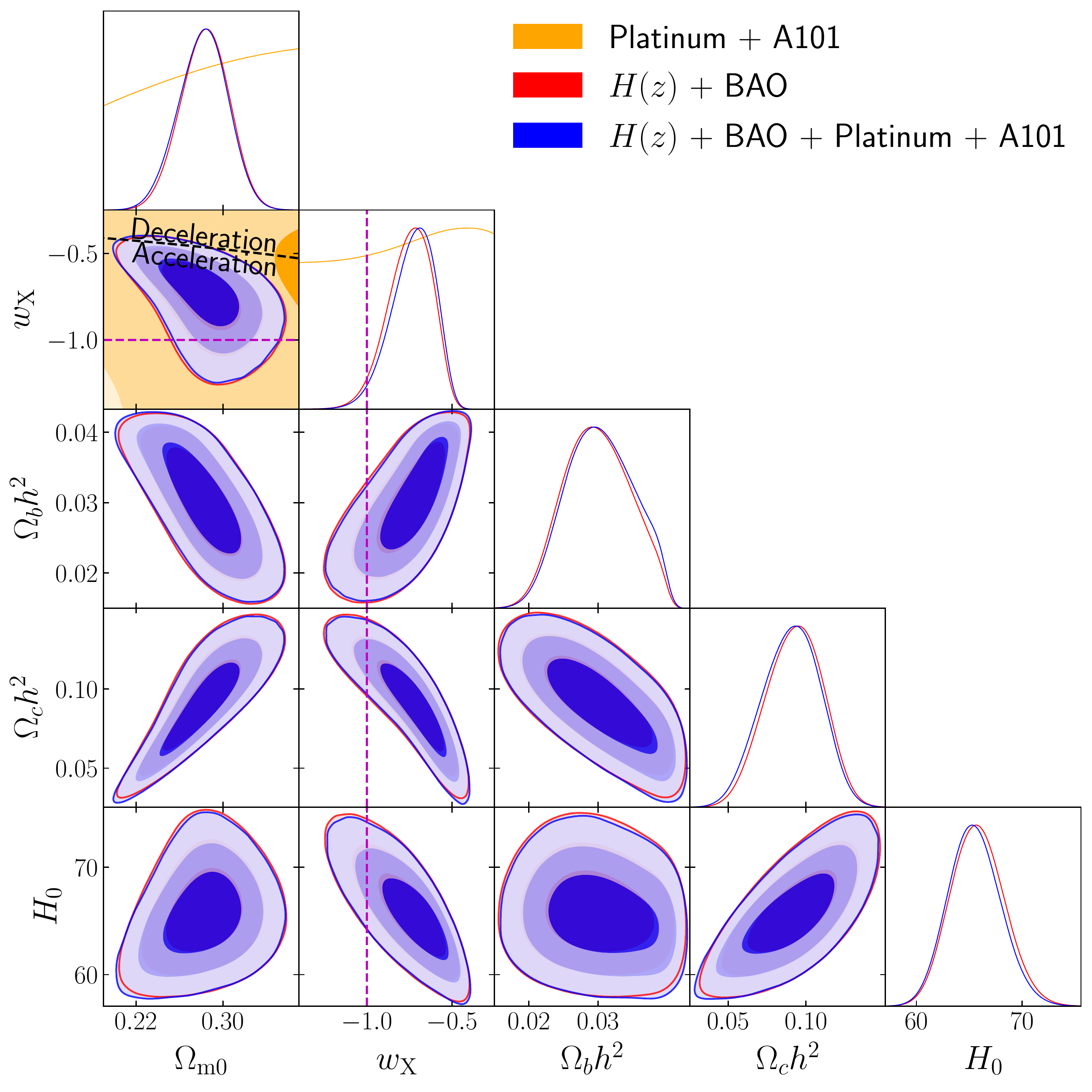}}\\
\caption{One-dimensional likelihood distributions and 1$\sigma$, 2$\sigma$, and 3$\sigma$ two-dimensional likelihood confidence contours for flat XCDM from various combinations of data. The zero-acceleration black dashed lines divide the parameter space into regions associated with currently-accelerating (either below left or below) and currently-decelerating (either above right or above) cosmological expansion. The magenta dashed lines represent $w_{\rm X}=-1$, i.e.\ flat \lcdm.}
\label{fig3}
\end{figure*}

\begin{figure*}
\centering
 \subfloat[]{%
    \includegraphics[width=0.5\textwidth,height=0.5\textwidth]{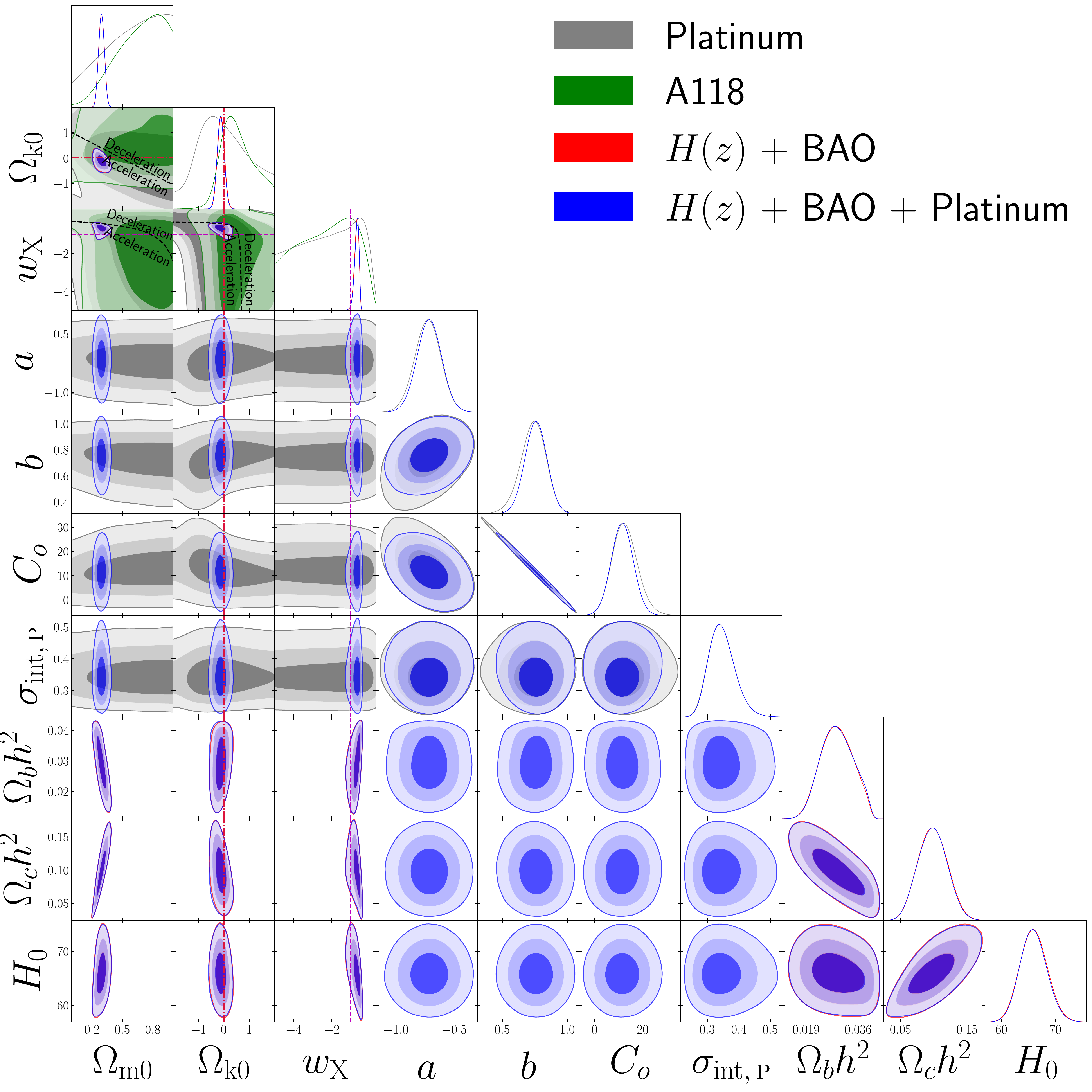}}
 \subfloat[]{%
    \includegraphics[width=0.5\textwidth,height=0.5\textwidth]{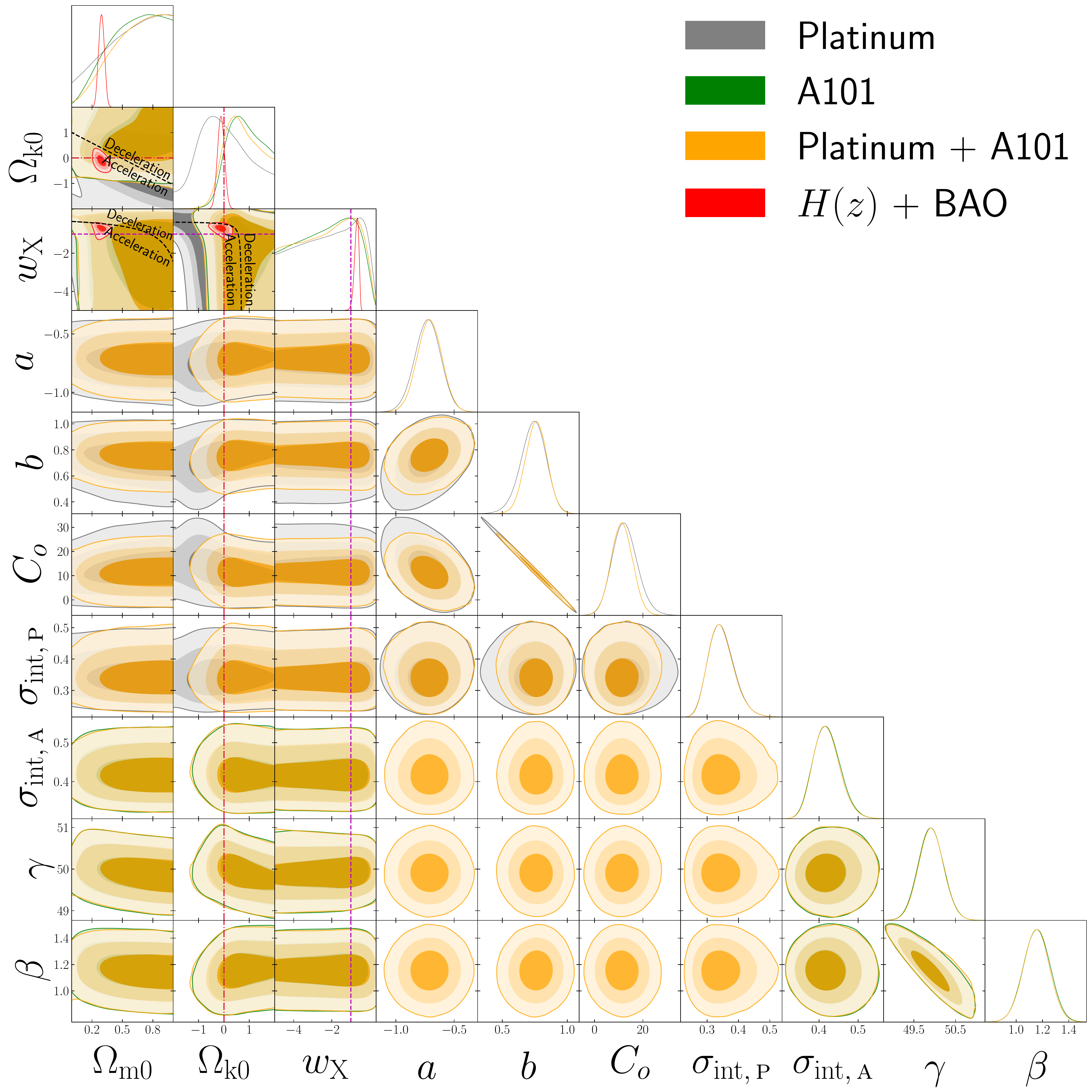}}\\
 \subfloat[]{%
    \includegraphics[width=0.5\textwidth,height=0.5\textwidth]{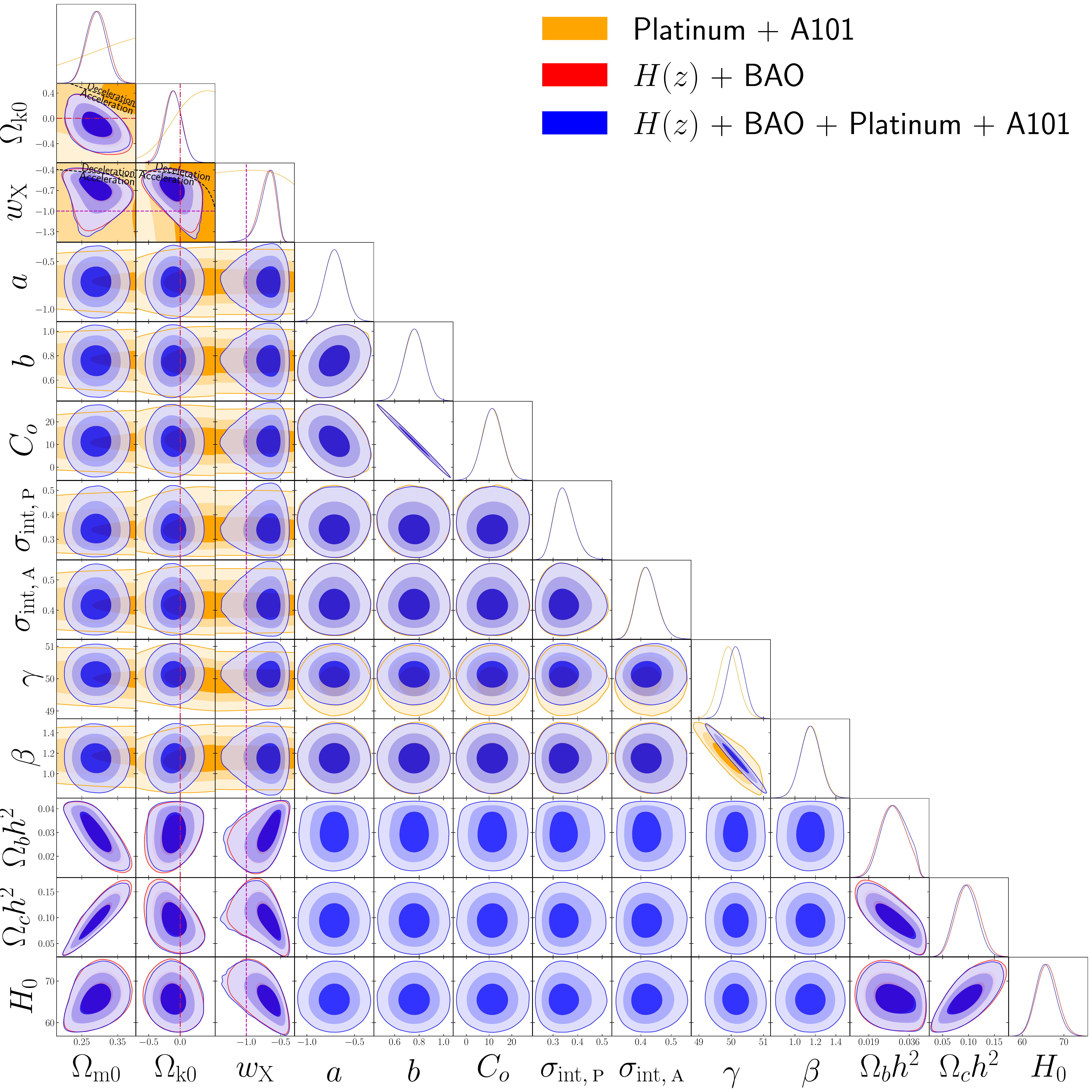}}
 \subfloat[Cosmological parameters zoom in]{%
    \includegraphics[width=0.5\textwidth,height=0.5\textwidth]{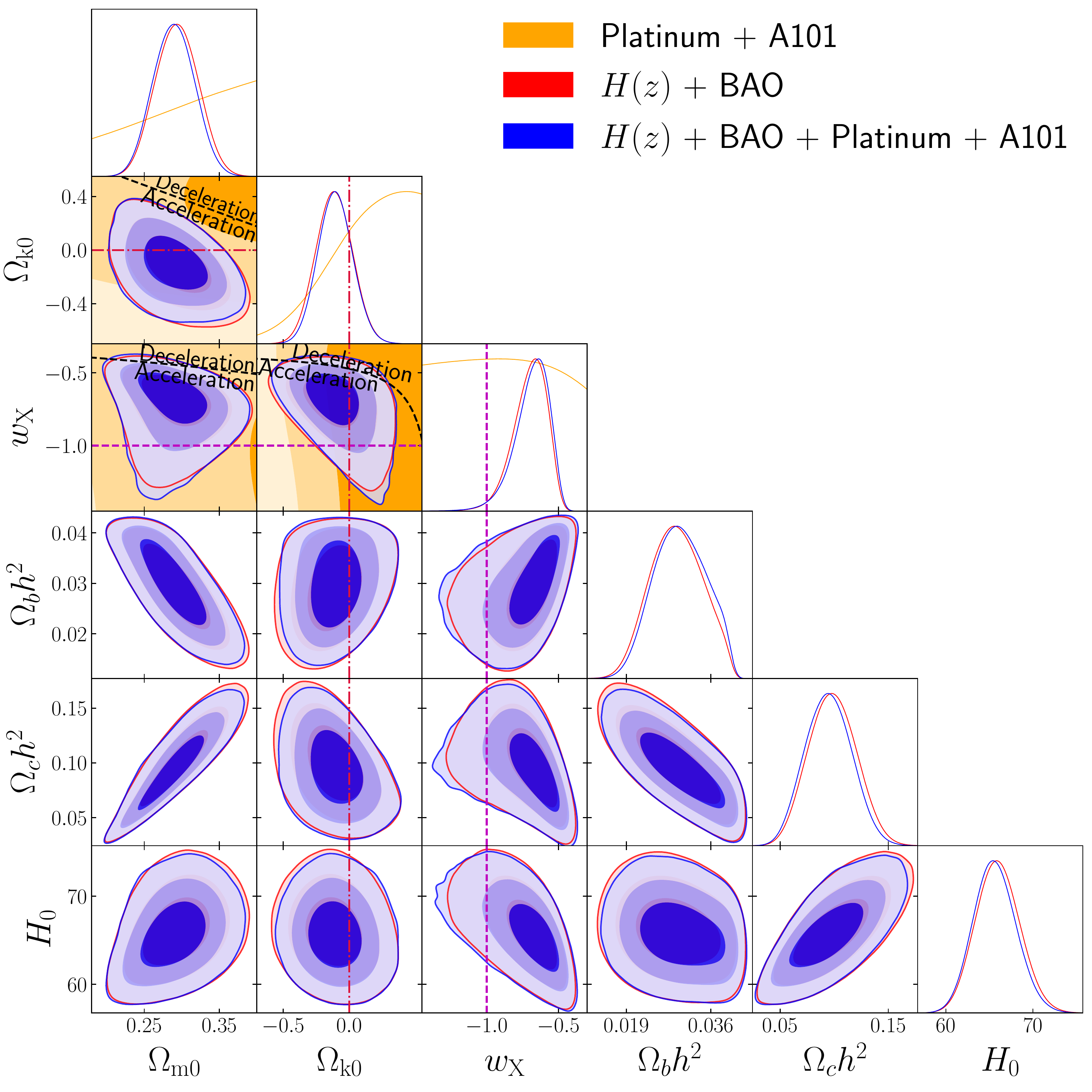}}\\
\caption{Same as Fig.\ \ref{fig3} but for non-flat XCDM. The zero-acceleration black dashed lines are computed for the third cosmological parameter set to the $H(z)$ + BAO data best-fitting values listed in Table \ref{tab:BFP}, and divide the parameter space into regions associated with currently-accelerating (either below left or below) and currently-decelerating (either above right or above) cosmological expansion. The crimson dash-dot lines represent flat hypersurfaces, with closed spatial hypersurfaces either below or to the left. The magenta dashed lines represent $w_{\rm X}=-1$, i.e.\ non-flat \lcdm.}
\label{fig4}
\end{figure*}

\begin{figure*}
\centering
\centering
 \subfloat[]{%
    \includegraphics[width=0.5\textwidth,height=0.5\textwidth]{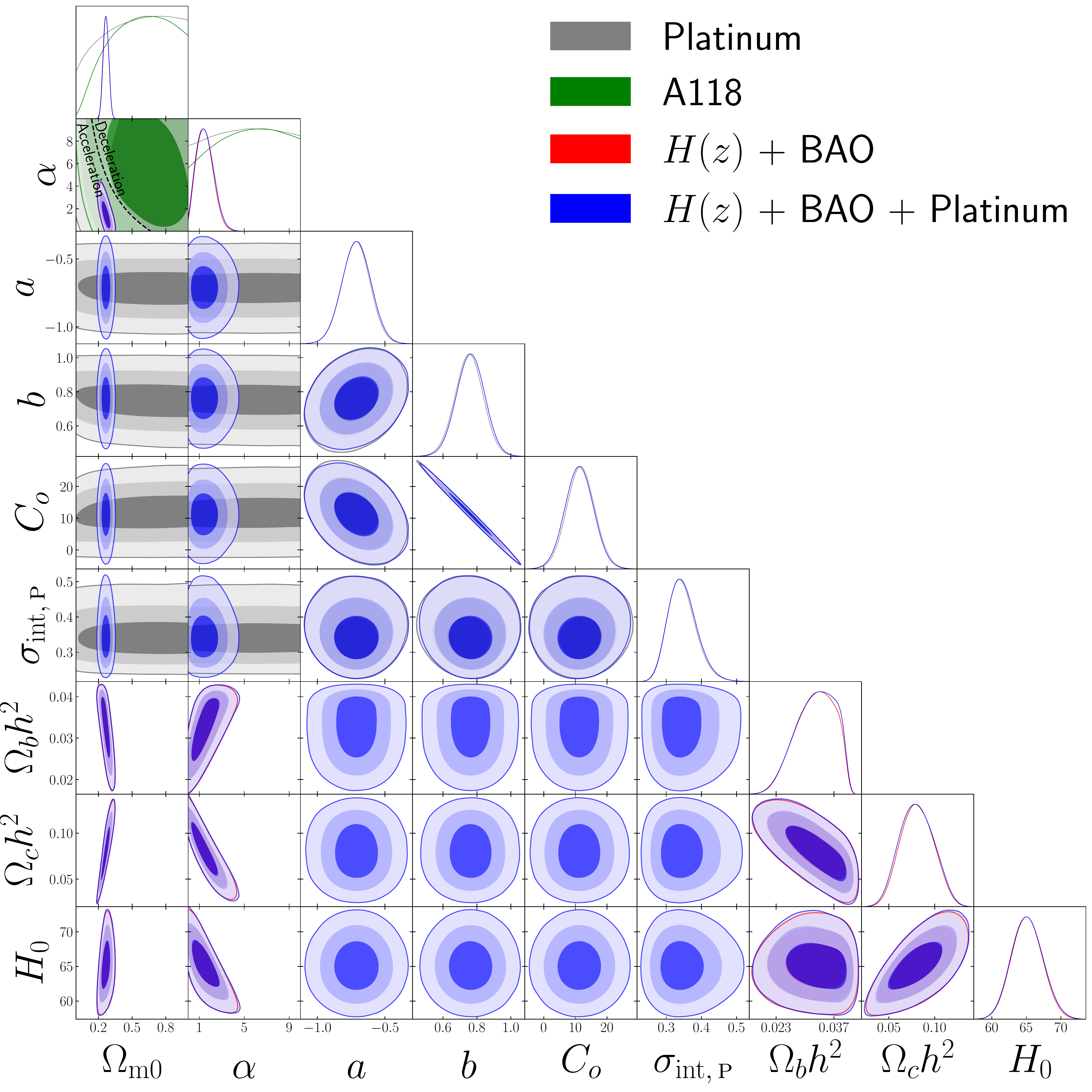}}
 \subfloat[]{%
    \includegraphics[width=0.5\textwidth,height=0.5\textwidth]{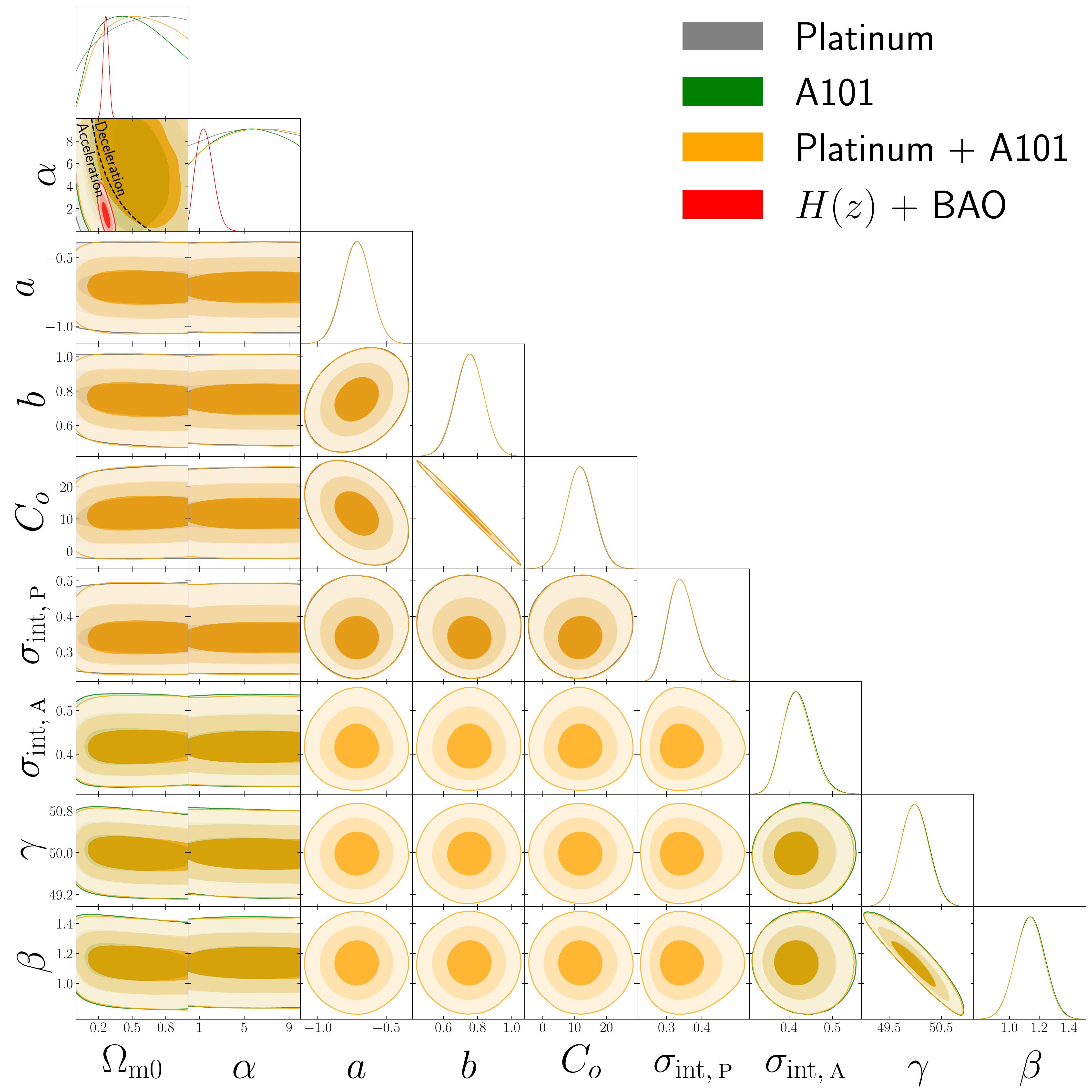}}\\
 \subfloat[]{%
    \includegraphics[width=0.5\textwidth,height=0.5\textwidth]{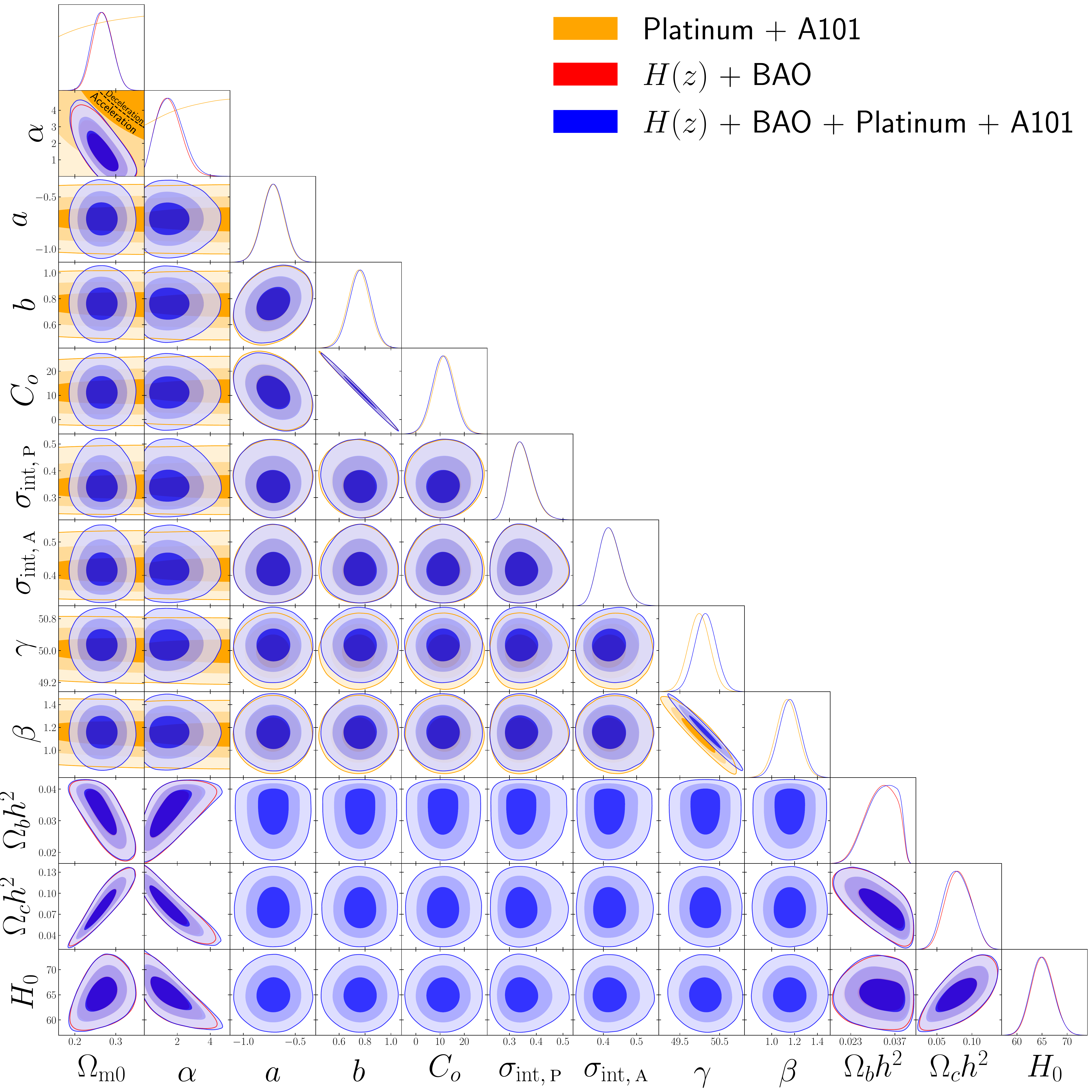}}
 \subfloat[Cosmological parameters zoom in]{%
    \includegraphics[width=0.5\textwidth,height=0.5\textwidth]{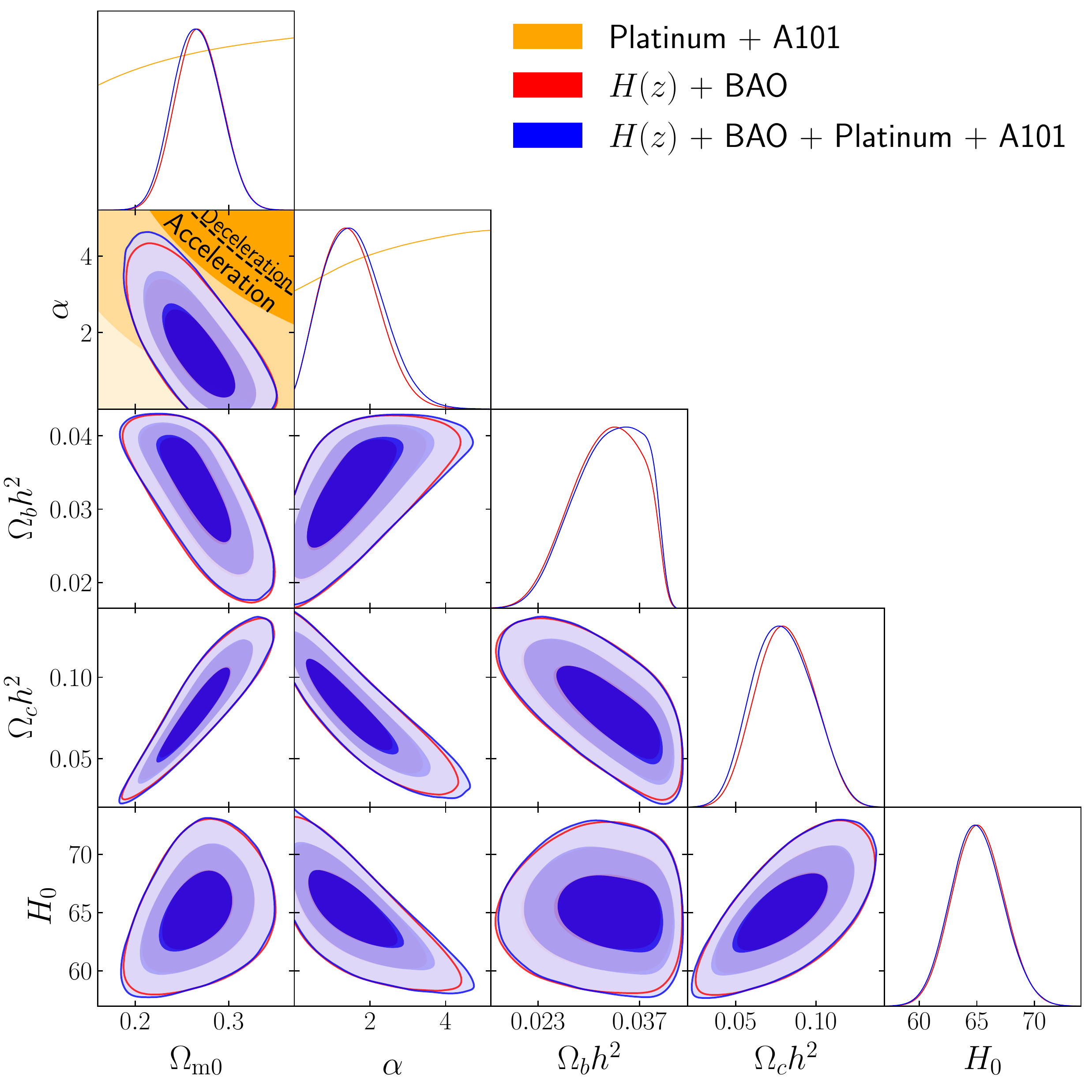}}\\
\caption{One-dimensional likelihood distributions and 1$\sigma$, 2$\sigma$, and 3$\sigma$ two-dimensional likelihood confidence contours for flat \pcdm\ from various combinations of data. The zero-acceleration black dashed lines divide the parameter space into regions associated with currently-accelerating (below left) and currently-decelerating (above right) cosmological expansion. The $\alpha = 0$ axes correspond to flat \lcdm.}
\label{fig5}
\end{figure*}

\begin{figure*}
\centering
 \subfloat[]{%
    \includegraphics[width=0.5\textwidth,height=0.5\textwidth]{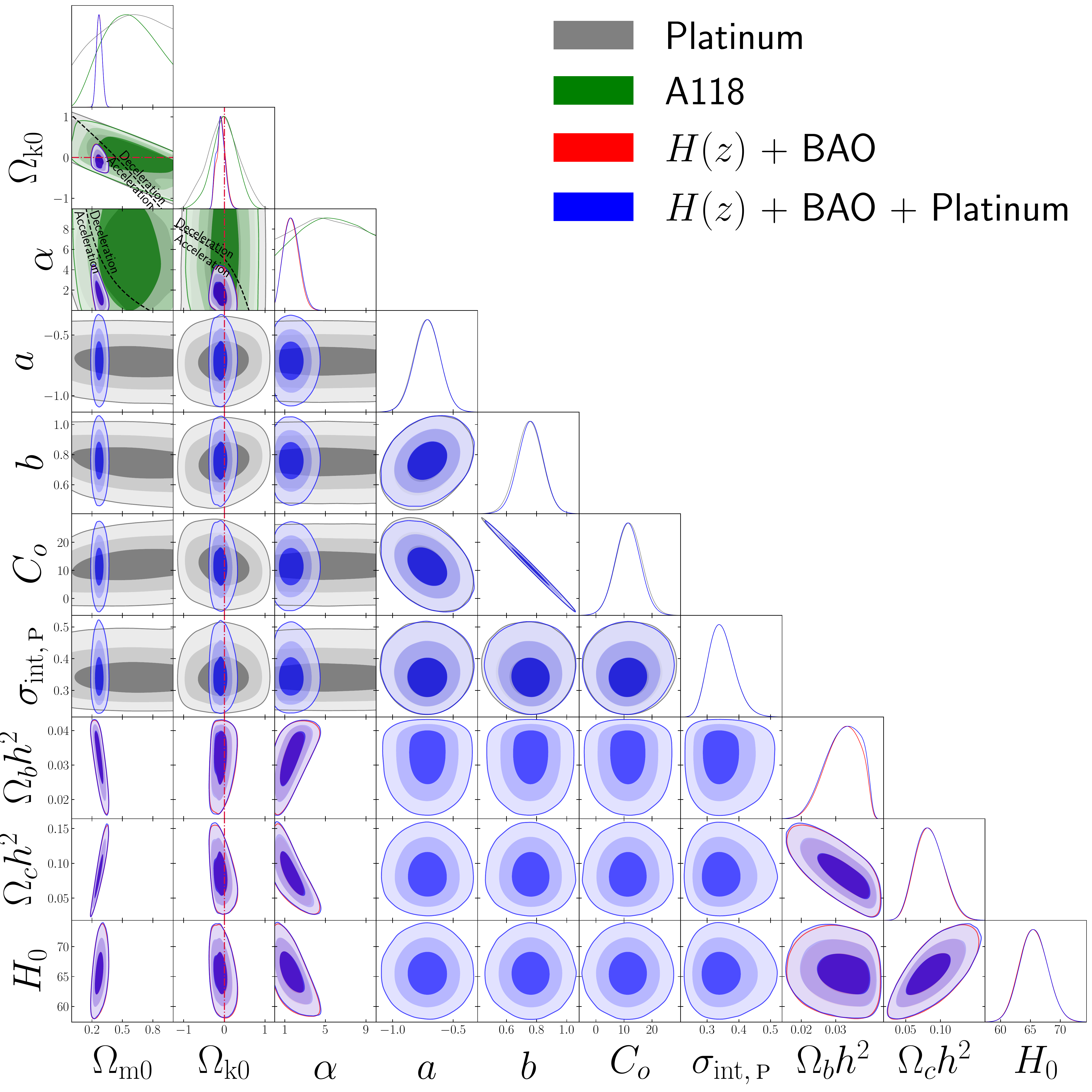}}
 \subfloat[]{%
    \includegraphics[width=0.5\textwidth,height=0.5\textwidth]{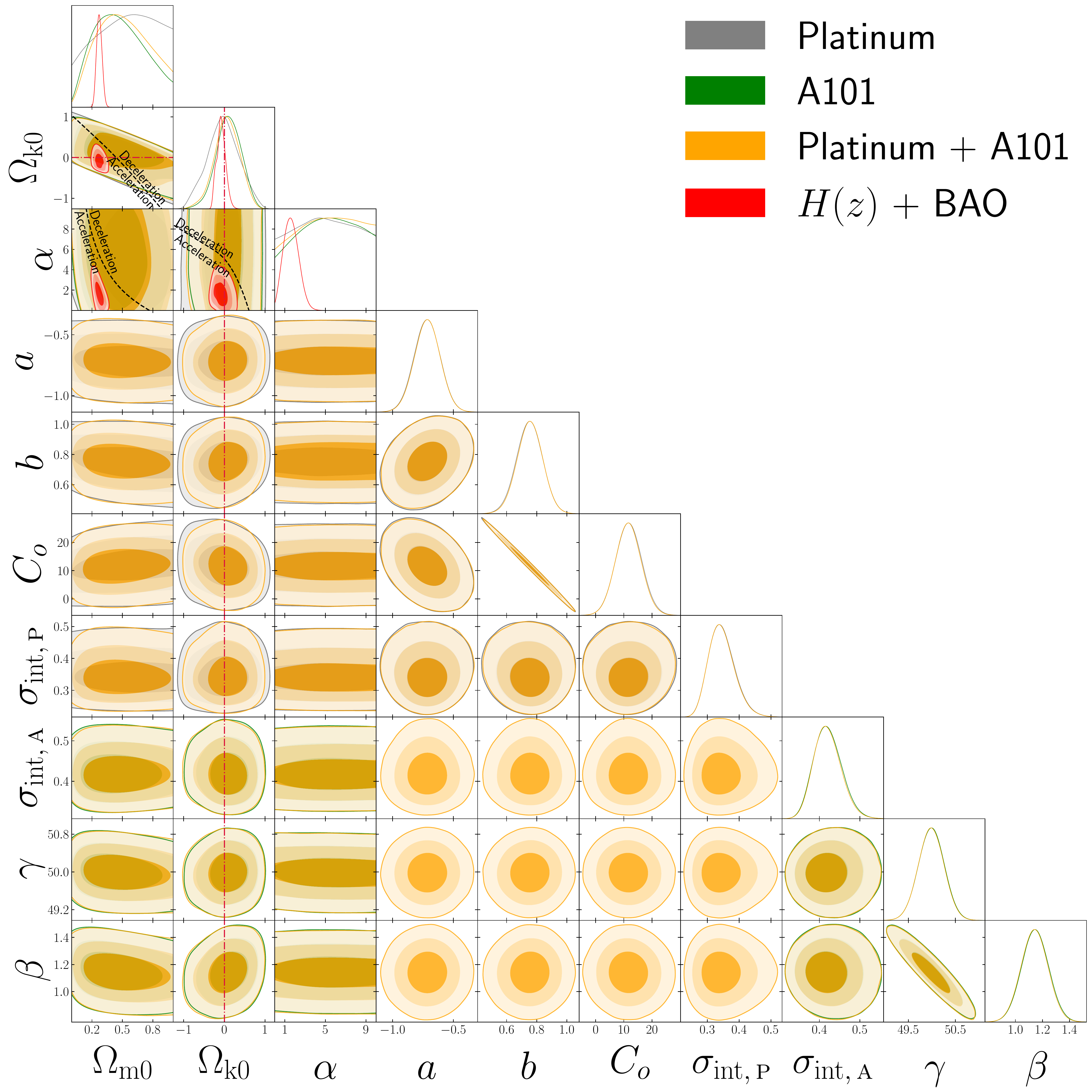}}\\
 \subfloat[]{%
    \includegraphics[width=0.5\textwidth,height=0.5\textwidth]{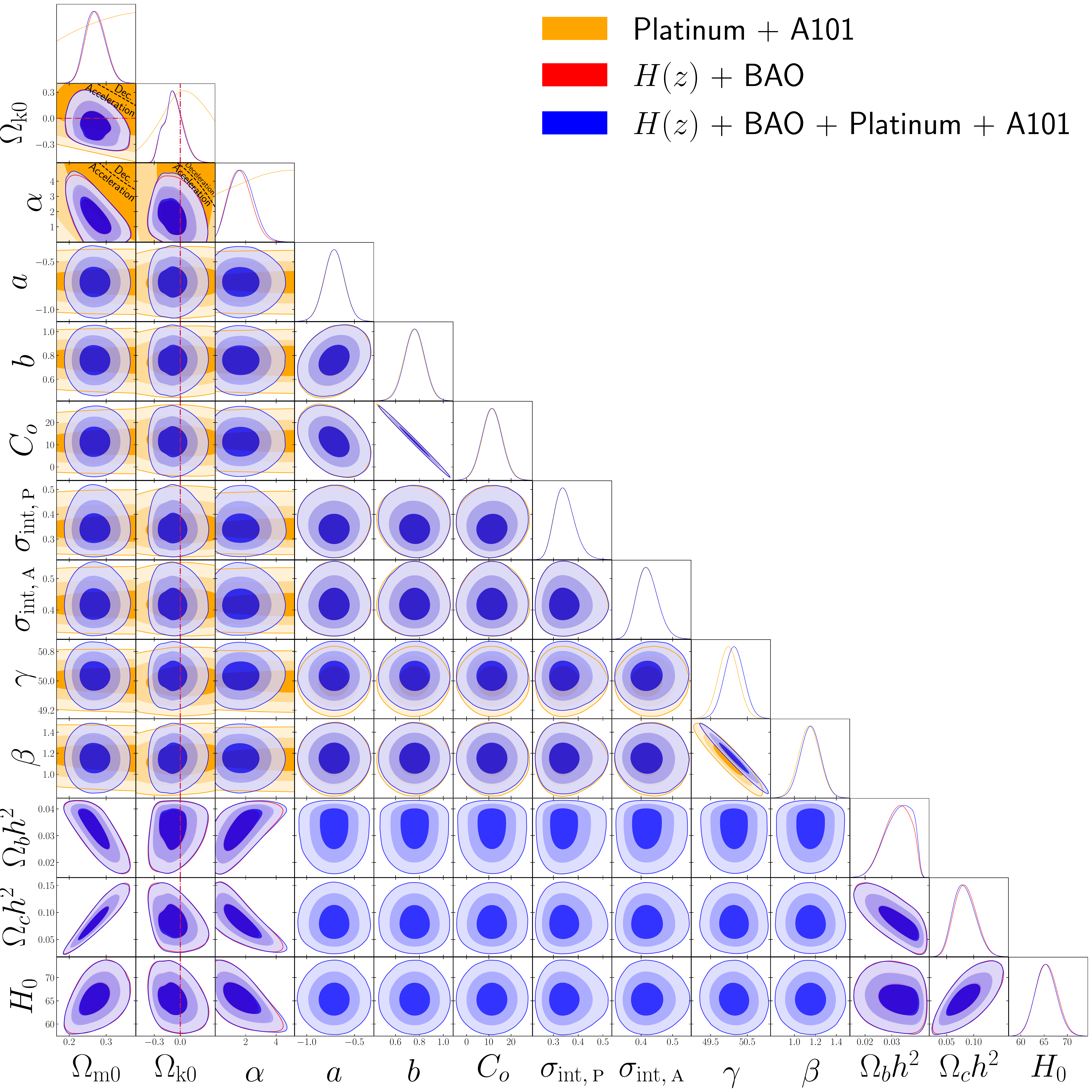}}
 \subfloat[Cosmological parameters zoom in]{%
    \includegraphics[width=0.5\textwidth,height=0.5\textwidth]{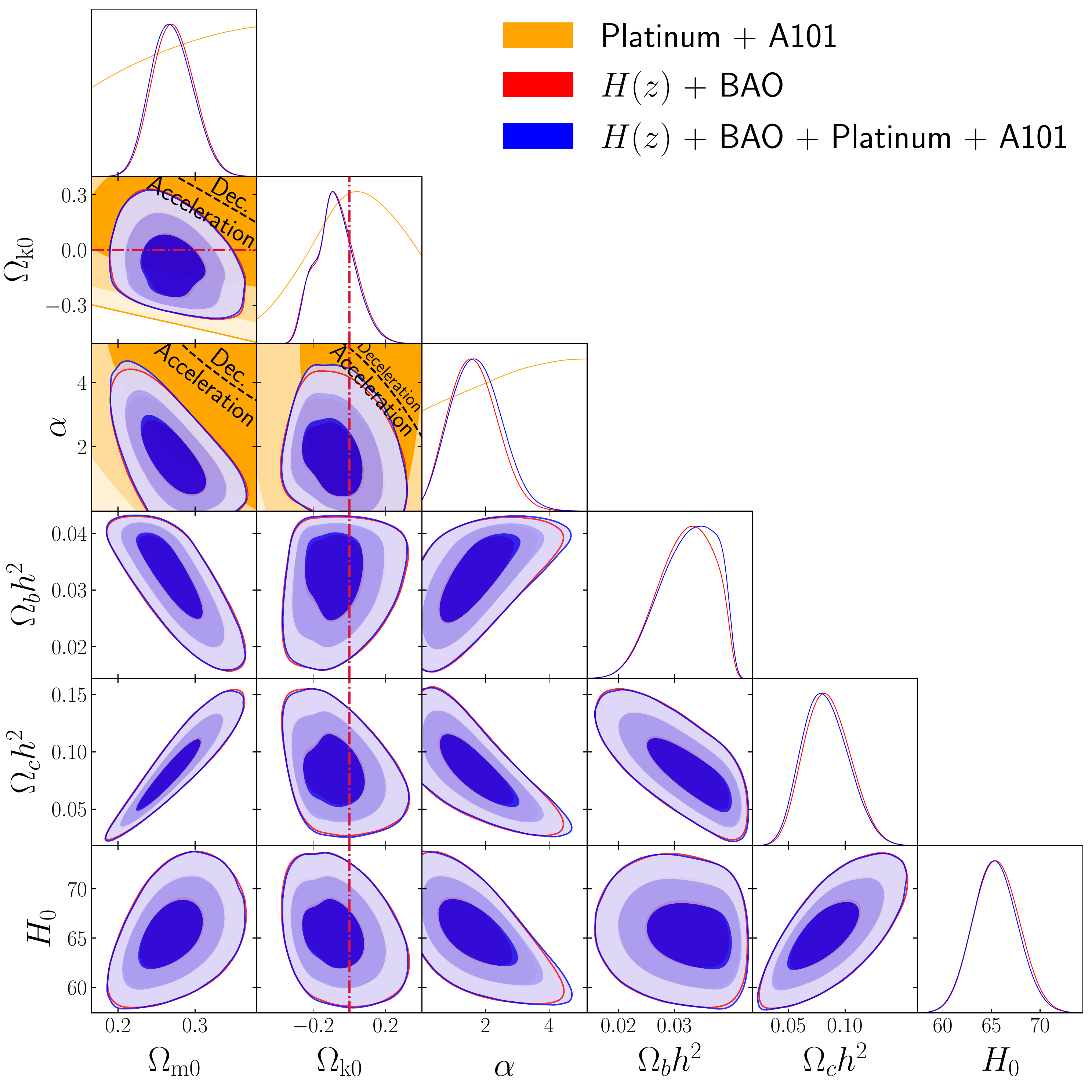}}\\
\caption{Same as Fig.\ \ref{fig5} but for non-flat \pcdm. The zero-acceleration black dashed lines are computed for the third cosmological parameter set to the $H(z)$ + BAO data best-fitting values listed in Table \ref{tab:BFP}, and divide the parameter space into regions associated with currently-accelerating (below left) and currently-decelerating (above right) cosmological expansion. The crimson dash-dot lines represent flat hypersurfaces, with closed spatial hypersurfaces either below or to the left. The $\alpha = 0$ axes correspond to non-flat \lcdm.}
\label{fig6}
\end{figure*}

\begin{figure*}
\centering
 \subfloat[Flat \lcdm]{%
    \includegraphics[width=0.48\textwidth,height=0.382\textwidth]{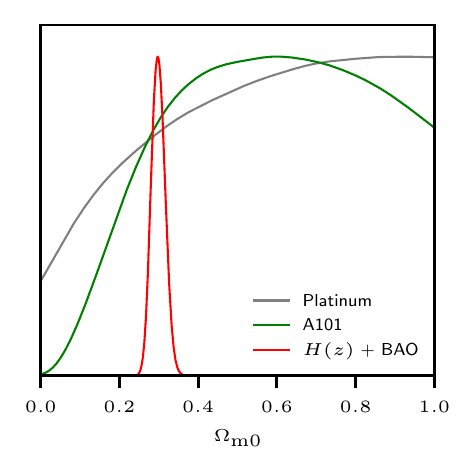}}
 \subfloat[Non-flat \lcdm]{%
    \includegraphics[width=0.48\textwidth,height=0.382\textwidth]{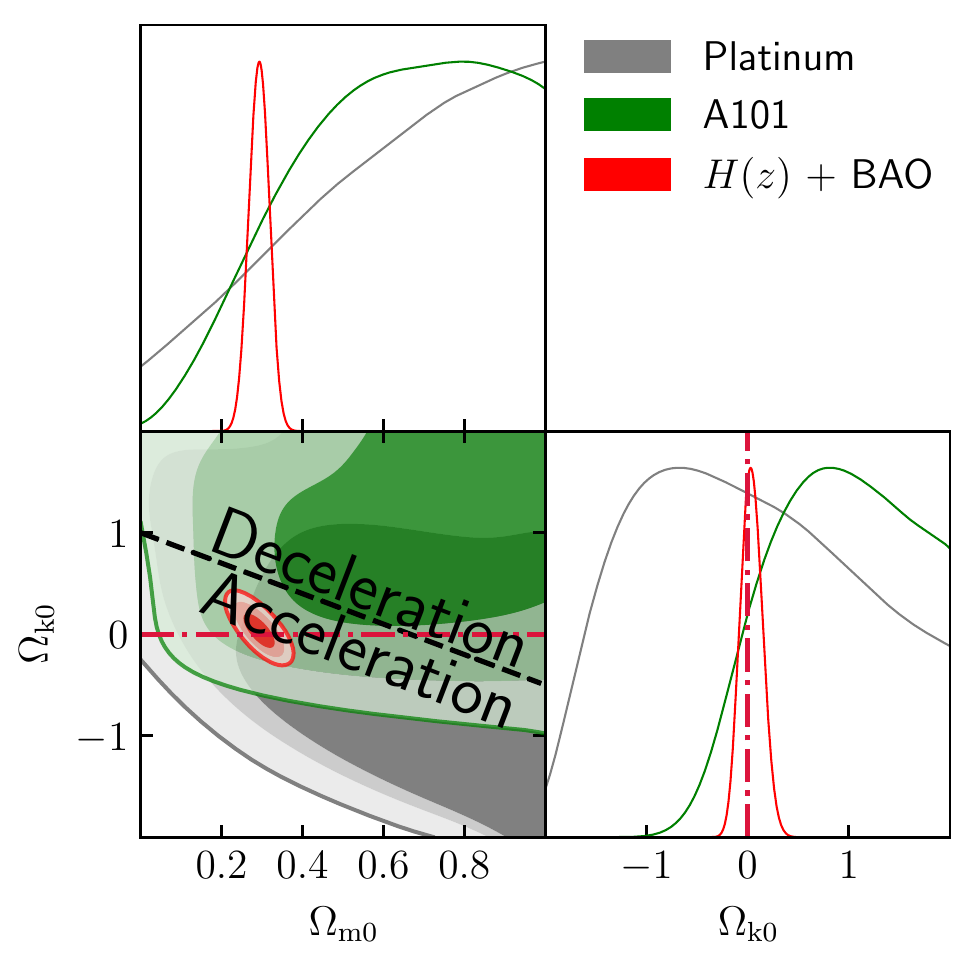}}\\
 \subfloat[Flat XCDM]{%
    \includegraphics[width=0.48\textwidth,height=0.382\textwidth]{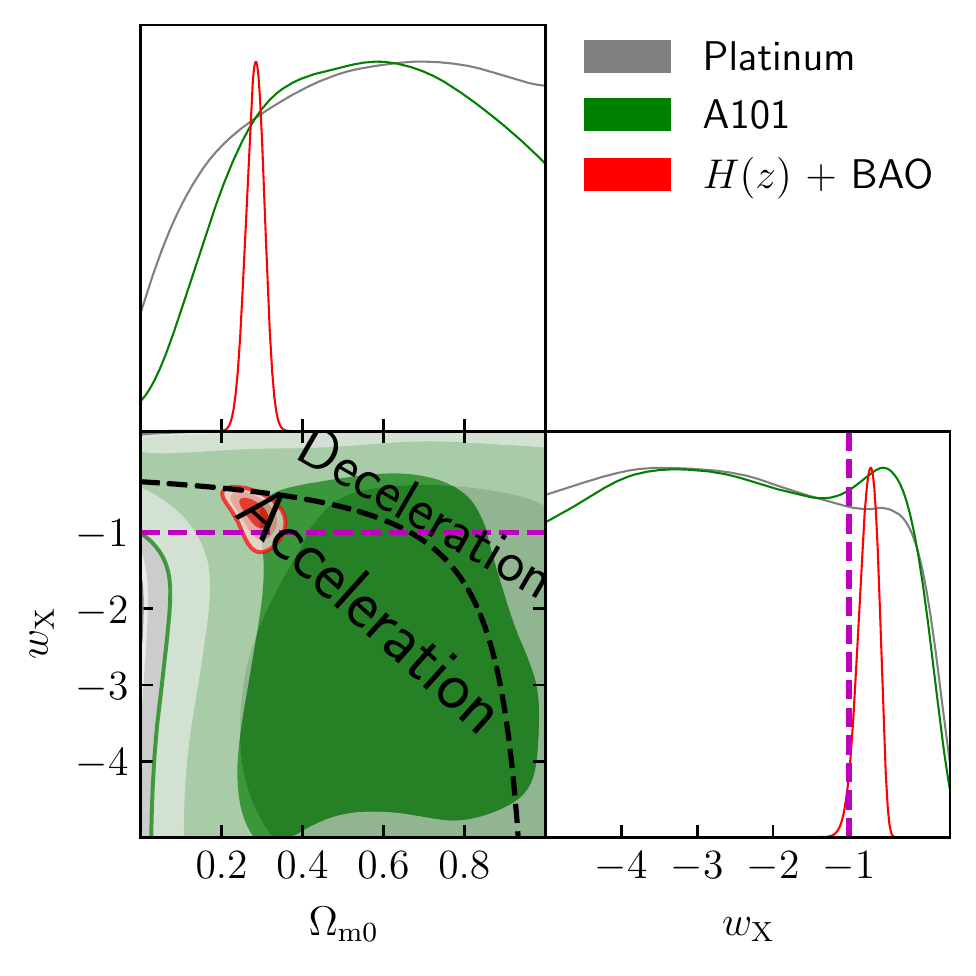}}
 \subfloat[Non-flat XCDM]{%
    \includegraphics[width=0.48\textwidth,height=0.382\textwidth]{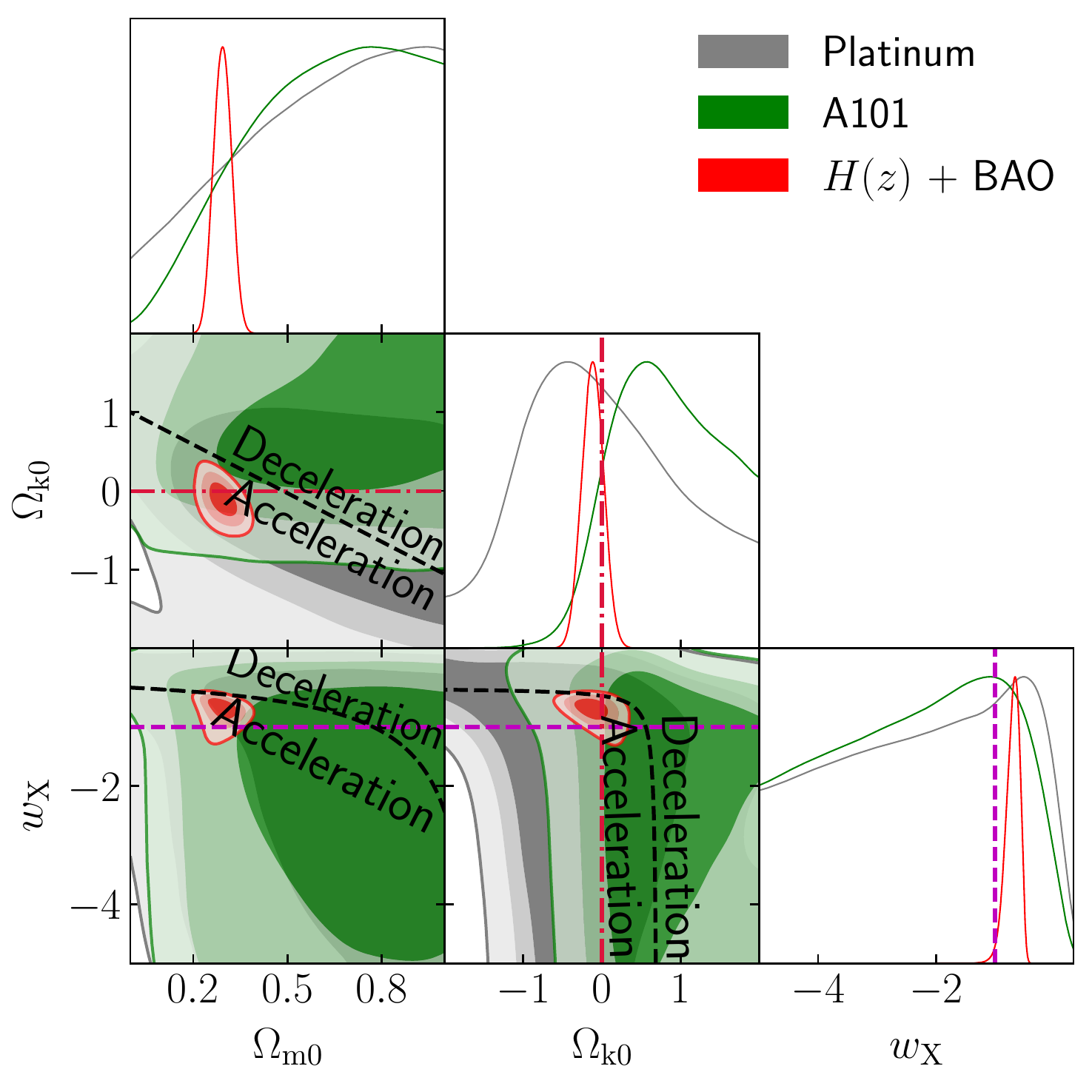}}\\
 \subfloat[Flat \pcdm]{%
    \includegraphics[width=0.48\textwidth,height=0.382\textwidth]{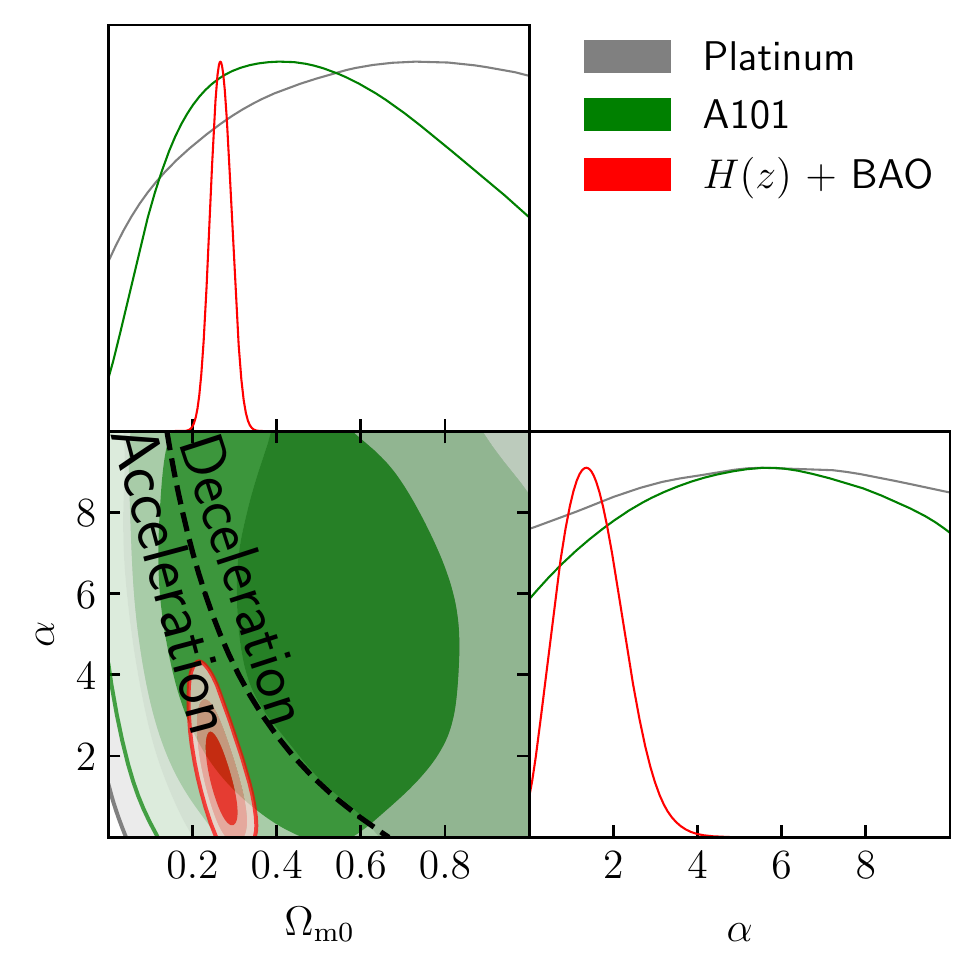}}
 \subfloat[Non-flat \pcdm]{%
    \includegraphics[width=0.48\textwidth,height=0.382\textwidth]{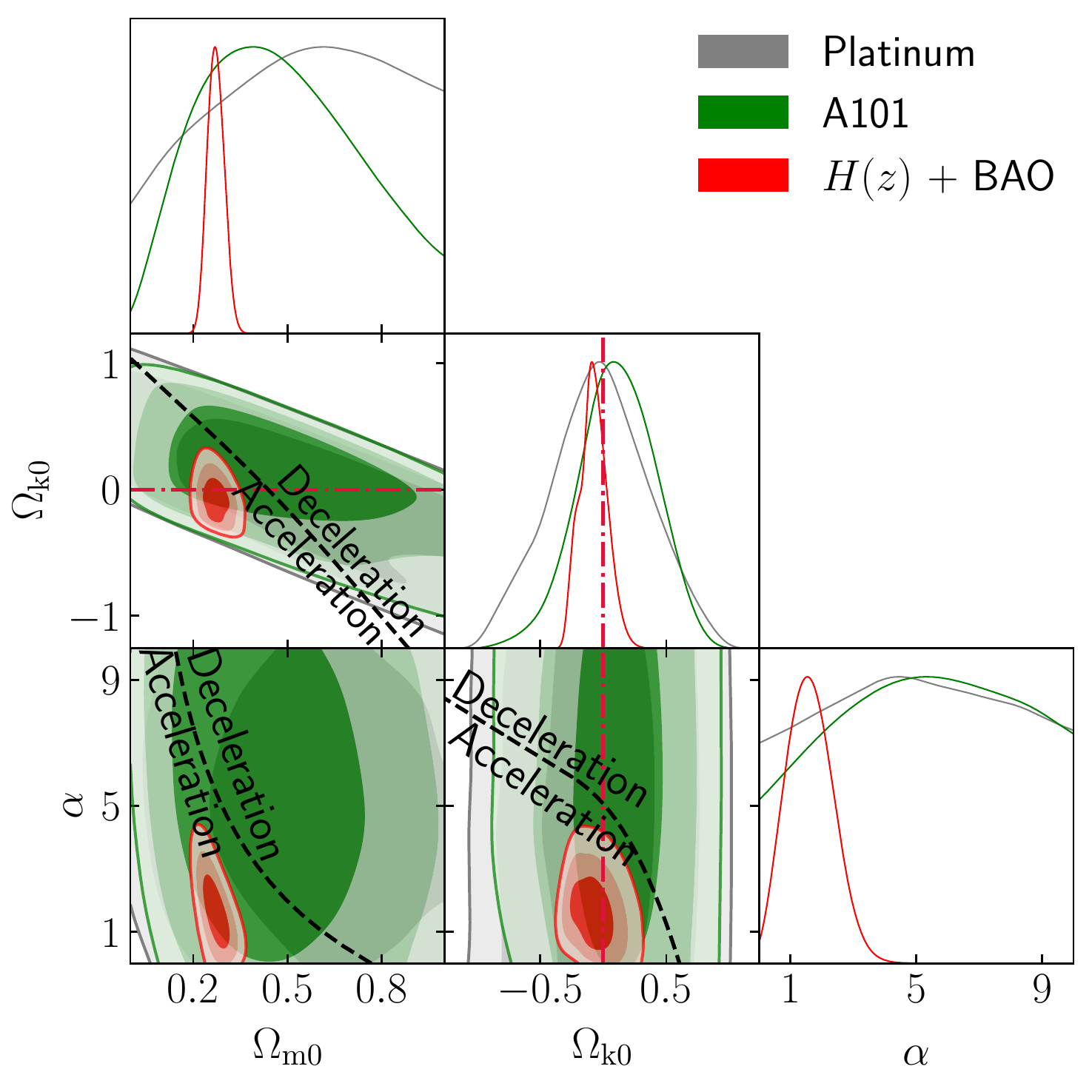}}\\
\caption{One-dimensional likelihood distributions and 1$\sigma$, 2$\sigma$, and 3$\sigma$ two-dimensional likelihood confidence contours for some cosmological parameters of the six cosmological models from Platinum data (gray), A101 data (green), and $H(z)$ + BAO data (red), which more clearly show the cosmological parameter overlaps between Platinum/A101 data and $H(z)$ + BAO data.}
\label{fig7}
\end{figure*}

We show the posterior one-dimensional (1D) probability distributions and two-dimensional (2D) confidence regions of cosmological-model and GRB-correlation parameters for the six cosmological models in Figs. \ref{fig1}--\ref{fig7}, in gray (Platinum), green (A118 and A101), orange (Platinum + A101), red [$H(z)$ + BAO], and blue [$H(z)$ + BAO + Platinum, in short HzBP, and $H(z)$ + BAO + Platinum + A101, in short HzBPA101]. The unmarginalized best-fitting parameter values, as well as the values of maximum likelihood $\mathcal{L}_{\rm max}$, $AIC$, $BIC$, $DIC$, $\Delta AIC$, $\Delta BIC$, and $\Delta DIC$, for all models and data combinations, are listed in Table \ref{tab:BFP}. We list the marginalized posterior mean parameter values and uncertainties ($\pm 1\sigma$ error bars and 1 or 2$\sigma$ limits), for all models and data combinations, in Table \ref{tab:1d_BFP}.

\begin{sidewaystable*}
\centering
\resizebox{\columnwidth}{!}{%
\begin{threeparttable}
\caption{Unmarginalized best-fitting parameter values for all models from various combinations of data.}\label{tab:BFP}
\begin{tabular}{lccccccccccccccccccccc}
\toprule
Model & Data set & $\Omega_{b}h^2$\,\tnote{a} & $\Omega_{c}h^2$ & $\Omega_{\mathrm{m0}}$ & $\Omega_{\mathrm{k0}}$ & $w_{\mathrm{X}}$/$\alpha$\tnote{b} & $H_0$\tnote{c} & $\sigma_{\mathrm{int,\,\textsc{p}}}$ & $a$ & $b$ & $C_{o}$ & $\sigma_{\mathrm{int,\,\textsc{a}}}$ & $\gamma$ & $\beta$ & $-2\ln\mathcal{L}_{\mathrm{max}}$ & $AIC$ & $BIC$ & $DIC$ & $\Delta AIC$ & $\Delta BIC$ & $\Delta DIC$ \\
\midrule
 & Platinum & -- & 0.4560 & 0.982 & -- & -- & -- & 0.322 & $-0.716$ & 0.743 & 12.19 & -- & -- & -- & 32.99 & 42.99 & 52.55 & 42.80 & 0.00 & 0.00 & 0.00\\
 & A118 & -- & 0.4088 & 0.886 & -- & -- & -- & -- & -- & -- & -- & 0.402 & 50.02 & 1.099 & 128.72 & 136.72 & 147.81 & 136.05 & 0.00 & 0.00 & 0.00\\
 & A101 & -- & 0.2547 & 0.571 & -- & -- & -- & -- & -- & -- & -- & 0.409 & 50.04 & 1.135 & 112.81 & 120.81 & 131.27 & 120.16 & 0.00 & 0.00 & 0.00\\
Flat & Plat. + A101 & -- & 0.3505 & 0.767 & -- & -- & -- & 0.324 & $-0.694$ & 0.753 & 11.60 & 0.413 & 49.98 & 1.125 & 146.08 & 162.08 & 186.22 & 162.07 & 0.00 & 0.00 & 0.00\\
\lcdm & $H(z)$ + BAO & 0.0240 & 0.1177 & 0.298 & -- & -- & 69.11 & -- & -- & -- & -- & -- & -- & -- & 23.66 & 29.66 & 34.87 & 30.23 & 0.00 & 0.00 & 0.00\\
 & HzBP\tnote{d} & 0.0243 & 0.1169 & 0.298 & -- & -- & 69.04 & 0.325 & $-0.715$ & 0.757 & 11.53 & -- & -- & -- & 57.33 & 73.33 & 88.99 & 72.58 & 0.00 & 0.00 & 0.00\\
 & HzBPA101\tnote{e} & 0.0236 & 0.1155 & 0.296 & -- & -- & 68.71 & 0.324 & $-0.707$ & 0.760 & 11.35 & 0.412 & 50.16 & 1.153 & 170.63 & 190.63 & 223.26 & 191.94 & 0.00 & 0.00 & 0.00\\
\midrule
 & Platinum & -- & 0.1178 & 0.292 & $-0.992$ & -- & -- & 0.292 & $-0.700$ & 0.829 & 7.59 & -- & -- & -- & 24.51 & 36.51 & 47.98 & 51.00 & $-6.48$ & $-4.57$ & 8.20\\
 & A118 & -- & 0.4625 & 0.995 & 1.094 & -- & -- & -- & -- & -- & -- & 0.401 & 49.90 & 1.117 & 127.96 & 137.96 & 151.82 & 136.72 & 1.24 & 4.01 & 0.67\\
 & A101 & -- & 0.4606 & 0.991 & 1.993 & -- & -- & -- & -- & -- & -- & 0.405 & 49.78 & 1.151 & 111.08 & 121.08 & 134.16 & 120.04 & 0.27 & 2.89 & $-0.12$\\
Non-flat & Plat. + A101 & -- & 0.4300 & 0.929 & 1.857 & -- & -- & 0.326 & $-0.712$ & 0.746 & 11.99 & 0.400 & 49.75 & 1.167 & 144.65 & 162.65 & 189.80 & 161.75 & 0.57 & 3.58 & $-0.32$\\
\lcdm  & $H(z)$ + BAO & 0.0247 & 0.1141 & 0.295 & 0.023 & -- & 68.78 & -- & -- & -- & -- & -- & -- & -- & 23.59 & 31.59 & 38.54 & 32.21 & 1.93 & 3.67 & 1.99\\
 & HzBP\tnote{d} & 0.0245 & 0.1149 & 0.294 & 0.010 & -- & 68.99 & 0.328 & $-0.719$ & 0.752 & 11.81 & -- & -- & -- & 57.31 & 73.31 & 93.48 & 74.65 & $-0.02$ & 4.49 & 2.07\\
 & HzBPA101\tnote{e} & 0.0237 & 0.1115 & 0.295 & 0.032 & -- & 67.91 & 0.328 & $-0.726$ & 0.750 & 11.91 & 0.410 & 50.13 & 1.164 & 170.70 & 192.70 & 228.59 & 193.50 & 2.07 & 5.33 & 1.56\\
\midrule
 & Platinum & -- & 0.0809 & 0.216 & -- & 0.137 & -- & 0.323 & $-0.724$ & 0.735 & 12.59 & -- & -- & -- & 32.81 & 44.81 & 56.28 & 42.83 & 1.82 & 3.73 & 0.04\\
 & A118 & -- & $-0.0198$ & 0.011 & -- & $-0.139$ & -- & -- & -- & -- & -- & 0.402 & 50.04 & 1.112 & 128.42 & 138.42 & 152.28 & 136.65 & 1.70 & 4.47 & 0.61\\
 & A101 & -- & $-0.0221$ & 0.006 & -- & $-0.251$ & -- & -- & -- & -- & -- & 0.406 & 50.01 & 1.148 & 111.99 & 121.99 & 135.06 & 121.23 & 1.18 & 3.79 & 1.08\\
Flat & Plat. + A101 & -- & 0.1570 & 0.372 & -- & $-0.193$ & -- & 0.317 & $-0.712$ & 0.753 & 11.65 & 0.405 & 50.02 & 1.115 & 145.74 & 163.74 & 190.90 & 162.40 & 1.66 & 4.68 & 0.33\\
XCDM & $H(z)$ + BAO & 0.0309 & 0.0870 & 0.280 & -- & $-0.694$ & 65.11 & -- & -- & -- & -- & -- & -- & -- & 19.65 & 27.65 & 34.60 & 28.11 & $-2.01$ & $-0.27$ & $-2.11$\\
 & HzBP\tnote{d} & 0.0302 & 0.0899 & 0.284 & -- & $-0.711$ & 65.22 & 0.320 & $-0.709$ & 0.758 & 11.46 & -- & -- & -- & 53.26 & 69.26 & 89.44 & 70.25 & $-4.07$ & 0.45 & $-2.33$\\
 & HzBPA101\tnote{e} & 0.0301 & 0.0891 & 0.282 & -- & $-0.709$ & 69.13 & 0.325 & $-0.724$ & 0.756 & 11.63 & 0.408 & 50.17 & 1.150 & 166.30 & 188.30 & 224.19 & 188.98 & $-2.33$ & 0.93 & $-2.96$\\
\midrule
 & Platinum & -- & 0.0918 & 0.239 & $-0.760$ & $-1.123$ & -- & 0.286 & $-0.761$ & 0.788 & 10.04 & -- & -- & -- & 24.58 & 38.58 & 51.97 & 51.01 & $-4.41$ & $-0.58$ & 8.21\\
 & A118 & -- & 0.4644 & 0.999 & 0.998 & $-1.137$ & -- & -- & -- & -- & -- & 0.398 & 49.91 & 1.113 & 127.97 & 139.97 & 156.60 & 137.16 & 3.25 & 8.79 & 1.11\\
 & A101 & -- & 0.4444 & 0.958 & 1.849 & $-1.222$ & -- & -- & -- & -- & -- & 0.410 & 49.73 & 1.164 & 111.14 & 123.14 & 138.83 & 120.59 & 2.33 & 7.56 & 0.44\\
Non-flat & Plat. + A101 & -- & 0.4120 & 0.892 & 1.526 & $-1.208$ & -- & 0.335 & $-0.707$ & 0.758 & 11.34 & 0.405 & 49.77 & 1.163 & 144.69 & 164.69 & 194.86 & 162.75 & 2.61 & 8.64 & 0.67\\
XCDM & $H(z)$ + BAO & 0.0295 & 0.0962 & 0.294 & $-0.159$ & $-0.648$ & 65.62 & -- & -- & -- & -- & -- & -- & -- & 18.31 & 28.31 & 37.00 & 28.96 & $-1.35$ & 2.13 & $-1.26$\\
 & HzBP\tnote{d} & 0.0301 & 0.0933 & 0.290 & $-0.164$ & $-0.640$ & 65.39 & 0.324 & $-0.709$ & 0.746 & 12.04 & -- & -- & -- & 51.90 & 69.90 & 92.60 & 70.74 & $-3.43$ & 3.61 & $-1.85$\\
 & HzBPA101\tnote{e} & 0.0299 & 0.0883 & 0.284 & $-0.112$ & $-0.640$ & 64.65 & 0.319 & $-0.714$ & 0.767 & 11.01 & 0.414 & 50.19 & 1.137 & 165.30 & 189.30 & 228.45 & 189.85 & $-1.33$ & 5.19 & $-2.09$\\
\midrule
 & Platinum & -- & 0.4615 & 0.993 & -- & 4.896 & -- & 0.322 & $-0.723$ & 0.744 & 12.14 & -- & -- & -- & 32.99 & 44.99 & 56.46 & 42.37 & 2.00 & 3.91 & $-0.43$\\
 & A118 & -- & 0.2119 & 0.484 & -- & 9.617 & -- & -- & -- & -- & -- & 0.400 & 50.04 & 1.105 & 128.56 & 138.56 & 152.41 & 135.74 & 1.84 & 4.60 & $-0.31$\\
 & A101 & -- & 0.0764 & 0.207 & -- & 9.922 & -- & -- & -- & -- & -- & 0.409 & 49.99 & 1.152 & 112.26 & 122.26 & 135.33 & 119.96 & 1.45 & 4.06 & $-0.20$\\
Flat & Plat. + A101 & -- & 0.0936 & 0.242 & -- & 7.330 & -- & 0.326 & $-0.723$ & 0.755 & 11.62 & 0.410 & 50.04 & 1.135 & 145.75 & 163.75 & 190.90 & 161.62 & 1.67 & 4.68 & $-0.46$\\
\pcdm  & $H(z)$ + BAO & 0.0332 & 0.0789 & 0.265 & -- & 1.455 & 65.24 & -- & -- & -- & -- & -- & -- & -- & 19.49 & 27.49 & 34.44 & 26.96 & $-2.17$ & $-0.43$ & $-3.27$\\
 & HzBP\tnote{d} & 0.0343 & 0.0746 & 0.259 & -- & 1.633 & 64.98 & 0.331 & $-0.714$ & 0.765 & 11.10 & -- & -- & -- & 53.12 & 69.12 & 89.29 & 69.20 & $-4.21$ & 0.30 & $-3.39$\\
 & HzBPA101\tnote{e} & 0.0374 & 0.0663 & 0.246 & -- & 1.968 & 65.20 & 0.327 & $-0.703$ & 0.759 & 11.38 & 0.412 & 50.10 & 1.172 & 166.28 & 188.28 & 224.17 & 187.99 & $-2.35$ & 0.91 & $-3.95$\\
\midrule
 & Platinum & -- & 0.4366 & 0.942 & $-0.905$ & 0.061 & -- & 0.325 & $-0.735$ & 0.725 & 13.16 & -- & -- & -- & 32.53 & 46.53 & 59.92 & 43.06 & 3.54 & 7.37 & 0.26\\
 & A118 & -- & 0.3331 & 0.731 & 0.234 & 5.269 & -- & -- & -- & -- & -- & 0.402 & 50.02 & 1.111 & 128.42 & 140.42 & 157.04 & 136.67 & 3.70 & 9.23 & 0.63\\
 & A101 & -- & 0.2367 & 0.534 & 0.460 & 8.680 & -- & -- & -- & -- & -- & 0.406 & 49.99 & 1.151 & 111.89 & 123.89 & 139.58 & 120.48 & 3.08 & 8.31 & 0.33\\
Non-flat & Plat. + A101 & -- & 0.3063 & 0.677 & 0.317 & 9.746 & -- & 0.319 & $-0.728$ & 0.742 & 12.27 & 0.412 & 49.96 & 1.140 & 145.40 & 165.40 & 195.57 & 162.44 & 3.32 & 9.35 & 0.36\\
\pcdm  & $H(z)$ + BAO & 0.0344 & 0.0786 & 0.263 & $-0.149$ & 2.014 & 65.71 & -- & -- & -- & -- & -- & -- & -- & 18.15 & 28.15 & 36.84 & 27.39 & $-1.51$ & 1.97 & $-2.84$\\
 & HzBP\tnote{d} & 0.0326 & 0.0851 & 0.271 & $-0.176$ & 1.826 & 66.10 & 0.322 & $-0.727$ & 0.752 & 11.78 & -- & -- & -- & 51.79 & 69.79 & 92.49 & 69.44 & $-3.54$ & 3.50 & $-3.14$\\
 & HzBPA101\tnote{e} & 0.0353 & 0.0730 & 0.257 & $-0.135$ & 2.199 & 65.09 & 0.321 & $-0.729$ & 0.758 & 11.52 & 0.403 & 50.16 & 1.144 & 165.08 & 189.08 & 228.23 & 188.65 & $-1.55$ & 4.97 & $-3.29$\\
\bottomrule
\end{tabular}
\begin{tablenotes}[flushleft]
\item [a] In the four GRB-only cases $\Omega_{b}h^2$ is set to 0.0245.
\item [b] \wx\ corresponds to flat/non-flat XCDM and $\alpha$ corresponds to flat/non-flat \pcdm.
\item [c] \hunit. In the four GRB-only cases $H_0$ is set to 70 \hunit.
\item [d] $H(z)$ + BAO + Platinum.
\item [e] $H(z)$ + BAO + Platinum + A101.
\end{tablenotes}
\end{threeparttable}%
}
\end{sidewaystable*}

\begin{sidewaystable*}
\centering
\resizebox*{\columnwidth}{0.74\columnwidth}{%
\begin{threeparttable}
\caption{One-dimensional marginalized posterior mean values and uncertainties ($\pm 1\sigma$ error bars or $2\sigma$ limits) of the parameters for all models from various combinations of data.}\label{tab:1d_BFP}
\begin{tabular}{lcccccccccccccc}
\toprule
Model & Data set & $\Omega_{b}h^2$\,\tnote{a} & $\Omega_{c}h^2$ & $\Omega_{\mathrm{m0}}$ & $\Omega_{\mathrm{k0}}$ & $w_{\mathrm{X}}$/$\alpha$\tnote{b} & $H_0$\tnote{c} & $\sigma_{\mathrm{int,\,\textsc{p}}}$ & $a$ & $b$ & $C_{o}$ & $\sigma_{\mathrm{int,\,\textsc{a}}}$ & $\gamma$ & $\beta$ \\
\midrule
 & Platinum & -- & -- & $>0.411$\tnote{d} & -- & -- & -- & $0.347^{+0.033}_{-0.046}$ & $-0.714\pm0.104$ & $0.756\pm0.083$ & $11.52\pm4.45$ & -- & -- & -- \\
 & A118 & -- & -- & $>0.256$ & -- & -- & -- & -- & -- & -- & -- & $0.412^{+0.027}_{-0.033}$ & $50.09\pm0.25$ & $1.109\pm0.089$ \\
 & A101 & -- & -- & $0.584^{+0.298}_{-0.240}$ & -- & -- & -- & -- & -- & -- & -- & $0.422^{+0.030}_{-0.037}$ & $50.03\pm0.28$ & $1.140\pm0.098$ \\
Flat & Plat. + A101 & -- & -- & $0.614^{+0.380}_{-0.130}$ & -- & -- & -- & $0.347^{+0.033}_{-0.046}$ & $-0.717\pm0.102$ & $0.751\pm0.083$ & $11.80\pm4.45$ & $0.421^{+0.029}_{-0.036}$ & $50.03\pm0.27$ & $1.138\pm0.095$ \\
\lcdm & $H(z)$ + BAO & $0.0243\pm0.0029$ & $0.1184^{+0.0077}_{-0.0084}$ & $0.299^{+0.016}_{-0.018}$ & -- & -- & $69.27\pm1.85$ & -- & -- & -- & -- & -- & -- & -- \\
 & HzBP\tnote{e} & $0.0242\pm0.0029$ & $0.1184^{+0.0076}_{-0.0084}$ & $0.299^{+0.016}_{-0.018}$ & -- & -- & $69.24\pm1.81$ & $0.347^{+0.033}_{-0.045}$ & $-0.711\pm0.104$ & $0.762\pm0.081$ & $11.23\pm4.35$ & -- & -- & -- \\
 & HzBPA101\tnote{f} & $0.0242^{+0.0027}_{-0.0030}$ & $0.1187^{+0.0075}_{-0.0083}$ & $0.300^{+0.016}_{-0.018}$ & -- & -- & $69.20\pm1.79$ & $0.347^{+0.032}_{-0.045}$ & $-0.711\pm0.104$ & $0.762\pm0.080$ & $11.23\pm4.32$ & $0.421^{+0.029}_{-0.036}$ & $50.13\pm0.26$ & $1.159\pm0.094$ \\
\midrule
 & Platinum & -- & -- & $>0.491$\tnote{d} & $-0.023^{+0.869}_{-1.401}$ & -- & -- & $0.347^{+0.034}_{-0.046}$ & $-0.727\pm0.109$ & $0.738^{+0.102}_{-0.088}$ & $12.49^{+4.71}_{-5.46}$ & -- & -- & -- \\
 & A118 & -- & -- & $>0.295$ & $0.701^{+0.636}_{-0.844}$ & -- & -- & -- & -- & -- & -- & $0.411^{+0.027}_{-0.033}$ & $50.00\pm0.26$ & $1.123\pm0.088$ \\
 & A101 & -- & -- & $>0.212$ & $0.886^{+0.819}_{-0.614}$ & -- & -- & -- & -- & -- & -- & $0.419^{+0.030}_{-0.037}$ & $49.93\pm0.27$ & $1.156\pm0.096$ \\
Non-flat & Plat. + A101 & -- & -- & $>0.235$ & $0.840^{+0.758}_{-0.720}$ & -- & -- & $0.346^{+0.031}_{-0.044}$ & $-0.711\pm0.100$ & $0.759\pm0.080$ & $11.33^{+4.28}_{-4.27}$ & $0.418^{+0.029}_{-0.036}$ & $49.93\pm0.27$ & $1.154\pm0.095$ \\
\lcdm & $H(z)$ + BAO & $0.0255^{+0.0040}_{-0.0048}$ & $0.1127\pm0.0195$ & $0.293\pm0.025$ & $0.039^{+0.100}_{-0.114}$ & -- & $68.76\pm2.36$ & -- & -- & -- & -- & -- & -- & -- \\
 & HzBP\tnote{e} & $0.0253^{+0.0039}_{-0.0048}$ & $0.1135^{+0.0192}_{-0.0193}$ & $0.294\pm0.024$ & $0.034^{+0.098}_{-0.113}$ & -- & $68.80\pm2.32$ & $0.347^{+0.033}_{-0.046}$ & $-0.710\pm0.104$ & $0.763\pm0.080$ & $11.18\pm4.30$ & -- & -- & -- \\
 & HzBPA101\tnote{f} & $0.0255^{+0.0038}_{-0.0047}$ & $0.1121^{+0.0181}_{-0.0182}$ & $0.293\pm0.023$ & $0.043^{+0.094}_{-0.107}$ & -- & $68.62\pm2.24$ & $0.347^{+0.033}_{-0.045}$ & $-0.710\pm0.102$ & $0.764\pm0.079$ & $11.15\pm4.27$ & $0.421^{+0.029}_{-0.036}$ & $50.13\pm0.26$ & $1.161\pm0.093$ \\
\midrule
 & Platinum & -- & -- & $>0.386$\tnote{d} & -- & $<-0.078$ & -- & $0.347^{+0.033}_{-0.046}$ & $-0.717\pm0.103$ & $0.749\pm0.083$ & $11.90\pm4.45$ & -- & -- & -- \\
 & A118 & -- & -- & $0.599^{+0.360}_{-0.166}$ & -- & $-2.412^{+1.790}_{-1.730}$ & -- & -- & -- & -- & -- & $0.412^{+0.028}_{-0.034}$ & $50.15^{+0.27}_{-0.30}$ & $1.106\pm0.090$ \\
 & A101 & -- & -- & $0.558^{+0.285}_{-0.280}$ & -- & $-2.423^{+1.765}_{-1.716}$ & -- & -- & -- & -- & -- & $0.422^{+0.030}_{-0.037}$ & $50.10^{+0.29}_{-0.33}$ & $1.134\pm0.097$ \\
Flat & Plat. + A101 & -- & -- & $0.576^{+0.382}_{-0.182}$ & -- & $<-0.089$ & -- & $0.346^{+0.033}_{-0.046}$ & $-0.718\pm0.105$ & $0.749\pm0.084$ & $11.94\pm4.54$ & $0.421^{+0.029}_{-0.036}$ & $50.08^{+0.28}_{-0.32}$ & $1.133\pm0.094$ \\
XCDM & $H(z)$ + BAO & $0.0299^{+0.0046}_{-0.0052}$ & $0.0926^{+0.0192}_{-0.0170}$ & $0.283^{+0.023}_{-0.021}$ & -- & $-0.750^{+0.149}_{-0.104}$ & $65.83^{+2.34}_{-2.63}$ & -- & -- & -- & -- & -- & -- & -- \\
 & HzBP\tnote{e} & $0.0301^{+0.0046}_{-0.0053}$ & $0.0917^{+0.0196}_{-0.0171}$ & $0.283^{+0.023}_{-0.021}$ & -- & $-0.743^{+0.153}_{-0.101}$ & $65.73^{+2.31}_{-2.63}$ & $0.347^{+0.033}_{-0.046}$ & $-0.711\pm0.104$ & $0.763\pm0.081$ & $11.24\pm4.32$ & -- & -- & -- \\
 & HzBPA101\tnote{f} & $0.0303^{+0.0047}_{-0.0053}$ & $0.0904^{+0.0196}_{-0.0173}$ & $0.282^{+0.023}_{-0.021}$ & -- & $-0.731^{+0.150}_{-0.096}$ & $65.54^{+2.26}_{-2.58}$ & $0.347^{+0.032}_{-0.046}$ & $-0.711\pm0.102$ & $0.762\pm0.080$ & $11.21\pm4.38$ & $0.420^{+0.029}_{-0.036}$ & $50.14\pm0.26$ & $1.159\pm0.093$ \\
\midrule
 & Platinum & -- & -- & $>0.457$\tnote{d} & $0.011^{+0.821}_{-1.158}$ & $-2.212^{+2.279}_{-1.046}$ & -- & $0.346^{+0.034}_{-0.046}$ & $-0.727\pm0.109$ & $0.736^{+0.102}_{-0.087}$ & $12.62^{+4.65}_{-5.51}$ & -- & -- & -- \\
 & A118 & -- & -- & $>0.257$ & $0.569^{+0.474}_{-0.790}$ & $-2.289^{+1.998}_{-1.087}$ & -- & -- & -- & -- & -- & $0.412^{+0.027}_{-0.033}$ & $50.01\pm0.27$ & $1.120\pm0.088$ \\
 & A101 & -- & -- & $>0.193$ & $0.824^{+0.644}_{-0.749}$ & $-2.338^{+2.038}_{-1.083}$ & -- & -- & -- & -- & -- & $0.421^{+0.030}_{-0.037}$ & $49.91\pm0.30$ & $1.157\pm0.098$ \\
Non-flat & Plat. + A101 & -- & -- & $>0.207$ & $0.743^{+0.618}_{-0.800}$ & $-2.305^{+2.139}_{-1.049}$ & -- & $0.346^{+0.033}_{-0.046}$ & $-0.711\pm0.103$ & $0.759\pm0.081$ & $11.33^{+4.38}_{-4.34}$ & $0.420^{+0.029}_{-0.036}$ & $49.91\pm0.29$ & $1.154\pm0.096$ \\
XCDM & $H(z)$ + BAO & $0.0291^{+0.0055}_{-0.0053}$ & $0.0986^{+0.0218}_{-0.0220}$ & $0.294\pm0.028$ & $-0.116^{+0.135}_{-0.136}$ & $-0.702^{+0.139}_{-0.083}$ & $65.98^{+2.36}_{-2.58}$ & -- & -- & -- & -- & -- & -- & -- \\
 & HzBP\tnote{e} & $0.0292^{+0.0051}_{-0.0057}$ & $0.0983^{+0.0215}_{-0.0218}$ & $0.294\pm0.028$ & $-0.128^{+0.135}_{-0.134}$ & $-0.695^{+0.141}_{-0.080}$ & $65.93^{+2.27}_{-2.53}$ & $0.347^{+0.033}_{-0.046}$ & $-0.713\pm0.103$ & $0.758\pm0.082$ & $11.46\pm4.44$ & -- & -- & -- \\
 & HzBPA101\tnote{f} & $0.0298^{+0.0053}_{-0.0055}$ & $0.0948^{+0.0212}_{-0.0214}$ & $0.290\pm0.027$ & $-0.106\pm0.128$ & $-0.690^{+0.144}_{-0.076}$ & $65.65^{+2.24}_{-2.53}$ & $0.346^{+0.033}_{-0.045}$ & $-0.713\pm0.103$ & $0.758\pm0.081$ & $11.44\pm4.38$ & $0.421^{+0.029}_{-0.036}$ & $50.14\pm0.26$ & $1.151\pm0.094$ \\
\midrule
 & Platinum & -- & -- & $>0.379$\tnote{d} & -- & -- & -- & $0.346^{+0.032}_{-0.044}$ & $-0.715\pm0.102$ & $0.753\pm0.081$ & $11.68\pm4.33$ & -- & -- & -- \\
 & A118 & -- & -- & $0.576^{+0.355}_{-0.202}$ & -- & -- & -- & -- & -- & -- & -- & $0.411^{+0.027}_{-0.033}$ & $50.05\pm0.25$ & $1.110\pm0.089$ \\
 & A101 & -- & -- & $0.506^{+0.239}_{-0.348}$ & -- & -- & -- & -- & -- & -- & -- & $0.421^{+0.030}_{-0.037}$ & $49.99\pm0.27$ & $1.140\pm0.097$ \\
Flat & Plat. + A101 & -- & -- & $0.541^{+0.301}_{-0.297}$ & -- & -- & -- & $0.346^{+0.033}_{-0.045}$ & $-0.716\pm0.103$ & $0.751\pm0.082$ & $11.77\pm4.39$ & $0.420^{+0.029}_{-0.036}$ & $49.99\pm0.26$ & $1.137\pm0.095$ \\
\pcdm & $H(z)$ + BAO & $0.0324^{+0.0060}_{-0.0034}$ & $0.0809^{+0.0180}_{-0.0178}$ & $0.268\pm0.024$ & -- & $1.492^{+0.623}_{-0.854}$ & $65.12\pm2.18$ & -- & -- & -- & -- & -- & -- & -- \\
 & HzBP\tnote{e} & $0.0325^{+0.0071}_{-0.0030}$ & $0.0805^{+0.0177}_{-0.0195}$ & $0.267\pm0.024$ & -- & $1.509^{+0.639}_{-0.904}$ & $65.15^{+2.13}_{-2.34}$ & $0.347^{+0.033}_{-0.045}$ & $-0.711\pm0.104$ & $0.762\pm0.081$ & $11.24\pm4.37$ & -- & -- & -- \\
 & HzBPA101\tnote{f} & $0.0328^{+0.0069}_{-0.0023}$ & $0.0794^{+0.0174}_{-0.0198}$ & $0.266^{+0.024}_{-0.025}$ & -- & $1.568^{+0.652}_{-0.911}$ & $65.03^{+2.10}_{-2.30}$ & $0.347^{+0.033}_{-0.046}$ & $-0.711\pm0.103$ & $0.762\pm0.082$ & $11.23^{+4.39}_{-4.38}$ & $0.420^{+0.029}_{-0.036}$ & $50.14\pm0.26$ & $1.158\pm0.093$ \\
\midrule
 & Platinum & -- & -- & $0.537^{+0.427}_{-0.187}$ & $-0.026^{+0.391}_{-0.384}$ & -- & -- & $0.346^{+0.033}_{-0.046}$ & $-0.717\pm0.105$ & $0.751\pm0.084$ & $11.78^{+4.51}_{-4.52}$ & -- & -- & -- \\
 & A118 & -- & -- & $0.559^{+0.258}_{-0.249}$ & $-0.002^{+0.300}_{-0.292}$ & $5.185^{+3.757}_{-2.564}$ & -- & -- & -- & -- & -- & $0.412^{+0.027}_{-0.033}$ & $50.05\pm0.25$ & $1.110\pm0.090$ \\
 & A101 & -- & -- & $0.477^{+0.199}_{-0.310}$ & $0.107^{+0.306}_{-0.294}$ & $5.205^{+3.974}_{-2.507}$ & -- & -- & -- & -- & -- & $0.421^{+0.030}_{-0.037}$ & $49.99\pm0.27$ & $1.146\pm0.098$ \\
Non-flat & Plat. + A101 & -- & -- & $0.514^{+0.225}_{-0.300}$ & $0.061^{+0.308}_{-0.297}$ & -- & -- & $0.346^{+0.033}_{-0.045}$ & $-0.715\pm0.103$ & $0.754\pm0.083$ & $11.64\pm4.42$ & $0.420^{+0.030}_{-0.036}$ & $49.98\pm0.26$ & $1.142\pm0.097$ \\
\pcdm & $H(z)$ + BAO & $0.0320^{+0.0060}_{-0.0037}$ & $0.0847^{+0.0179}_{-0.0218}$ & $0.272^{+0.025}_{-0.028}$ & $-0.076^{+0.106}_{-0.114}$ & $1.623^{+0.670}_{-0.821}$ & $65.53^{+2.30}_{-2.29}$ & -- & -- & -- & -- & -- & -- & -- \\
 & HzBP\tnote{e} & $0.0321^{+0.0069}_{-0.0031}$ & $0.0847^{+0.0179}_{-0.0226}$ & $0.272^{+0.024}_{-0.029}$ & $-0.090^{+0.092}_{-0.130}$ & $1.674^{+0.697}_{-0.850}$ & $65.57\pm2.27$ & $0.347^{+0.033}_{-0.046}$ & $-0.713\pm0.104$ & $0.760\pm0.081$ & $11.37\pm4.35$ & -- & -- & -- \\
 & HzBPA101\tnote{f} & $0.0324^{+0.0073}_{-0.0024}$ & $0.0829^{+0.0175}_{-0.0223}$ & $0.270^{+0.024}_{-0.029}$ & $-0.079^{+0.104}_{-0.112}$ & $1.714^{+0.713}_{-0.856}$ & $65.39\pm2.24$ & $0.346^{+0.032}_{-0.045}$ & $-0.712\pm0.103$ & $0.760\pm0.082$ & $11.36\pm4.40$ & $0.420^{+0.029}_{-0.036}$ & $50.14\pm0.26$ & $1.153\pm0.094$ \\
\bottomrule
\end{tabular}
\begin{tablenotes}[flushleft]
\item [a] In the four GRB-only cases $\Omega_{b}h^2$ is set to 0.0245.
\item [b] \wx\ corresponds to flat/non-flat XCDM and $\alpha$ corresponds to flat/non-flat \pcdm.
\item [c] \hunit. In the four GRB-only cases $H_0$ is set to 70 \hunit.
\item [d] This is the 1$\sigma$ limit. The 2$\sigma$ limit is set by the prior and not shown here.
\item [e] $H(z)$ + BAO + Platinum.
\item [f] $H(z)$ + BAO + Platinum + A101.
\end{tablenotes}
\end{threeparttable}%
}
\end{sidewaystable*}

\subsection{Constraints from Platinum, A118, A101, and Platinum + A101 data}
\label{PA}

As in \cite{Khadkaetal_2021b} and \cite{CaoKhadkaRatra2021}, in the four GRB-only cases here we set $H_0=70$ \hunit\ and $\Omega_{b}=0.05$.

The constraints on the Platinum GRB correlation parameters in the six different cosmological models are mutually consistent, so the three-dimensional Dainotti correlation Platinum data set is standardizable. The constraints on the intrinsic scatter parameter $\sigma_{\rm int,\,\textsc{p}}$ are almost identical, ranging from $0.346^{+0.032}_{-0.044}$ (flat \pcdm) to $0.347^{+0.034}_{-0.046}$ (non-flat \lcdm). The constraints on the slope $a$ range from a low of $-0.727\pm0.109$ (non-flat \lcdm\ and non-flat XCDM) to a high of $-0.714\pm0.104$ (flat \lcdm), the constraints on the slope $b$ range from a low of $0.736^{+0.102}_{-0.087}$ (non-flat XCDM) to a high of $0.756\pm0.083$ (flat \lcdm), and the constraints on the intercept $C_{o}$ range from a low of $11.52\pm4.45$ (flat \lcdm) to a high of $12.62^{+4.65}_{-5.51}$ (non-flat XCDM), with central values of each pair being $0.09\sigma$, $0.15\sigma$, and $0.16\sigma$ away from each other, respectively. We note that a compilation of (the two-parameter) Dainotti-correlation GRBs, the 31 MD-LGRBs of \cite{Wangetal_2021}, have a somewhat smaller intrinsic dispersion, $\sigma_{\rm int} \sim 0.303-0.306$, Table 5 of \cite{CaoKhadkaRatra2021}, than the $\sigma_{\rm int,\,\textsc{p}} \sim 0.346-0.347$ of the 50 Platinum GRBs.\footnote{There are 12 common GRBs between the Platinum and MD-LGRB data sets (060605, 060906, 061222A, 070306, 080310, 081008, 120404A, 160121A, 160227A, 180329B, 190106A, and 190114A). There are also common GRBs between the Platinum and the (two-parameter) Dainotti-correlated GW-LGRB compilation of \cite{Huetal_2021} [091029, 120118B, 131105A, 170202A, 170705A, and 151027A (same name but different redshifts)]. Because of the significant number of overlapping GRBs in these cases, we believe it would not be that useful to perform joint analyses of the Platinum and truncated (to remove the overlapping GRBs) MD-LGRB or GW-LGRB data sets.} A possible reason for the smaller scatter of the 2-parameter MD-LGRB sample is due to how the sample was chosen. In principle one can retain fewer GRBs that lie exactly on the plane, or are much closer to the plane, thus reducing the scatter. However, further investigation is needed in order to draw a definite conclusion. Probably as a consequence of the smaller $\sigma_{\rm int}$, the MD-LGRB constraints on \om\ and \ok\ are more restrictive than the constraints from Platinum data.

Compared with the analyses in \cite{CaoKhadkaRatra2021} where we neglect massive neutrinos (setting $\onh=0$), here we include a non-zero \onhs. The constraints on the A118 GRB correlation parameters here are almost identical to those listed in Table 8 of \cite{CaoKhadkaRatra2021} and are cosmological-model-independent implying that the A118 GRBs are standardizable. For the 118 GRBs A118 data $\sigma_{\rm int,\,\textsc{a}} \sim 0.411-0.412$, larger than the $\sigma_{\rm int,\,\textsc{p}} \sim 0.346-0.347$ of the 50 Platinum GRBs. The constraints on the A101 GRB correlation parameters are also cosmological-model-independent, so A101 GRBs are also standardizable. The constraints on the A101 intrinsic scatter parameter $\sigma_{\rm int,\,\textsc{a}}$ range from a low of $0.419^{+0.030}_{-0.037}$ (non-flat \lcdm) to a high of $0.422^{+0.030}_{-0.037}$ (flat \lcdm\ and flat XCDM), which are slightly ($0.2\sigma$ at most) higher than those from A118 data; the constraints on the slope $\beta$ range from a low of $1.140\pm0.098$ (flat \lcdm) to a high of $1.157\pm0.098$ (non-flat XCDM), which are slightly ($0.28\sigma$ at most) higher than those from A118 data; and the constraints on the intercept $\gamma$ range from a low of $49.91\pm0.30$ (non-flat XCDM) to a high of $50.10^{+0.29}_{-0.33}$ (flat XCDM), which are slightly ($0.25\sigma$ at most) lower than those from A118 data; with central values of each pair being $0.06\sigma$, $0.12\sigma$, and $0.43\sigma$ away from each other, respectively.

Below we discuss cosmological parameter constraints in more detail. From Fig.\ \ref{fig7} we see that the overlap of the constraints on the cosmological parameters indicate that Platinum data and A101 data cosmological constraints are mutually consistent and so these two data sets can be jointly analyzed. The Platinum + A101 data combination is also standardizable with cosmological-model-independent constraints on GRB correlation parameters that are consistent (well within $1\sigma$) with those from both Platinum data and A101 data individually.

We find that in the flat \lcdm\ model and the flat XCDM parametrization all GRB data more favor currently accelerating cosmological expansion. In the non-flat \lcdm\ model, all but the Platinum GRBs more favor currently decelerating cosmological expansion. In the non-flat XCDM parametrization, in the $\Ok-\Om$ parameter subspace, all but the Platinum GRBs more favor currently decelerating cosmological expansion, while in the $\wX-\Om$ ($\wX-\Ok$) parameter subspace, all (all but the A101) GRBs more favor currently accelerating cosmological expansion. In the flat and non-flat \pcdm\ models, currently decelerating cosmological expansion is slightly more favored.

Cosmological constraints from Platinum data are less restrictive than those from A118 and A101 data, which contain a little more than twice as many GRBs than the Platinum compilation. Comparing the Platinum + A101 constraints with the Platinum constraints and with the A101 constraints, we see that A101 data, with double the number of GRBs compared to Platinum, play a more dominant role in the combination.

In the flat and non-flat \lcdm\ models, the Platinum, A118, A101, and Platinum + A101 $2\sigma$ constraints on \om\ are mutually consistent and also consistent with those of $H(z)$ + BAO, with the $2\sigma$ lower limits in the flat (non-flat) \lcdm\ model being None (None), $>0.256$ ($>0.295$), $>0.191$ ($>0.212$), and $>0.216$ ($>0.235$), respectively. In the flat \lcdm\ model, the A101 constraint on \om\ is more consistent (than the Platinum and Platinum + A101 constraints) with that from $H(z)$ + BAO data, differing by only $1.18\sigma$. In both models Platinum + A101 data favor a higher $1\sigma$ lower limit of \om\ than do Platinum data or A101 data. In the non-flat \lcdm\ model, the constraints on \ok\ from the four GRB data sets are mutually consistent within $1\sigma$, with all but Platinum data slightly favoring open spatial hypersurfaces. Platinum and A118 data are consistent with flat hypersurfaces within $1\sigma$, while A101 and Platinum + A101 data are $1.44\sigma$ and $1.17\sigma$, respectively, away from flat. The posterior mean value of \ok\ from A101 data is $1.36\sigma$ away from that from $H(z)$ + BAO data, while those from the other three GRB data sets are consistent with that from $H(z)$ + BAO data within $1\sigma$.

In the flat and non-flat XCDM parametrizations, the Platinum, A118, A101, and Platinum + A101 $2\sigma$ constraints on \om\ are mutually consistent and also consistent with those of $H(z)$ + BAO, with the $2\sigma$ lower limits in the flat (non-flat) XCDM parametrization being None (None), $>0.181$ ($>0.257$), $>0.148$ ($>0.193$), and $>0.151$ ($>0.207$), respectively. In flat XCDM, the A101 constraint on \om\ is consistent with that from $H(z)$ + BAO within $1\sigma$. Platinum + A101 data also favor a higher $1\sigma$ lower limit of \om\ than do Platinum or A101 data in both XCDM parametrizations. The constraints on \wx\ are weak, thus affected by the \wx\ prior, and consistent with each other. They mildly favor phantom dark energy, but $\Lambda$ is less than $1\sigma$ away, except for the flat XCDM Platinum and Platinum + A101 cases (where $\Lambda$ is still within $2\sigma$), as is the case for the \wx\ constraints from $H(z)$ + BAO data.\footnote{These are computed 2$\sigma$ constraints, not necessarily twice the 1$\sigma$ ones, and are not shown in the table.} In non-flat XCDM, the constraints on \ok\ from the four GRB data sets are mutually consistent within $1\sigma$, and they slightly favor open hypersurfaces. Platinum, A118, and Platinum + A101 data are consistent with flat hypersurfaces within $1\sigma$, while A101 data are $1.10\sigma$ away from flat. The posterior mean value of \ok\ from A101 data is $1.24\sigma$ away from that from $H(z)$ + BAO data, while those from the other three GRB data sets are consistent with that from $H(z)$ + BAO data within $1\sigma$.

In the flat and non-flat \pcdm\ models, the Platinum, A118, A101, and Platinum + A101 constraints on \om\ are mutually consistent within $1\sigma$ and the A101 and Platinum + A101 (Platinum and A118) constraints are within 1$\sigma$ (2$\sigma$) of those from $H(z)$ + BAO data. Only A118 and A101 data provide (very weak) constraints on $\alpha$ with $\Lambda$ being more than $1\sigma$ away but within $2\sigma$. In non-flat \pcdm, the constraints on \ok\ from the four GRB data sets are mutually consistent and consistent with that from $H(z)$ + BAO data and flat hypersurfaces are within $1\sigma$.

From the $\Delta AIC$ and $\Delta BIC$ values listed in Table \ref{tab:BFP}, in the Platinum case, non-flat \lcdm\ is the most favored model, while the evidence against other models are positive, strong, and very strong (non-flat \pcdm). In the A118, A101, and Platinum + A101 cases, the flat \lcdm\ model is the most favored model and, except for non-flat XCDM and non-flat \pcdm\ (with strong $BIC$ evidence against them), the evidence against other models are either weak or positive. However, based on $\Delta DIC$, in theses cases, flat \pcdm\ is the most favored model; in the A118, A101, and Platinum + A101 cases, the evidence against other models are weak; whereas in the Platinum case, the evidence against other models are either weak or strong (non-flat \lcdm\ and non-flat XCDM).

\subsection{Constraints from $H(z)$ + BAO data in combination with Platinum (HzBP) and Platinum + A101 (HzBPA101) data}
\label{HzBPA}

Since the $H(z)$ + BAO, Platinum, and Platinum + A101 cosmological constraints are mutually consistent, we jointly analyze combinations of these data to determine more restrictive constraints on cosmological and GRB correlation parameters. We find that the new constraints on GRB correlation parameters are more restrictive but change less than $1\sigma$ compared to those from the GRB-only analyses. The correlation parameter constraints remain cosmological-model-independent, indicating again that these GRBs are standardizable. In what follows we discuss how the $H(z)$ + BAO data cosmological parameter constraints are altered when these GRB data are jointly analyzed with $H(z)$ + BAO data.

Compared to the $H(z)$ + BAO constraints, the HzBP constraints on \om\ are almost unchanged in all models, but there are small changes in other cosmological parameter constraints. The constraints on $H_0$ are slightly tightened, with posterior means being $0.01-0.03\sigma$ smaller or larger from model to model; the constraints on \wx\ are slightly shifted away from $\Lambda$, with posterior means being $\sim0.04\sigma$ larger; the constraints on $\alpha$ are slightly shifted away from $\Lambda$, with posterior means being $0.02\sigma$ ($0.05\sigma$) larger in flat (non-flat) \pcdm; and the constraints on \ok\ are slightly shifted away from flat, with posterior means being $0.06\sigma$ ($0.10\sigma$) smaller in non-flat XCDM (\pcdm).

Compared to $H(z)$ + BAO, the HzBPA101 data provide mildly different constraints on \om\ in all models, but the effects on other cosmological parameter constraints are more noticeable. The constraints on $H_0$ are slightly tightened, with posterior means being $0.03-0.10\sigma$ smaller from model to model; the constraints on \wx\ are slightly shifted away from $\Lambda$, with posterior means being $0.11\sigma$ ($0.08\sigma$) larger in flat (non-flat) XCDM; the constraints on $\alpha$ are slightly shifted away from $\Lambda$, with posterior means being $0.07\sigma$ ($0.08\sigma$) larger in flat (non-flat) \pcdm; and the constraints on \ok\ are slightly shifted towards (away) from flat, with posterior means being $0.05\sigma$ ($0.02\sigma$) larger (smaller) in non-flat XCDM (\pcdm), respectively.

From the $\Delta AIC$, $\Delta BIC$, and $\Delta DIC$ values listed in Table \ref{tab:BFP}, in both the HzBP and the HzBPA101 cases, following the patterns of the $H(z)$ + BAO case, flat \pcdm\ is the most favored model but the evidence against other models are only either weak or positive. As expected, the better-established $H(z)$ + BAO data play the dominant role in these combined analyses and therefore currently accelerating cosmological expansion is favored.

\section{Conclusion}
\label{sec:conclusion}

We have analyzed the 50 Platinum GRBs, that obey the three-parameter fundamental plane (or Dainotti) relation, using six different cosmological models. By simultaneously constraining cosmological model and GRB correlation parameters our approach circumvents the circularity problem. We find that the Platinum GRB correlation parameters are cosmological-model-independent so the Platinum sample is standardizable through the three-parameter Dainotti correlation and can be used to constrain cosmological parameters. Since the cosmological constraints from Platinum data are consistent with those from $H(z)$ + BAO data, we have combined Platinum and $H(z)$ + BAO data to perform a joint (HzBP) analysis and find mild changes of the cosmological parameter constraints relative to those from $H(z)$ + BAO data, with the central values in the two cases agreeing to two significant figures.

We have also reanalyzed the A118 GRBs that obey the Amati ($E_{\rm p}-E_{\rm iso}$) correlation \citep{Khadkaetal_2021b,CaoKhadkaRatra2021}, because of different, improved modeling of neutrino physics here, and, by excluding the 17 overlapping (with Platinum) GRBs from the larger A118 data set to form the truncated A101 data set, we have also performed a joint analysis of Platinum and A101 data. The twice as large A101 data set plays a dominant role in the joint Platinum + A101 analysis, and the cosmological constraints from Platinum + A101 data are closer to those from A101 data. The results show that the Platinum + A101 GRBs are also standardizable and their cosmological constraints are also more consistent (than the A118 ones) with those from $H(z)$ + BAO data. The joint analyses of $H(z)$ + BAO and Platinum + A101 (HzBPA101) data show that Platinum + A101 data do have more impact on the joint constraints than do Platinum data, however, $H(z)$ + BAO data still play the dominant role.

Current compilations of GRBs provide some improvements on the cosmological constraints determined using $H(z)$ + BAO data. Importantly, GRBs are the only currently reliable probes of the $z \sim 3-8$ part of cosmological redshift space and so are well worth improving upon. With the upcoming SVOM mission \citep{Cordier2019} in 2023, and possibly Theseus \citep{Amatietal2021} in 2037 if approved in the new ESA call, there soon will be larger GRB data sets over a wider range of redshifts that will allow for improved GRB cosmological constraints.

\section*{Acknowledgements}
This research was supported in part by DOE grant DE-SC0011840. The computations for this project were performed on the Beocat Research Cluster at Kansas State University, which is funded in part by NSF grants CNS-1006860, EPS-1006860, EPS-0919443, ACI-1440548, CHE-1726332, and NIH P20GM113109.

\section*{Data availability}
The data underlying this article are listed in Table \ref{tab:P50}.



\bibliographystyle{mnras}
\bibliography{mybibfile} 

\begin{thebibliography}{}
\makeatletter
\relax
\def\mn@urlcharsother{\let\do\@makeother \do\$\do\&\do\#\do\^\do\_\do\%\do\~}
\def\mn@doi{\begingroup\mn@urlcharsother \@ifnextchar [ {\mn@doi@}
  {\mn@doi@[]}}
\def\mn@doi@[#1]#2{\def\@tempa{#1}\ifx\@tempa\@empty \href
  {http://dx.doi.org/#2} {doi:#2}\else \href {http://dx.doi.org/#2} {#1}\fi
  \endgroup}
\def\mn@eprint#1#2{\mn@eprint@#1:#2::\@nil}
\def\mn@eprint@arXiv#1{\href {http://arxiv.org/abs/#1} {{\tt arXiv:#1}}}
\def\mn@eprint@dblp#1{\href {http://dblp.uni-trier.de/rec/bibtex/#1.xml}
  {dblp:#1}}
\def\mn@eprint@#1:#2:#3:#4\@nil{\def\@tempa {#1}\def\@tempb {#2}\def\@tempc
  {#3}\ifx \@tempc \@empty \let \@tempc \@tempb \let \@tempb \@tempa \fi \ifx
  \@tempb \@empty \def\@tempb {arXiv}\fi \@ifundefined
  {mn@eprint@\@tempb}{\@tempb:\@tempc}{\expandafter \expandafter \csname
  mn@eprint@\@tempb\endcsname \expandafter{\@tempc}}}

\bibitem[\protect\citeauthoryear{{Amati}, {Guidorzi}, {Frontera}, {Della
  Valle}, {Finelli}, {Landi}  \& {Montanari}}{{Amati} et~al.}{2008}]{Amati2008}
{Amati} L.,  {Guidorzi} C.,  {Frontera} F.,  {Della Valle} M.,  {Finelli} F.,
  {Landi} R.,   {Montanari} E.,  2008, \mn@doi [\mnras]
  {10.1111/j.1365-2966.2008.13943.x}, \href
  {https://ui.adsabs.harvard.edu/abs/2008MNRAS.391..577A} {391, 577}

\bibitem[\protect\citeauthoryear{{Amati}, {D'Agostino}, {Luongo}, {Muccino}  \&
  {Tantalo}}{{Amati} et~al.}{2019}]{Amati2019}
{Amati} L.,  {D'Agostino} R.,  {Luongo} O.,  {Muccino} M.,   {Tantalo} M.,
  2019, \mn@doi [\mnras] {10.1093/mnrasl/slz056}, \href
  {https://ui.adsabs.harvard.edu/abs/2019MNRAS.486L..46A} {486, L46}

\bibitem[\protect\citeauthoryear{{Amati} et~al.,}{{Amati}
  et~al.}{2021}]{Amatietal2021}
{Amati} L.,  et~al., 2021, \mn@doi [Experimental Astronomy]
  {10.1007/s10686-021-09807-8}, \href
  {https://ui.adsabs.harvard.edu/abs/2021ExA....52..183A} {52, 183}

\bibitem[\protect\citeauthoryear{{Arjona} \& {Nesseris}}{{Arjona} \&
  {Nesseris}}{2021}]{ArjonaNesseris2021}
{Arjona} R.,  {Nesseris} S.,  2021, \mn@doi [\prd]
  {10.1103/PhysRevD.103.103539}, \href
  {https://ui.adsabs.harvard.edu/abs/2021PhRvD.103j3539A} {103, 103539}

\bibitem[\protect\citeauthoryear{{Blas}, {Lesgourgues}  \& {Tram}}{{Blas}
  et~al.}{2011}]{class}
{Blas} D.,  {Lesgourgues} J.,   {Tram} T.,  2011, \mn@doi [\jcap]
  {10.1088/1475-7516/2011/07/034}, \href
  {https://ui.adsabs.harvard.edu/abs/2011JCAP...07..034B} {2011, 034}

\bibitem[\protect\citeauthoryear{{Brinckmann} \& {Lesgourgues}}{{Brinckmann} \&
  {Lesgourgues}}{2019}]{Brinckmann2019}
{Brinckmann} T.,  {Lesgourgues} J.,  2019, \mn@doi [Physics of the Dark
  Universe] {10.1016/j.dark.2018.100260}, \href
  {https://ui.adsabs.harvard.edu/abs/2019PDU....24..260B} {24, 100260}

\bibitem[\protect\citeauthoryear{{Cao}, {Biesiada}, {Jackson}, {Zheng}, {Zhao}
  \& {Zhu}}{{Cao} et~al.}{2017}]{Cao_et_al2017a}
{Cao} S.,  {Biesiada} M.,  {Jackson} J.,  {Zheng} X.,  {Zhao} Y.,   {Zhu}
  Z.-H.,  2017, \mn@doi [\jcap] {10.1088/1475-7516/2017/02/012}, \href
  {http://ads.bao.ac.cn/abs/2017JCAP...02..012C} {2, 012}

\bibitem[\protect\citeauthoryear{{Cao}, {Ryan}  \& {Ratra}}{{Cao}
  et~al.}{2020}]{Caoetal_2020}
{Cao} S.,  {Ryan} J.,   {Ratra} B.,  2020, \mn@doi [\mnras]
  {10.1093/mnras/staa2190}, \href
  {https://ui.adsabs.harvard.edu/abs/2020MNRAS.tmp.2278C} {497, 3191}

\bibitem[\protect\citeauthoryear{{Cao}, {Ryan}, {Khadka}  \& {Ratra}}{{Cao}
  et~al.}{2021a}]{Caoetal_2021a}
{Cao} S.,  {Ryan} J.,  {Khadka} N.,   {Ratra} B.,  2021a, \mn@doi [\mnras]
  {10.1093/mnras/staa3748}, \href
  {https://ui.adsabs.harvard.edu/abs/2020MNRAS.tmp.3537C} {501, 1520}

\bibitem[\protect\citeauthoryear{{Cao}, {Ryan}  \& {Ratra}}{{Cao}
  et~al.}{2021b}]{Caoetal_2021b}
{Cao} S.,  {Ryan} J.,   {Ratra} B.,  2021b, \mn@doi [\mnras]
  {10.1093/mnras/stab942}, \href
  {https://ui.adsabs.harvard.edu/abs/2021MNRAS.504..300C} {504, 300}

\bibitem[\protect\citeauthoryear{{Cao}, {Ryan}  \& {Ratra}}{{Cao}
  et~al.}{2022a}]{Caoetal_2021c}
{Cao} S.,  {Ryan} J.,   {Ratra} B.,  2022a, \mn@doi [\mnras]
  {10.1093/mnras/stab3304}, \href
  {https://ui.adsabs.harvard.edu/abs/2022MNRAS.509.4745C} {509, 4745}

\bibitem[\protect\citeauthoryear{{Cao}, {Khadka}  \& {Ratra}}{{Cao}
  et~al.}{2022b}]{CaoKhadkaRatra2021}
{Cao} S.,  {Khadka} N.,   {Ratra} B.,  2022b, \mn@doi [\mnras]
  {10.1093/mnras/stab3559}, \href
  {https://ui.adsabs.harvard.edu/abs/2021MNRAS.tmp.3230C} {510, 2928}

\bibitem[\protect\citeauthoryear{{Cardone}, {Capozziello}  \&
  {Dainotti}}{{Cardone} et~al.}{2009}]{CardoneCapozzielloDainotti2009}
{Cardone} V.~F.,  {Capozziello} S.,   {Dainotti} M.~G.,  2009, \mn@doi [\mnras]
  {10.1111/j.1365-2966.2009.15456.x}, \href
  {https://ui.adsabs.harvard.edu/abs/2009MNRAS.400..775C} {400, 775}

\bibitem[\protect\citeauthoryear{{Cardone}, {Dainotti}, {Capozziello}  \&
  {Willingale}}{{Cardone} et~al.}{2010}]{Cardoneetal2010}
{Cardone} V.~F.,  {Dainotti} M.~G.,  {Capozziello} S.,   {Willingale} R.,
  2010, \mn@doi [\mnras] {10.1111/j.1365-2966.2010.17197.x}, \href
  {https://ui.adsabs.harvard.edu/abs/2010MNRAS.408.1181C} {408, 1181}

\bibitem[\protect\citeauthoryear{{Ch{\'a}vez}, {Terlevich}, {Terlevich},
  {Bresolin}, {Melnick}, {Plionis}  \& {Basilakos}}{{Ch{\'a}vez}
  et~al.}{2014}]{Chavez_2014}
{Ch{\'a}vez} R.,  {Terlevich} R.,  {Terlevich} E.,  {Bresolin} F.,  {Melnick}
  J.,  {Plionis} M.,   {Basilakos} S.,  2014, \mn@doi [\mnras]
  {10.1093/mnras/stu987}, \href
  {https://ui.adsabs.harvard.edu/abs/2014MNRAS.442.3565C} {442, 3565}

\bibitem[\protect\citeauthoryear{{Chen}, {Ratra}, {Biesiada}, {Li}  \&
  {Zhu}}{{Chen} et~al.}{2016}]{Chenetal2016}
{Chen} Y.,  {Ratra} B.,  {Biesiada} M.,  {Li} S.,   {Zhu} Z.-H.,  2016, \mn@doi
  [\apj] {10.3847/0004-637X/829/2/61}, \href
  {https://ui.adsabs.harvard.edu/abs/2016ApJ...829...61C} {829, 61}

\bibitem[\protect\citeauthoryear{{Chen}, {Kumar}  \& {Ratra}}{{Chen}
  et~al.}{2017}]{chen_etal_2017}
{Chen} Y.,  {Kumar} S.,   {Ratra} B.,  2017, \mn@doi [\apj]
  {10.3847/1538-4357/835/1/86}, \href
  {http://adsabs.harvard.edu/abs/2017ApJ...835...86C} {835, 86}

\bibitem[\protect\citeauthoryear{Cordier}{Cordier}{2019}]{Cordier2019}
Cordier B.,  2019, Mem. Soc. Ast. It., 90, 242

\bibitem[\protect\citeauthoryear{{Cucchiara} et~al.,}{{Cucchiara}
  et~al.}{2011}]{Cucchiaraetal2011}
{Cucchiara} A.,  et~al., 2011, \mn@doi [\apj] {10.1088/0004-637X/736/1/7},
  \href {https://ui.adsabs.harvard.edu/abs/2011ApJ...736....7C} {736, 7}

\bibitem[\protect\citeauthoryear{{Czerny} et~al.,}{{Czerny}
  et~al.}{2021}]{Czernyetal2021}
{Czerny} B.,  et~al., 2021, \mn@doi [Acta Physica Polonica A]
  {10.12693/APhysPolA.139.389}, \href
  {https://ui.adsabs.harvard.edu/abs/2021AcPPA.139..389C} {139, 389}

\bibitem[\protect\citeauthoryear{{D'Agostini}}{{D'Agostini}}{2005}]{D'Agostini_2005}
{D'Agostini} G.,  2005, preprint, \href
  {https://ui.adsabs.harvard.edu/abs/2005physics..11182D} {} (\mn@eprint
  {arXiv} {physics/0511182})

\bibitem[\protect\citeauthoryear{{DES Collaboration}}{{DES
  Collaboration}}{2019}]{DESCollaboration2019}
{DES Collaboration} 2019, \mn@doi [\prd] {10.1103/PhysRevD.99.123505}, \href
  {https://ui.adsabs.harvard.edu/abs/2019PhRvD..99l3505A} {99, 123505}

\bibitem[\protect\citeauthoryear{{Dai}, {Zheng}, {Li}, {Gao}  \& {Zhu}}{{Dai}
  et~al.}{2021}]{Daietal_2021}
{Dai} Y.,  {Zheng} X.-G.,  {Li} Z.-X.,  {Gao} H.,   {Zhu} Z.-H.,  2021, \mn@doi
  [\aap] {10.1051/0004-6361/202140895}, \href
  {https://ui.adsabs.harvard.edu/abs/2021A&A...651L...8D} {651, L8}

\bibitem[\protect\citeauthoryear{{Dainotti}, {Cardone}  \&
  {Capozziello}}{{Dainotti} et~al.}{2008}]{Dainottietal2008}
{Dainotti} M.~G.,  {Cardone} V.~F.,   {Capozziello} S.,  2008, \mn@doi [\mnras]
  {10.1111/j.1745-3933.2008.00560.x}, \href
  {https://ui.adsabs.harvard.edu/abs/2008MNRAS.391L..79D} {391, L79}

\bibitem[\protect\citeauthoryear{{Dainotti}, {Willingale}, {Capozziello},
  {Fabrizio Cardone}  \& {Ostrowski}}{{Dainotti}
  et~al.}{2010}]{Dainottietal2010}
{Dainotti} M.~G.,  {Willingale} R.,  {Capozziello} S.,  {Fabrizio Cardone} V.,
   {Ostrowski} M.,  2010, \mn@doi [\apjl] {10.1088/2041-8205/722/2/L215}, \href
  {https://ui.adsabs.harvard.edu/abs/2010ApJ...722L.215D} {722, L215}

\bibitem[\protect\citeauthoryear{{Dainotti}, {Fabrizio Cardone}, {Capozziello},
  {Ostrowski}  \& {Willingale}}{{Dainotti} et~al.}{2011}]{Dainottietal2011}
{Dainotti} M.~G.,  {Fabrizio Cardone} V.,  {Capozziello} S.,  {Ostrowski} M.,
  {Willingale} R.,  2011, \mn@doi [\apj] {10.1088/0004-637X/730/2/135}, \href
  {https://ui.adsabs.harvard.edu/abs/2011ApJ...730..135D} {730, 135}

\bibitem[\protect\citeauthoryear{{Dainotti}, {Cardone}, {Piedipalumbo}  \&
  {Capozziello}}{{Dainotti} et~al.}{2013a}]{Dainottietal2013a}
{Dainotti} M.~G.,  {Cardone} V.~F.,  {Piedipalumbo} E.,   {Capozziello} S.,
  2013a, \mn@doi [\mnras] {10.1093/mnras/stt1516}, \href
  {https://ui.adsabs.harvard.edu/abs/2013MNRAS.436...82D} {436, 82}

\bibitem[\protect\citeauthoryear{{Dainotti}, {Petrosian}, {Singal}  \&
  {Ostrowski}}{{Dainotti} et~al.}{2013b}]{Dainottietal2013b}
{Dainotti} M.~G.,  {Petrosian} V.,  {Singal} J.,   {Ostrowski} M.,  2013b,
  \mn@doi [\apj] {10.1088/0004-637X/774/2/157}, \href
  {https://ui.adsabs.harvard.edu/abs/2013ApJ...774..157D} {774, 157}

\bibitem[\protect\citeauthoryear{{Dainotti}, {Petrosian}, {Willingale},
  {O'Brien}, {Ostrowski}  \& {Nagataki}}{{Dainotti}
  et~al.}{2015}]{Dainottietal2015}
{Dainotti} M.,  {Petrosian} V.,  {Willingale} R.,  {O'Brien} P.,  {Ostrowski}
  M.,   {Nagataki} S.,  2015, \mn@doi [\mnras] {10.1093/mnras/stv1229}, \href
  {https://ui.adsabs.harvard.edu/abs/2015MNRAS.451.3898D} {451, 3898}

\bibitem[\protect\citeauthoryear{{Dainotti}, {Postnikov}, {Hernandez}  \&
  {Ostrowski}}{{Dainotti} et~al.}{2016}]{Dainottietal2016}
{Dainotti} M.~G.,  {Postnikov} S.,  {Hernandez} X.,   {Ostrowski} M.,  2016,
  \mn@doi [\apjl] {10.3847/2041-8205/825/2/L20}, \href
  {https://ui.adsabs.harvard.edu/abs/2016ApJ...825L..20D} {825, L20}

\bibitem[\protect\citeauthoryear{{Dainotti}, {Nagataki}, {Maeda}, {Postnikov}
  \& {Pian}}{{Dainotti} et~al.}{2017}]{Dainottietal2017}
{Dainotti} M.~G.,  {Nagataki} S.,  {Maeda} K.,  {Postnikov} S.,   {Pian} E.,
  2017, \mn@doi [\aap] {10.1051/0004-6361/201628384}, \href
  {https://ui.adsabs.harvard.edu/abs/2017A&A...600A..98D} {600, A98}

\bibitem[\protect\citeauthoryear{{Dainotti}, {Lenart}, {Sarracino}, {Nagataki},
  {Capozziello}  \& {Fraija}}{{Dainotti} et~al.}{2020}]{Dainottietal2020}
{Dainotti} M.~G.,  {Lenart} A.~{\L}.,  {Sarracino} G.,  {Nagataki} S.,
  {Capozziello} S.,   {Fraija} N.,  2020, \mn@doi [\apj]
  {10.3847/1538-4357/abbe8a}, \href
  {https://ui.adsabs.harvard.edu/abs/2020ApJ...904...97D} {904, 97}

\bibitem[\protect\citeauthoryear{{Dainotti}, {Lenart}, {Fraija}, {Nagataki},
  {Warren}, {De Simone}, {Srinivasaragavan}  \& {Mata}}{{Dainotti}
  et~al.}{2021a}]{Dainottietal2021}
{Dainotti} M.~G.,  {Lenart} A.~{\L}.,  {Fraija} N.,  {Nagataki} S.,  {Warren}
  D.~C.,  {De Simone} B.,  {Srinivasaragavan} G.,   {Mata} A.,  2021a, \mn@doi
  [\pasj] {10.1093/pasj/psab057}, \href
  {https://ui.adsabs.harvard.edu/abs/2021PASJ...73..970D} {73, 970}

\bibitem[\protect\citeauthoryear{{Dainotti}, {De Simone}, {Schiavone},
  {Montani}, {Rinaldi}  \& {Lambiase}}{{Dainotti}
  et~al.}{2021b}]{Dainottietal2021a}
{Dainotti} M.~G.,  {De Simone} B.,  {Schiavone} T.,  {Montani} G.,  {Rinaldi}
  E.,   {Lambiase} G.,  2021b, \mn@doi [\apj] {10.3847/1538-4357/abeb73}, \href
  {https://ui.adsabs.harvard.edu/abs/2021ApJ...912..150D} {912, 150}

\bibitem[\protect\citeauthoryear{{Dainotti}, {De Simone}, {Schiavone},
  {Montani}, {Rinaldi}, {Lambiase}, {Bogdan}  \& {Ugale}}{{Dainotti}
  et~al.}{2022}]{Dainottietal2022}
{Dainotti} M.~G.,  {De Simone} B.~D.,  {Schiavone} T.,  {Montani} G.,
  {Rinaldi} E.,  {Lambiase} G.,  {Bogdan} M.,   {Ugale} S.,  2022, \mn@doi
  [Galaxies] {10.3390/galaxies10010024}, \href
  {https://ui.adsabs.harvard.edu/abs/2022Galax..10...24D} {10, 24}

\bibitem[\protect\citeauthoryear{{\MakeLowercase{D}e Cruz Perez}, {Sola
  Peracaula}, {Gomez-Valent}  \& {Moreno-Pulido}}{{\MakeLowercase{D}e Cruz
  Perez} et~al.}{2021}]{deCruzetal2021}
{\MakeLowercase{D}e Cruz Perez} J.,  {Sola Peracaula} J.,  {Gomez-Valent} A.,
  {Moreno-Pulido} C.,  2021, preprint, \href
  {https://ui.adsabs.harvard.edu/abs/2021arXiv211007569D} {} (\mn@eprint {}
  {2110.07569})

\bibitem[\protect\citeauthoryear{Demianski, Piedipalumbo, Sawant  \&
  Amati}{Demianski et~al.}{2021}]{Demianskietal_2021}
Demianski M.,  Piedipalumbo E.,  Sawant D.,   Amati L.,  2021, \mn@doi [\mnras]
  {10.1093/mnras/stab1669}, 506, 903

\bibitem[\protect\citeauthoryear{{Dhawan}, {Alsing}  \& {Vagnozzi}}{{Dhawan}
  et~al.}{2021}]{Dhawanetal2021}
{Dhawan} S.,  {Alsing} J.,   {Vagnozzi} S.,  2021, \mn@doi [\mnras]
  {10.1093/mnrasl/slab058}, \href
  {https://ui.adsabs.harvard.edu/abs/2021MNRAS.506L...1D} {506, L1}

\bibitem[\protect\citeauthoryear{{Di Valentino} et~al.,}{{Di Valentino}
  et~al.}{2021a}]{DiValentinoetal2021b}
{Di Valentino} E.,  et~al., 2021a, \mn@doi [Classical and Quantum Gravity]
  {10.1088/1361-6382/ac086d}, \href
  {https://ui.adsabs.harvard.edu/abs/2021CQGra..38o3001D} {38, 153001}

\bibitem[\protect\citeauthoryear{{Di Valentino}, {Melchiorri}  \& {Silk}}{{Di
  Valentino} et~al.}{2021b}]{DiValentinoetal2021a}
{Di Valentino} E.,  {Melchiorri} A.,   {Silk} J.,  2021b, \mn@doi [\apjl]
  {10.3847/2041-8213/abe1c4}, \href
  {https://ui.adsabs.harvard.edu/abs/2021ApJ...908L...9D} {908, L9}

\bibitem[\protect\citeauthoryear{{\MakeLowercase{E}BOSS
  Collaboration}}{{\MakeLowercase{E}BOSS Collaboration}}{2021}]{eBOSS_2020}
{\MakeLowercase{E}BOSS Collaboration} 2021, \mn@doi [\prd]
  {10.1103/PhysRevD.103.083533}, \href
  {https://ui.adsabs.harvard.edu/abs/2021PhRvD.103h3533A} {103, 083533}

\bibitem[\protect\citeauthoryear{{Efstathiou} \& {Gratton}}{{Efstathiou} \&
  {Gratton}}{2020}]{EfstathiouGratton2020}
{Efstathiou} G.,  {Gratton} S.,  2020, \mn@doi [\mnras]
  {10.1093/mnrasl/slaa093}, \href
  {https://ui.adsabs.harvard.edu/abs/2020MNRAS.496L..91E} {496, L91}

\bibitem[\protect\citeauthoryear{{Fana Dirirsa} et~al.,}{{Fana Dirirsa}
  et~al.}{2019}]{Dirirsa2019}
{Fana Dirirsa} F.,  et~al., 2019, \mn@doi [\apj] {10.3847/1538-4357/ab4e11},
  \href {https://ui.adsabs.harvard.edu/abs/2019ApJ...887...13F} {887, 13}

\bibitem[\protect\citeauthoryear{{Farooq}, {Ranjeet Madiyar}, {Crandall}  \&
  {Ratra}}{{Farooq} et~al.}{2017}]{Farooq_Ranjeet_Crandall_Ratra_2017}
{Farooq} O.,  {Ranjeet Madiyar} F.,  {Crandall} S.,   {Ratra} B.,  2017,
  \mn@doi [\apj] {10.3847/1538-4357/835/1/26}, \href
  {http://adsabs.harvard.edu/abs/2017ApJ...835...26F} {835, 26}

\bibitem[\protect\citeauthoryear{{Geng}, {Hsu}  \& {Lu}}{{Geng}
  et~al.}{2022}]{Gengetal2021}
{Geng} C.-Q.,  {Hsu} Y.-T.,   {Lu} J.-R.,  2022, \mn@doi [\apj]
  {10.3847/1538-4357/ac4495}, \href
  {https://ui.adsabs.harvard.edu/abs/2022ApJ...926...74G} {926, 74}

\bibitem[\protect\citeauthoryear{{Gonz{\'a}lez-Mor{\'a}n}
  et~al.,}{{Gonz{\'a}lez-Mor{\'a}n} et~al.}{2019}]{G-M_2019}
{Gonz{\'a}lez-Mor{\'a}n} A.~L.,  et~al., 2019, \mn@doi [\mnras]
  {10.1093/mnras/stz1577}, \href
  {https://ui.adsabs.harvard.edu/abs/2019MNRAS.487.4669G} {487, 4669}

\bibitem[\protect\citeauthoryear{{Gonz{\'a}lez-Mor{\'a}n}
  et~al.,}{{Gonz{\'a}lez-Mor{\'a}n} et~al.}{2021}]{GM2021}
{Gonz{\'a}lez-Mor{\'a}n} A.~L.,  et~al., 2021, \mn@doi [\mnras]
  {10.1093/mnras/stab1385}, \href
  {https://ui.adsabs.harvard.edu/abs/2021MNRAS.505.1441G} {505, 1441}

\bibitem[\protect\citeauthoryear{{Handley}}{{Handley}}{2019}]{Handley2019}
{Handley} W.,  2019, \mn@doi [\prd] {10.1103/PhysRevD.100.123517}, \href
  {https://ui.adsabs.harvard.edu/abs/2019PhRvD.100l3517H} {100, 123517}

\bibitem[\protect\citeauthoryear{{Hu}, {Wang}  \& {Dai}}{{Hu}
  et~al.}{2021}]{Huetal_2021}
{Hu} J.~P.,  {Wang} F.~Y.,   {Dai} Z.~G.,  2021, \mn@doi [\mnras]
  {10.1093/mnras/stab2180}, \href
  {https://ui.adsabs.harvard.edu/abs/2021MNRAS.507..730H} {507, 730}

\bibitem[\protect\citeauthoryear{{Jesus}, {Valentim}, {Escobal}, {Pereira}  \&
  {Benndorf}}{{Jesus} et~al.}{2021}]{Jesusetal2021}
{Jesus} J.~F.,  {Valentim} R.,  {Escobal} A.~A.,  {Pereira} S.~H.,   {Benndorf}
  D.,  2021, preprint, \href
  {https://ui.adsabs.harvard.edu/abs/2021arXiv211209722J} {} (\mn@eprint {}
  {2112.09722})

\bibitem[\protect\citeauthoryear{{Johnson}, {Sangwan}  \&
  {Shankaranarayanan}}{{Johnson} et~al.}{2022}]{Johnsonetal2021}
{Johnson} J.~P.,  {Sangwan} A.,   {Shankaranarayanan} S.,  2022, \mn@doi
  [\jcap] {10.1088/1475-7516/2022/01/024}, \href
  {https://ui.adsabs.harvard.edu/abs/2022JCAP...01..024J} {2022, 024}

\bibitem[\protect\citeauthoryear{{Khadka} \& {Ratra}}{{Khadka} \&
  {Ratra}}{2020a}]{KhadkaRatra2020a}
{Khadka} N.,  {Ratra} B.,  2020a, \mn@doi [\mnras] {10.1093/mnras/staa101},
  \href {https://ui.adsabs.harvard.edu/abs/2020MNRAS.492.4456K} {492, 4456}

\bibitem[\protect\citeauthoryear{{Khadka} \& {Ratra}}{{Khadka} \&
  {Ratra}}{2020b}]{KhadkaRatra2020b}
{Khadka} N.,  {Ratra} B.,  2020b, \mn@doi [\mnras] {10.1093/mnras/staa1855},
  \href {https://ui.adsabs.harvard.edu/abs/2020MNRAS.497..263K} {497, 263}

\bibitem[\protect\citeauthoryear{{Khadka} \& {Ratra}}{{Khadka} \&
  {Ratra}}{2020c}]{KhadkaRatra2020c}
{Khadka} N.,  {Ratra} B.,  2020c, \mn@doi [\mnras] {10.1093/mnras/staa2779},
  \href {https://ui.adsabs.harvard.edu/abs/2020MNRAS.499..391K} {499, 391}

\bibitem[\protect\citeauthoryear{{Khadka} \& {Ratra}}{{Khadka} \&
  {Ratra}}{2021}]{KhadkaRatra2021}
{Khadka} N.,  {Ratra} B.,  2021, \mn@doi [\mnras] {10.1093/mnras/stab486},
  \href {https://ui.adsabs.harvard.edu/abs/2021MNRAS.502.6140K} {502, 6140}

\bibitem[\protect\citeauthoryear{{Khadka} \& {Ratra}}{{Khadka} \&
  {Ratra}}{2022}]{KhadkaRatra2022}
{Khadka} N.,  {Ratra} B.,  2022, \mn@doi [\mnras] {10.1093/mnras/stab3678},
  \href {https://ui.adsabs.harvard.edu/abs/2021MNRAS.tmp.3383K} {510, 2753}

\bibitem[\protect\citeauthoryear{{Khadka}, {Mart{\'\i}nez-Aldama},
  {Zaja{\v{c}}ek}, {Czerny}  \& {Ratra}}{{Khadka}
  et~al.}{2021a}]{Khadkaetal2021c}
{Khadka} N.,  {Mart{\'\i}nez-Aldama} M.~L.,  {Zaja{\v{c}}ek} M.,  {Czerny} B.,
   {Ratra} B.,  2021a, preprint, \href
  {https://ui.adsabs.harvard.edu/abs/2021arXiv211200052K} {} (\mn@eprint {}
  {2112.00052})

\bibitem[\protect\citeauthoryear{{Khadka}, {Yu}, {Zaja{\v{c}}ek},
  {Martinez-Aldama}, {Czerny}  \& {Ratra}}{{Khadka}
  et~al.}{2021b}]{Khadkaetal_2021a}
{Khadka} N.,  {Yu} Z.,  {Zaja{\v{c}}ek} M.,  {Martinez-Aldama} M.~L.,  {Czerny}
  B.,   {Ratra} B.,  2021b, \mn@doi [\mnras] {10.1093/mnras/stab2807}, \href
  {https://ui.adsabs.harvard.edu/abs/2021MNRAS.508.4722K} {508, 4722}

\bibitem[\protect\citeauthoryear{{Khadka}, {Luongo}, {Muccino}  \&
  {Ratra}}{{Khadka} et~al.}{2021c}]{Khadkaetal_2021b}
{Khadka} N.,  {Luongo} O.,  {Muccino} M.,   {Ratra} B.,  2021c, \mn@doi [\jcap]
  {10.1088/1475-7516/2021/09/042}, \href
  {https://ui.adsabs.harvard.edu/abs/2021JCAP...09..042K} {2021, 042}

\bibitem[\protect\citeauthoryear{{KiDS Collaboration}}{{KiDS
  Collaboration}}{2021}]{KiDSCollaboration2021}
{KiDS Collaboration} 2021, \mn@doi [\aap] {10.1051/0004-6361/202039805}, \href
  {https://ui.adsabs.harvard.edu/abs/2021A&A...649A..88T} {649, A88}

\bibitem[\protect\citeauthoryear{{Kunz}, {Trotta}  \& {Parkinson}}{{Kunz}
  et~al.}{2006}]{KunzTrottaParkinson2006}
{Kunz} M.,  {Trotta} R.,   {Parkinson} D.~R.,  2006, \mn@doi [\prd]
  {10.1103/PhysRevD.74.023503}, \href
  {https://ui.adsabs.harvard.edu/abs/2006PhRvD..74b3503K} {74, 023503}

\bibitem[\protect\citeauthoryear{{Lamb} \& {Reichart}}{{Lamb} \&
  {Reichart}}{2000}]{LambReichart2000}
{Lamb} D.~Q.,  {Reichart} D.~E.,  2000, \mn@doi [\apj] {10.1086/308918}, \href
  {https://ui.adsabs.harvard.edu/abs/2000ApJ...536....1L} {536, 1}

\bibitem[\protect\citeauthoryear{{Lewis}}{{Lewis}}{2019}]{Lewis_2019}
{Lewis} A.,  2019, preprint, \href
  {https://ui.adsabs.harvard.edu/abs/2019arXiv191013970L} {} (\mn@eprint
  {arXiv} {1910.13970})

\bibitem[\protect\citeauthoryear{{Li}, {Du}  \& {Xu}}{{Li}
  et~al.}{2020}]{Lietal2020}
{Li} E.-K.,  {Du} M.,   {Xu} L.,  2020, \mn@doi [\mnras]
  {10.1093/mnras/stz3308}, \href
  {https://ui.adsabs.harvard.edu/abs/2020MNRAS.491.4960L} {491, 4960}

\bibitem[\protect\citeauthoryear{{Li}, {Keeley}, {Shafieloo}, {Zheng}, {Cao},
  {Biesiada}  \& {Zhu}}{{Li} et~al.}{2021}]{Lietal2021}
{Li} X.,  {Keeley} R.~E.,  {Shafieloo} A.,  {Zheng} X.,  {Cao} S.,  {Biesiada}
  M.,   {Zhu} Z.-H.,  2021, \mn@doi [\mnras] {10.1093/mnras/stab2154}, \href
  {https://ui.adsabs.harvard.edu/abs/2021MNRAS.507..919L} {507, 919}

\bibitem[\protect\citeauthoryear{{Lian}, {Cao}, {Biesiada}, {Chen}, {Zhang}  \&
  {Guo}}{{Lian} et~al.}{2021}]{Lian_etal_2021}
{Lian} Y.,  {Cao} S.,  {Biesiada} M.,  {Chen} Y.,  {Zhang} Y.,   {Guo} W.,
  2021, \mn@doi [\mnras] {10.1093/mnras/stab1373}, \href
  {https://ui.adsabs.harvard.edu/abs/2021MNRAS.505.2111L} {505, 2111}

\bibitem[\protect\citeauthoryear{Luongo \& Muccino}{Luongo \&
  Muccino}{2021}]{galaxies9040077}
Luongo O.,  Muccino M.,  2021, \mn@doi [Galaxies] {10.3390/galaxies9040077}, 9

\bibitem[\protect\citeauthoryear{{Luongo}, {Muccino}, {Colg{\'a}in},
  {Sheikh-Jabbari}  \& {Yin}}{{Luongo} et~al.}{2021}]{Luongoetal2021}
{Luongo} O.,  {Muccino} M.,  {Colg{\'a}in} E.~{\'O}.,  {Sheikh-Jabbari} M.~M.,
   {Yin} L.,  2021, preprint, \href
  {https://ui.adsabs.harvard.edu/abs/2021arXiv210813228L} {} (\mn@eprint {}
  {2108.13228})

\bibitem[\protect\citeauthoryear{{Lusso} et~al.,}{{Lusso}
  et~al.}{2020}]{Lussoetal2020}
{Lusso} E.,  et~al., 2020, \mn@doi [\aap] {10.1051/0004-6361/202038899}, \href
  {https://ui.adsabs.harvard.edu/abs/2020A&A...642A.150L} {642, A150}

\bibitem[\protect\citeauthoryear{{Mania} \& {Ratra}}{{Mania} \&
  {Ratra}}{2012}]{Mania_2012}
{Mania} D.,  {Ratra} B.,  2012, \mn@doi [Physics Letters B]
  {10.1016/j.physletb.2012.07.011}, \href
  {https://ui.adsabs.harvard.edu/abs/2012PhLB..715....9M} {715, 9}

\bibitem[\protect\citeauthoryear{{Mehrabi} et~al.,}{{Mehrabi}
  et~al.}{2022}]{Mehrabietal2022}
{Mehrabi} A.,  et~al., 2022, \mn@doi [\mnras] {10.1093/mnras/stab2915}, \href
  {https://ui.adsabs.harvard.edu/abs/2022MNRAS.509..224M} {509, 224}

\bibitem[\protect\citeauthoryear{{Ooba}, {Ratra}  \& {Sugiyama}}{{Ooba}
  et~al.}{2018a}]{Oobaetal2018a}
{Ooba} J.,  {Ratra} B.,   {Sugiyama} N.,  2018a, \mn@doi [\apj]
  {10.3847/1538-4357/aad633}, \href
  {https://ui.adsabs.harvard.edu/abs/2018ApJ...864...80O} {864, 80}

\bibitem[\protect\citeauthoryear{{Ooba}, {Ratra}  \& {Sugiyama}}{{Ooba}
  et~al.}{2018b}]{ooba_etal_2018b}
{Ooba} J.,  {Ratra} B.,   {Sugiyama} N.,  2018b, \mn@doi [\apj]
  {10.3847/1538-4357/aadcf3}, \href
  {http://adsabs.harvard.edu/abs/2018ApJ...866...68O} {866, 68}

\bibitem[\protect\citeauthoryear{{Ooba}, {Ratra}  \& {Sugiyama}}{{Ooba}
  et~al.}{2018c}]{Oobaetal2018b}
{Ooba} J.,  {Ratra} B.,   {Sugiyama} N.,  2018c, \mn@doi [\apj]
  {10.3847/1538-4357/aaec6f}, \href
  {https://ui.adsabs.harvard.edu/abs/2018ApJ...869...34O} {869, 34}

\bibitem[\protect\citeauthoryear{{Ooba}, {Ratra}  \& {Sugiyama}}{{Ooba}
  et~al.}{2019}]{ooba_etal_2019}
{Ooba} J.,  {Ratra} B.,   {Sugiyama} N.,  2019, \mn@doi [\apss]
  {10.1007/s10509-019-3663-4}, \href
  {https://ui.adsabs.harvard.edu/abs/2019Ap&SS.364..176O} {364, 176}

\bibitem[\protect\citeauthoryear{{Park} \& {Ratra}}{{Park} \&
  {Ratra}}{2018}]{park_ratra_2018}
{Park} C.-G.,  {Ratra} B.,  2018, \mn@doi [\apj] {10.3847/1538-4357/aae82d},
  \href {http://adsabs.harvard.edu/abs/2018ApJ...868...83P} {868, 83}

\bibitem[\protect\citeauthoryear{{Park} \& {Ratra}}{{Park} \&
  {Ratra}}{2019a}]{ParkRatra2019b}
{Park} C.-G.,  {Ratra} B.,  2019a, \mn@doi [\apss] {10.1007/s10509-019-3567-3},
  \href {https://ui.adsabs.harvard.edu/abs/2019Ap&SS.364...82P} {364, 82}

\bibitem[\protect\citeauthoryear{{Park} \& {Ratra}}{{Park} \&
  {Ratra}}{2019b}]{park_ratra_2019b}
{Park} C.-G.,  {Ratra} B.,  2019b, \mn@doi [\apss] {10.1007/s10509-019-3627-8},
  \href {https://ui.adsabs.harvard.edu/abs/2019Ap&SS.364..134P} {364, 134}

\bibitem[\protect\citeauthoryear{{Park} \& {Ratra}}{{Park} \&
  {Ratra}}{2019c}]{ParkRatra2019a}
{Park} C.-G.,  {Ratra} B.,  2019c, \mn@doi [\apj] {10.3847/1538-4357/ab3641},
  \href {https://ui.adsabs.harvard.edu/abs/2019ApJ...882..158P} {882, 158}

\bibitem[\protect\citeauthoryear{{Park} \& {Ratra}}{{Park} \&
  {Ratra}}{2020}]{park_ratra_2020}
{Park} C.-G.,  {Ratra} B.,  2020, \mn@doi [\prd] {10.1103/PhysRevD.101.083508},
  \href {https://ui.adsabs.harvard.edu/abs/2020PhRvD.101h3508P} {101, 083508}

\bibitem[\protect\citeauthoryear{{Pavlov}, {Westmoreland}, {Saaidi}  \&
  {Ratra}}{{Pavlov} et~al.}{2013}]{pavlov13}
{Pavlov} A.,  {Westmoreland} S.,  {Saaidi} K.,   {Ratra} B.,  2013, \mn@doi
  [\prd] {10.1103/PhysRevD.88.123513}, \href
  {http://adsabs.harvard.edu/abs/2013PhRvD..88l3513P} {88, 123513}

\bibitem[\protect\citeauthoryear{{Peebles}}{{Peebles}}{1984}]{peeb84}
{Peebles} P.~J.~E.,  1984, \mn@doi [\apj] {10.1086/162425}, \href
  {http://adsabs.harvard.edu/abs/1984ApJ...284..439P} {284, 439}

\bibitem[\protect\citeauthoryear{{Peebles} \& {Ratra}}{{Peebles} \&
  {Ratra}}{1988}]{peebrat88}
{Peebles} P.~J.~E.,  {Ratra} B.,  1988, \mn@doi [\apjl] {10.1086/185100}, \href
  {http://adsabs.harvard.edu/abs/1988ApJ...325L..17P} {325, L17}

\bibitem[\protect\citeauthoryear{{Perivolaropoulos} \&
  {Skara}}{{Perivolaropoulos} \& {Skara}}{2021}]{PerivolaropoulosSkara2021}
{Perivolaropoulos} L.,  {Skara} F.,  2021, preprint, \href
  {https://ui.adsabs.harvard.edu/abs/2021arXiv210505208P} {} (\mn@eprint {}
  {2105.05208})

\bibitem[\protect\citeauthoryear{{Planck Collaboration}}{{Planck
  Collaboration}}{2020}]{planck2018b}
{Planck Collaboration} 2020, \mn@doi [\aap] {10.1051/0004-6361/201833910},
  \href {https://ui.adsabs.harvard.edu/abs/2020A&A...641A...6P} {641, A6}

\bibitem[\protect\citeauthoryear{{Postnikov}, {Dainotti}, {Hernandez}  \&
  {Capozziello}}{{Postnikov} et~al.}{2014}]{Postnikovetal2014}
{Postnikov} S.,  {Dainotti} M.~G.,  {Hernandez} X.,   {Capozziello} S.,  2014,
  \mn@doi [\apj] {10.1088/0004-637X/783/2/126}, \href
  {https://ui.adsabs.harvard.edu/abs/2014ApJ...783..126P} {783, 126}

\bibitem[\protect\citeauthoryear{{Rana}, {Jain}, {Mahajan}  \&
  {Mukherjee}}{{Rana} et~al.}{2017}]{Ranaetal2017}
{Rana} A.,  {Jain} D.,  {Mahajan} S.,   {Mukherjee} A.,  2017, \mn@doi [\jcap]
  {10.1088/1475-7516/2017/03/028}, \href
  {https://ui.adsabs.harvard.edu/abs/2017JCAP...03..028R} {2017, 028}

\bibitem[\protect\citeauthoryear{{Ratra} \& {Peebles}}{{Ratra} \&
  {Peebles}}{1988}]{ratpeeb88}
{Ratra} B.,  {Peebles} P.~J.~E.,  1988, \mn@doi [\prd]
  {10.1103/PhysRevD.37.3406}, \href
  {http://adsabs.harvard.edu/abs/1988PhRvD..37.3406R} {37, 3406}

\bibitem[\protect\citeauthoryear{{Renzi}, {Hogg}  \& {Giar{\`e}}}{{Renzi}
  et~al.}{2021}]{Renzietal2021}
{Renzi} F.,  {Hogg} N.~B.,   {Giar{\`e}} W.,  2021, preprint, \href
  {https://ui.adsabs.harvard.edu/abs/2021arXiv211205701R} {} (\mn@eprint {}
  {2112.05701})

\bibitem[\protect\citeauthoryear{Rezaei, Solà Peracaula  \& Malekjani}{Rezaei
  et~al.}{2021}]{Rezaeietal2021}
Rezaei M.,  Solà Peracaula J.,   Malekjani M.,  2021, \mn@doi [\mnras]
  {10.1093/mnras/stab3117}, 509, 2593

\bibitem[\protect\citeauthoryear{{Risaliti} \& {Lusso}}{{Risaliti} \&
  {Lusso}}{2015}]{RisalitiLusso2015}
{Risaliti} G.,  {Lusso} E.,  2015, \mn@doi [\apj] {10.1088/0004-637X/815/1/33},
  \href {https://ui.adsabs.harvard.edu/abs/2015ApJ...815...33R} {815, 33}

\bibitem[\protect\citeauthoryear{{Risaliti} \& {Lusso}}{{Risaliti} \&
  {Lusso}}{2019}]{RisalitiLusso2019}
{Risaliti} G.,  {Lusso} E.,  2019, \mn@doi [Nature Astronomy]
  {10.1038/s41550-018-0657-z}, \href
  {https://ui.adsabs.harvard.edu/abs/2019NatAs...3..272R} {3, 272}

\bibitem[\protect\citeauthoryear{{Ryan}, {Doshi}  \& {Ratra}}{{Ryan}
  et~al.}{2018}]{Ryan_1}
{Ryan} J.,  {Doshi} S.,   {Ratra} B.,  2018, \mn@doi [\mnras]
  {10.1093/mnras/sty1922}, \href
  {https://ui.adsabs.harvard.edu/abs/2018MNRAS.480..759R} {480, 759}

\bibitem[\protect\citeauthoryear{{Ryan}, {Chen}  \& {Ratra}}{{Ryan}
  et~al.}{2019}]{Ryanetal2019}
{Ryan} J.,  {Chen} Y.,   {Ratra} B.,  2019, \mn@doi [\mnras]
  {10.1093/mnras/stz1966}, \href
  {https://ui.adsabs.harvard.edu/abs/2019MNRAS.488.3844R} {488, 3844}

\bibitem[\protect\citeauthoryear{{Samushia} \& {Ratra}}{{Samushia} \&
  {Ratra}}{2010}]{samushia_ratra_2010}
{Samushia} L.,  {Ratra} B.,  2010, \mn@doi [\apj]
  {10.1088/0004-637X/714/2/1347}, \href
  {http://adsabs.harvard.edu/abs/2010ApJ...714.1347S} {714, 1347}

\bibitem[\protect\citeauthoryear{{Sangwan}, {Tripathi}  \& {Jassal}}{{Sangwan}
  et~al.}{2018}]{Sangwanetal2018}
{Sangwan} A.,  {Tripathi} A.,   {Jassal} H.~K.,  2018, preprint, \href
  {https://ui.adsabs.harvard.edu/abs/2018arXiv180409350S} {} (\mn@eprint {}
  {1804.09350})

\bibitem[\protect\citeauthoryear{{Scolnic} et~al.,}{{Scolnic}
  et~al.}{2018}]{scolnic_et_al_2018}
{Scolnic} D.~M.,  et~al., 2018, \mn@doi [\apj] {10.3847/1538-4357/aab9bb},
  \href {http://adsabs.harvard.edu/abs/2018ApJ...859..101S} {859, 101}

\bibitem[\protect\citeauthoryear{{Singh}, {Sangwan}  \& {Jassal}}{{Singh}
  et~al.}{2019}]{Singhetal2019}
{Singh} A.,  {Sangwan} A.,   {Jassal} H.~K.,  2019, \mn@doi [\jcap]
  {10.1088/1475-7516/2019/04/047}, \href
  {https://ui.adsabs.harvard.edu/abs/2019JCAP...04..047S} {2019, 047}

\bibitem[\protect\citeauthoryear{{Sinha} \& {Banerjee}}{{Sinha} \&
  {Banerjee}}{2021}]{SinhaBanerjee2021}
{Sinha} S.,  {Banerjee} N.,  2021, \mn@doi [\jcap]
  {10.1088/1475-7516/2021/04/060}, \href
  {https://ui.adsabs.harvard.edu/abs/2021JCAP...04..060S} {2021, 060}

\bibitem[\protect\citeauthoryear{{Sol{\`a} Peracaula}, {G{\'o}mez-Valent}  \&
  {de Cruz P{\'e}rez}}{{Sol{\`a} Peracaula}
  et~al.}{2019}]{SolaPercaulaetal2019}
{Sol{\`a} Peracaula} J.,  {G{\'o}mez-Valent} A.,   {de Cruz P{\'e}rez} J.,
  2019, \mn@doi [Physics of the Dark Universe] {10.1016/j.dark.2019.100311},
  \href {https://ui.adsabs.harvard.edu/abs/2019PDU....25..311S} {25, 100311}

\bibitem[\protect\citeauthoryear{{Ure{\~n}a-L{\'o}pez} \&
  {Roy}}{{Ure{\~n}a-L{\'o}pez} \& {Roy}}{2020}]{UrenaLopezRoy2020}
{Ure{\~n}a-L{\'o}pez} L.~A.,  {Roy} N.,  2020, \mn@doi [\prd]
  {10.1103/PhysRevD.102.063510}, \href
  {https://ui.adsabs.harvard.edu/abs/2020PhRvD.102f3510U} {102, 063510}

\bibitem[\protect\citeauthoryear{{Vagnozzi}, {Di Valentino}, {Gariazzo},
  {Melchiorri}, {Mena}  \& {Silk}}{{Vagnozzi} et~al.}{2021a}]{Vagnozzietal2020}
{Vagnozzi} S.,  {Di Valentino} E.,  {Gariazzo} S.,  {Melchiorri} A.,  {Mena}
  O.,   {Silk} J.,  2021a, \mn@doi [Physics of the Dark Universe]
  {10.1016/j.dark.2021.100851}, \href
  {https://ui.adsabs.harvard.edu/abs/2021PDU....3300851V} {33, 100851}

\bibitem[\protect\citeauthoryear{{Vagnozzi}, {Loeb}  \& {Moresco}}{{Vagnozzi}
  et~al.}{2021b}]{Vagnozzietal2021}
{Vagnozzi} S.,  {Loeb} A.,   {Moresco} M.,  2021b, \mn@doi [\apj]
  {10.3847/1538-4357/abd4df}, \href
  {https://ui.adsabs.harvard.edu/abs/2021ApJ...908...84V} {908, 84}

\bibitem[\protect\citeauthoryear{{Wang}, {Dai}  \& {Liang}}{{Wang}
  et~al.}{2015}]{Wangetal2015}
{Wang} F.~Y.,  {Dai} Z.~G.,   {Liang} E.~W.,  2015, \mn@doi [\nar]
  {10.1016/j.newar.2015.03.001}, \href
  {https://ui.adsabs.harvard.edu/abs/2015NewAR..67....1W} {67, 1}

\bibitem[\protect\citeauthoryear{{Wang}, {Wang}, {Cheng}  \& {Dai}}{{Wang}
  et~al.}{2016}]{Wang_2016}
{Wang} J.~S.,  {Wang} F.~Y.,  {Cheng} K.~S.,   {Dai} Z.~G.,  2016, \mn@doi
  [\aap] {10.1051/0004-6361/201526485}, \href
  {https://ui.adsabs.harvard.edu/abs/2016A&A...585A..68W} {585, A68}

\bibitem[\protect\citeauthoryear{{Wang}, {Hu}, {Zhang}  \& {Dai}}{{Wang}
  et~al.}{2022}]{Wangetal_2021}
{Wang} F.~Y.,  {Hu} J.~P.,  {Zhang} G.~Q.,   {Dai} Z.~G.,  2022, \mn@doi [\apj]
  {10.3847/1538-4357/ac3755}, \href
  {https://ui.adsabs.harvard.edu/abs/2022ApJ...924...97W} {924, 97}

\bibitem[\protect\citeauthoryear{{Wei}}{{Wei}}{2018}]{Wei2018}
{Wei} J.-J.,  2018, \mn@doi [\apj] {10.3847/1538-4357/aae696}, \href
  {https://ui.adsabs.harvard.edu/abs/2018ApJ...868...29W} {868, 29}

\bibitem[\protect\citeauthoryear{{Willingale} et~al.,}{{Willingale}
  et~al.}{2007}]{Willingaleetal2007}
{Willingale} R.,  et~al., 2007, \mn@doi [\apj] {10.1086/517989}, \href
  {https://ui.adsabs.harvard.edu/abs/2007ApJ...662.1093W} {662, 1093}

\bibitem[\protect\citeauthoryear{{Willingale}, {Genet}, {Granot}  \&
  {O'Brien}}{{Willingale} et~al.}{2010}]{Willingaleetal2010}
{Willingale} R.,  {Genet} F.,  {Granot} J.,   {O'Brien} P.~T.,  2010, \mn@doi
  [\mnras] {10.1111/j.1365-2966.2009.16187.x}, \href
  {https://ui.adsabs.harvard.edu/abs/2010MNRAS.403.1296W} {403, 1296}

\bibitem[\protect\citeauthoryear{{Xu}, {Chen}, {Xu}  \& {Cao}}{{Xu}
  et~al.}{2021}]{Xuetal2021}
{Xu} T.,  {Chen} Y.,  {Xu} L.,   {Cao} S.,  2021, preprint, \href
  {https://ui.adsabs.harvard.edu/abs/2021arXiv210902453X} {} (\mn@eprint {}
  {2109.02453})

\bibitem[\protect\citeauthoryear{{Yang}, {Banerjee}  \& {{\'O}
  Colg{\'a}in}}{{Yang} et~al.}{2020}]{Yangetal2020}
{Yang} T.,  {Banerjee} A.,   {{\'O} Colg{\'a}in} E.,  2020, \mn@doi [\prd]
  {10.1103/PhysRevD.102.123532}, \href
  {https://ui.adsabs.harvard.edu/abs/2020PhRvD.102l3532Y} {102, 123532}

\bibitem[\protect\citeauthoryear{{Yu}, {Ratra}  \& {Wang}}{{Yu}
  et~al.}{2018}]{Yuetal2018}
{Yu} H.,  {Ratra} B.,   {Wang} F.-Y.,  2018, \mn@doi [\apj]
  {10.3847/1538-4357/aab0a2}, \href
  {https://ui.adsabs.harvard.edu/abs/2018ApJ...856....3Y} {856, 3}

\bibitem[\protect\citeauthoryear{{Yu} et~al.,}{{Yu} et~al.}{2021}]{Yuetal2021}
{Yu} Z.,  et~al., 2021, \mn@doi [\mnras] {10.1093/mnras/stab2244}, \href
  {https://ui.adsabs.harvard.edu/abs/2021MNRAS.507.3771Y} {507, 3771}

\bibitem[\protect\citeauthoryear{{Zaja{\v{c}}ek} et~al.,}{{Zaja{\v{c}}ek}
  et~al.}{2021}]{Zajaceketal2021}
{Zaja{\v{c}}ek} M.,  et~al., 2021, \mn@doi [\apj] {10.3847/1538-4357/abe9b2},
  \href {https://ui.adsabs.harvard.edu/abs/2021ApJ...912...10Z} {912, 10}

\bibitem[\protect\citeauthoryear{{Zhai}, {Blanton}, {Slosar}  \&
  {Tinker}}{{Zhai} et~al.}{2017}]{Zhaietal2017}
{Zhai} Z.,  {Blanton} M.,  {Slosar} A.,   {Tinker} J.,  2017, \mn@doi [\apj]
  {10.3847/1538-4357/aa9888}, \href
  {https://ui.adsabs.harvard.edu/abs/2017ApJ...850..183Z} {850, 183}

\bibitem[\protect\citeauthoryear{{Zhang}, {Zhang}, {Yuan}, {Liu}, {Zhang}  \&
  {Sun}}{{Zhang} et~al.}{2014}]{73}
{Zhang} C.,  {Zhang} H.,  {Yuan} S.,  {Liu} S.,  {Zhang} T.-J.,   {Sun} Y.-C.,
  2014, \mn@doi [Research in Astronomy and Astrophysics]
  {10.1088/1674-4527/14/10/002}, \href
  {http://adsabs.harvard.edu/abs/2014RAA....14.1221Z} {14, 1221}

\bibitem[\protect\citeauthoryear{{Zhao} \& {Xia}}{{Zhao} \&
  {Xia}}{2021}]{ZhaoXia2021}
{Zhao} D.,  {Xia} J.-Q.,  2021, \mn@doi [European Physical Journal C]
  {10.1140/epjc/s10052-021-09491-0}, \href
  {https://ui.adsabs.harvard.edu/abs/2021EPJC...81..694Z} {81, 694}

\bibitem[\protect\citeauthoryear{{Zheng}, {Cao}, {Biesiada}, {Li}, {Liu}  \&
  {Liu}}{{Zheng} et~al.}{2021}]{Zhengetal2021}
{Zheng} X.,  {Cao} S.,  {Biesiada} M.,  {Li} X.,  {Liu} T.,   {Liu} Y.,  2021,
  \mn@doi [Science China Physics, Mechanics, and Astronomy]
  {10.1007/s11433-020-1664-9}, \href
  {https://ui.adsabs.harvard.edu/abs/2021SCPMA..6459511Z} {64, 259511}

\makeatother
\end{thebibliography}




\appendix

\section{Platinum GRB data}
\label{sec:appendix}

\begin{table*}
\centering
\setlength{\tabcolsep}{3mm}{
\begin{threeparttable}
\caption{50 Platinum GRB samples.}
\label{tab:P50}
\begin{tabular}{lccccc}
\toprule
GRB & $z$ & $\log T^{*}_{X}\ (\mathrm{s})$ & $\log F_{X}\ (\mathrm{erg\ cm}^{-2}\ \mathrm{s}^{-1})$ & $\beta^{\prime}$ & $F_{\rm peak}\ (10^{-8}\ \mathrm{erg\ cm}^{-2}\ \mathrm{s}^{-1})$ \\
\midrule

060418 & 1.49 & $3.11887^{+0.05725}_{-0.05724}$ & $-9.79296\pm0.04904$ & 1.98 & $49.9\pm1.63$\\
060605 & 3.8 & $4.04162^{+0.03738}_{-0.03737}$ & $-11.2609^{+0.0531}_{-0.0532}$ & 1.835 & $4.73\pm0.693$\\
060708 & 1.92 & $3.50569^{+0.06273}_{-0.06272}$ & $-10.8688^{+0.0611}_{-0.0612}$ & 2.485 & $6.89\pm0.796$\\
060714 & 2.71 & $3.73449^{+0.04676}_{-0.04677}$ & $-10.8586\pm0.0352$ & 1.87 & $9.13\pm0.549$\\
060814 & 0.84 & $4.22387^{+0.03485}_{-0.03486}$ & $-10.9108\pm0.0353$ & 1.97 & $60.6\pm1.48$\\
060906 & 3.685 & $4.33479^{+0.07214}_{-0.07215}$ & $-11.8833^{+0.0885}_{-0.0884}$ & 2.06 & $12.2\pm1.18$\\
061121 & 1.314 & $3.78753\pm0.01703$ & $-10.0299^{+0.0126}_{-0.0127}$ & 1.9485 & $196\pm2.43$\\
061222A & 2.088 & $3.92091^{+0.02444}_{-0.02443}$ & $-9.95369\pm0.02311$ & 1.86 & $73.3\pm1.51$\\
070110 & 2.352 & $4.26825^{+0.05399}_{-0.05400}$ & $-11.0488^{+0.0547}_{-0.0548}$ & 2.186 & $4.73\pm0.648$\\
070306 & 1.4959 & $4.86926\pm0.02824$ & $-11.2737\pm0.0335$ & 1.831 & $20.1\pm0.899$\\
070508 & 0.82 & $3.02525\pm0.01189$ & $-9.16118^{+0.01080}_{-0.01079}$ & 1.764 & $224\pm3.07$\\
070521 & 0.553 & $3.55243\pm0.04844$ & $-10.01250^{+0.04902}_{-0.04900}$ & 1.98 & $8.71\pm1.2$\\
070529 & 2.4996 & $3.09323^{+0.06805}_{-0.06804}$ & $-10.2468\pm0.0552$ & 1.76 & $11.1\pm1.84$\\
080310 & 2.4266 & $4.326080559^{+0.037986760}_{-0.037986759}$ & $-11.39909559\pm0.04247343$ & 1.878 & $7.52\pm0.893$\\
080430 & 0.767 & $4.217629106^{+0.038388523}_{-0.038388522}$ & $-11.17223461^{+0.02818096}_{-0.02818097}$ & 4.18 & $18.2\pm0.808$\\
080721 & 2.6 & $2.878762860^{+0.007554815}_{-0.007554816}$ & $-8.713485538^{+0.007019298}_{-0.007019299}$ & 1.735 & $193\pm10.3$\\
081008 & 1.967 & $3.855508334\pm0.056285684$ & $-10.79325557\pm0.06052989$ & 1.84 & $10.5\pm0.813$\\
081221 & 2.26 & $2.931957486\pm0.031176651$ & $-9.354590372\pm0.024985500$ & 1.991 & $147\pm2.46$\\
090418A & 1.608 & $3.528136567^{+0.030759021}_{-0.030759020}$ & $-10.04734554^{+0.02865941}_{-0.02865940}$ & 1.98 & $15.8\pm1.66$\\
091018 & 0.971 & $2.896555345^{+0.034497877}_{-0.034497878}$ & $-9.626765539^{+0.024288604}_{-0.024288603}$ & 1.91 & $59.1\pm1.22$\\
091020 & 1.71 & $2.952650108\pm0.045271255$ & $-9.712875280\pm0.034480687$ & 1.895 & $36.4\pm1.71$\\
091029 & 2.752 & $4.303689975^{+0.034224003}_{-0.034224002}$ & $-11.34682089\pm0.02563352$ & 2.064 & $10.9\pm0.786$\\
100219A & 4.7 & $4.721899099\pm0.103773463$ & $-12.13530676^{+0.32639562}_{-0.32639563}$ & 1.46 & $3.15\pm0.721$\\
110213A & 1.46 & $3.84089^{+0.02576}_{-0.02127}$ & $-9.89698^{+0.03865}_{-0.03311}$ & 2.1905 & $7.15\pm1.88$\\
110818A & 3.36 & $3.846274153\pm0.059648416$ & $-11.28292779\pm0.05514905$ & 1.83 & $14.1\pm1.5$\\
111008A & 5 & $3.989933926\pm0.035676696$ & $-10.66765993^{+0.02851108}_{-0.02851109}$ & 1.829 & $54.1\pm3.91$\\
120118B & 2.943 & $3.68307^{+0.15437}_{-0.10916}$ & $-10.9560^{+0.1224}_{-0.1120}$ & 2.01 & $13.8\pm1.32$\\
120404A & 2.88 & $3.79948^{+0.09188}_{-0.04469}$ & $-11.0269^{+0.0927}_{-0.1414}$ & 1.69 & $7.62\pm0.109$\\
120811C & 2.67 & $3.159095907^{+0.084200302}_{-0.084200301}$ & $-10.14322956\pm0.05294686$ & 1.65 & $26\pm1.1$\\
120922A & 3.1 & $3.59799^{+0.21328}_{-0.05347}$ & $-10.5957^{+0.1158}_{-0.2115}$ & 2.17 & $10.6\pm0.757$\\
121128A & 2.2 & $3.233929220^{+0.029795567}_{-0.029795568}$ & $-9.618411717^{+0.028690024}_{-0.028690023}$ & 1.9455 & $104\pm2.1$\\
131030A & 1.29 & $2.84594^{+0.02758}_{-0.02757}$ & $-9.29414^{+0.02517}_{-0.02518}$ & 1.693 & $265\pm4.61$\\
131105A & 1.686 & $3.94880^{+0.04788}_{-0.04789}$ & $-10.9111^{+0.0333}_{-0.0334}$ & 1.94 & $26.8\pm0.204$\\
140206A & 2.7 & $3.59073^{+0.02132}_{-0.02133}$ & $-9.72980^{+0.01300}_{-0.01299}$ & 1.672 & $169\pm2.62$\\
140419A & 3.956 & $3.68383\pm0.02442$ & $-10.00560^{+0.02397}_{-0.02400}$ & 1.678 & $41.8\pm1.28$\\
140506A & 0.889 & $3.30686\pm0.06882$ & $-9.90438\pm0.05516$ & 1.9 & $86.7\pm4.56$\\
140509A & 2.4 & $3.58709^{+0.11327}_{-0.06416}$ & $-11.0977^{+0.0897}_{-0.0782}$ & 1.86 & $11.8\pm1.77$\\
140629A & 2.3 & $2.85608\pm0.06161$ & $-9.84891^{+0.04278}_{-0.04279}$ & 1.815 & $35\pm1.87$\\
150314A & 1.758 & $2.49217^{+0.01586}_{-0.01585}$ & $-8.60990^{+0.01503}_{-0.01502}$ & 1.73 & $381\pm6.06$\\
150403A & 2.06 & $3.19694^{+0.00698}_{-0.00697}$ & $-8.82118^{+0.00510}_{-0.00511}$ & 1.679 & $171\pm3.81$\\
150910A & 1.36 & $3.85419\pm0.02868$ & $-9.97065^{+1.37578}_{-0.03215}$ & 1.6645 & $8.28\pm2.18$\\
151027A & 0.81 & $3.99907\pm0.01731$ & $-9.93519\pm0.02221$ & 1.9235 & $58.2\pm2.88$\\
160121A & 1.96 & $3.82440^{+0.12038}_{-0.12039}$ & $-11.1726\pm0.0694$ & 2.02 & $7.79\pm0.823$\\
160227A & 2.38 & $4.47751\pm0.03885$ & $-10.9957^{+0.0305}_{-0.0306}$ & 1.679 & $4.88\pm0.688$\\
160327A & 4.99 & $3.76407\pm0.06033$ & $-11.2238\pm0.0673$ & 1.78 & $12.5\pm0.877$\\
170202A & 3.645 & $3.77629^{+0.06866}_{-0.06865}$ & $-10.6535^{+0.0449}_{-0.0448}$ & 2.04 & $39.2\pm1.79$\\
170705A & 2.01 & $3.63746\pm0.06603$ & $-10.1867^{+0.0406}_{-0.0405}$ & 1.66 & $113\pm2.41$\\
180329B & 1.998 & $4.02296\pm0.04314$ & $-11.1493^{+0.0483}_{-0.0482}$ & 1.77 & $9.09\pm2.02$\\
190106A & 1.86 & $4.19124^{+0.03109}_{-0.03110}$ & $-10.4325^{+0.0285}_{-0.0286}$ & 2.9365 & $33.6\pm1.3$\\
190114A & 3.37 & $3.74534^{+0.04221}_{-0.04222}$ & $-10.7527^{+0.0336}_{-0.0337}$ & 1.86 & $4.12\pm0.953$\\
\bottomrule
\end{tabular}
\end{threeparttable}%
}
\end{table*}



\bsp	
\label{lastpage}
\end{document}